\documentclass[letterpaper,twocolumn,10pt]{article}

\usepackage{usenix-2020-09}
\usepackage{tikz}
\usepackage{amsmath}

\usepackage{multirow}
\usepackage{xcolor}
\usepackage{colortbl}
\usepackage{xspace}
\usepackage{caption}
\usepackage{subcaption}
\usepackage[linesnumbered,vlined,ruled,commentsnumbered]{algorithm2e}
\usepackage{mathtools}
\usepackage{tabu}
\usepackage{graphicx}
\usepackage[framemethod=TikZ]{mdframed}
\usepackage[frozencache,cachedir=.]{minted}  
\definecolor{LightGray}{gray}{0.9}
\usepackage{soul}
\usepackage{enumitem}
\usepackage{authblk}

\newcommand{\oursystemsoze}{{S\"{o}ze}\xspace}

\newcommand*\circled[1]{\tikz[baseline=(char.base)]{
            \node[shape=circle,fill,inner sep=0.5pt] (char) {\textcolor{white}{#1}};}}

\newcommand{\secref}[1]{{\S\ref{#1}}}
\newcommand{\fref}[1]{{Figure~\ref{#1}}}

\newcommand{\eref}[1]{{Equation~\ref{#1}}}

\newcounter{defn}[section]
\renewcommand{\thedefn}{\arabic{section}.\arabic{defn}}
\newenvironment{defn}[2][]{%
\refstepcounter{defn}%
\ifstrempty{#1}%
{\mdfsetup{%
frametitle={%
\tikz[baseline=(current bounding box.east),outer sep=0pt]
\node[anchor=east,rectangle,fill=purple!20]
{\strut Definition~\thedefn};}}
}%
{\mdfsetup{%
frametitle={%
\tikz[baseline=(current bounding box.east),outer sep=0pt]
\node[anchor=east,rectangle,fill=purple!20]
{\strut Definition~\thedefn:~#1};}}%
}%
\mdfsetup{innertopmargin=2pt,linecolor=purple!20,%
linewidth=2pt,topline=true,%
frametitleaboveskip=\dimexpr-\ht\strutbox\relax
}
\begin{mdframed}[]\relax%
\label{#2}}{\end{mdframed}}

\newcounter{theo}[section]
\renewcommand{\thetheo}{\arabic{section}.\arabic{theo}}
\newenvironment{theo}[2][]{%
\refstepcounter{theo}%
\ifstrempty{#1}%
{\mdfsetup{%
frametitle={%
\tikz[baseline=(current bounding box.east),outer sep=0pt]
\node[anchor=east,rectangle,fill=blue!20]
{\strut Theorem~\thetheo};}}
}%
{\mdfsetup{%
frametitle={%
\tikz[baseline=(current bounding box.east),outer sep=0pt]
\node[anchor=east,rectangle,fill=blue!20]
{\strut Theorem~\thetheo:~#1};}}%
}%
\mdfsetup{innertopmargin=2pt,linecolor=blue!20,%
linewidth=2pt,topline=true,%
frametitleaboveskip=\dimexpr-\ht\strutbox\relax
}
\begin{mdframed}[]\relax%
\label{#2}}{\end{mdframed}}

\newcounter{lem}[section]
\renewcommand{\thelem}{\arabic{section}.\arabic{lem}}
\newenvironment{lem}[2][]{%
\refstepcounter{lem}%
\ifstrempty{#1}%
{\mdfsetup{%
frametitle={%
\tikz[baseline=(current bounding box.east),outer sep=0pt]
\node[anchor=east,rectangle,fill=green!20]
{\strut Lemma~\thelem};}}
}%
{\mdfsetup{%
frametitle={%
\tikz[baseline=(current bounding box.east),outer sep=0pt]
\node[anchor=east,rectangle,fill=green!20]
{\strut Lemma~\thelem:~#1};}}%
}%
\mdfsetup{innertopmargin=2pt,linecolor=green!20,%
linewidth=2pt,topline=true,%
frametitleaboveskip=\dimexpr-\ht\strutbox\relax
}
\begin{mdframed}[]\relax%
\label{#2}}{\end{mdframed}}

\newcounter{prf}[section]
\renewcommand{\theprf}{\arabic{section}.\arabic{prf}}

\newcounter{defnapp}[section]
\renewcommand{\thedefnapp}{}  
\newenvironment{defnapp}[2][]{%
\refstepcounter{defnapp}%
\ifstrempty{#1}%
{\mdfsetup{%
frametitle={%
\tikz[baseline=(current bounding box.east),outer sep=0pt]
\node[anchor=east,rectangle,fill=purple!20]
{\strut Definition~\thedefnapp};}}
}%
{\mdfsetup{%
frametitle={%
\tikz[baseline=(current bounding box.east),outer sep=0pt]
\node[anchor=east,rectangle,fill=purple!20]
{\strut Definition~\thedefnapp:~#1};}}%
}%
\mdfsetup{innertopmargin=2pt,linecolor=purple!20,%
linewidth=2pt,topline=true,%
frametitleaboveskip=\dimexpr-\ht\strutbox\relax
}
\begin{mdframed}[]\relax%
\label{#2}}{\end{mdframed}}

\newcounter{theoapp}[section]
\renewcommand{\thetheoapp}{}
\newenvironment{theoapp}[2][]{%
\refstepcounter{theoapp}%
\ifstrempty{#1}%
{\mdfsetup{%
frametitle={%
\tikz[baseline=(current bounding box.east),outer sep=0pt]
\node[anchor=east,rectangle,fill=blue!20]
{\strut Theorem~\thetheoapp};}}
}%
{\mdfsetup{%
frametitle={%
\tikz[baseline=(current bounding box.east),outer sep=0pt]
\node[anchor=east,rectangle,fill=blue!20]
{\strut Theorem~\thetheoapp:~#1};}}%
}%
\mdfsetup{innertopmargin=2pt,linecolor=blue!20,%
linewidth=2pt,topline=true,%
frametitleaboveskip=\dimexpr-\ht\strutbox\relax
}
\begin{mdframed}[]\relax%
\label{#2}}{\end{mdframed}}

\newcounter{lemapp}[section]
\renewcommand{\thelemapp}{}  
\newenvironment{lemapp}[2][]{%
\refstepcounter{lemapp}%
\ifstrempty{#1}%
{\mdfsetup{%
frametitle={%
\tikz[baseline=(current bounding box.east),outer sep=0pt]
\node[anchor=east,rectangle,fill=green!20]
{\strut Lemma~\thelemapp};}}
}%
{\mdfsetup{%
frametitle={%
\tikz[baseline=(current bounding box.east),outer sep=0pt]
\node[anchor=east,rectangle,fill=green!20]
{\strut Lemma~\thelemapp:~#1};}}%
}%
\mdfsetup{innertopmargin=2pt,linecolor=green!20,%
linewidth=2pt,topline=true,%
frametitleaboveskip=\dimexpr-\ht\strutbox\relax
}
\begin{mdframed}[]\relax%
\label{#2}}{\end{mdframed}}

\newcounter{prfapp}[section]
\renewcommand{\theprfapp}{}  

\begin{document}

\date{}



\title{\oursystemsoze: One Network Telemetry Is All You Need \\ for Per-flow Weighted Bandwidth Allocation at Scale}

\author{Weitao Wang \qquad \qquad T. S. Eugene Ng}
\affil{Rice University}

\pagestyle{empty}

\maketitle

\begin{abstract}



Weighted bandwidth allocation is a powerful abstraction that has a wide range of use cases in modern data center networks. However, realizing highly agile and precise weighted bandwidth allocation for large-scale cloud environments is fundamentally challenging. In this paper, we propose S\"{o}ze, a lightweight decentralized weighted bandwidth allocation system that leverages simple network telemetry features of commodity Ethernet switches. Given the flow weights, S\"{o}ze can effectively use the telemetry information to compute and enforce the weighted bandwidth allocations without per-flow, topology, or routing knowledge. We demonstrate the effectiveness of S\"{o}ze through simulations and testbed experiments, improving TPC-H jobs completion time by up to $0.59\times$ and $0.79\times$ on average.

\end{abstract}

\section{Introduction} \label{sec:introduction}
\vspace{-0.1in}
As a fundamental building block of the modern data center, the transport layer provides accurate and reliable data delivery between machines. 
Transport solutions in use by data center networks today~\cite{zhu2015congestion,alizadeh2010data,kumar2020swift,li2019hpcc} aim to achieve a fixed goal, where the bandwidth resource sharing is generally fair and the actual allocation only depends on the traffic pattern.
Unfortunately, since modern cloud applications' performance goals~\cite{gpugaming,tpugoogle,tpugoogle2,grosvenor2015queues, nishtala2013scaling} for data transmissions are diverse and may change over time~\cite{roy2015inside}, such a rigid resource allocation strategy clearly cannot suit every application's need perfectly.

To better serve the various and evolving applications, weighted allocation can be a powerful abstraction~\cite{oudghiri1992global, ma2014resource}.
For example, when a flow with weight 3 shares a bottleneck link with another flow with weight 1, the bandwidth is allocated 75\% and 25\%.
    Weighted allocation has many potential use cases. For example, important jobs or latency-sensitive jobs can be prioritized for shorter waiting time~\cite{gujarati2020serving, casini2020predictable}; within a job, resources can be prioritized for critical paths to reduce the job completion time~\cite{ramezani2021dynamic}; service-level objectives can be achieved by carefully assigning weights~\cite{nagaraj2016numfabric}.
    

Although weighted bandwidth allocation can be extremely helpful and provide unique benefits for applications~\cite{oudghiri1992global, ma2014resource}, the reason that blocks its wide deployment is the scale of the cloud data center~\cite{abts2022high, barroso2022datacenter}. 
Realizing the flow rates that conform to a weighted bandwidth allocation policy at scale is challenging.
    Existing systems try to implement and enforce the policies through either packet scheduling~\cite{demers1989analysis,shreedhar1995efficient} on switches or using a logically centralized bandwidth allocator together with rate shapers~\cite{kumar2015bwe,shieh2011sharing}. 
        On one hand, packet scheduling-based implementation is limited by the small number of per-flow queues and the coarse granularity of weight parameters supported by switch hardware~\cite{moon2000scalable,yu2021programmable}.
        On the other hand, a logically centralized bandwidth allocator-based implementation has high communication latency to gather and dispatch information and high computation cost to calculate the optimal allocation~\cite{kumar2015bwe,guo2010secondnet}.
    Thus, realizing weighted bandwidth allocation for cloud networks with fine granularity, high agility, and high scalability remains an open challenge.

In this paper, we provide all the above properties for weighted bandwidth allocation with our approach called \oursystemsoze.
    Our insight is that we need only one network telemetry along a flow's path, regardless of the number of hops, to solve the weighted bandwidth allocation problem at the bottleneck. Furthermore, instead of using this telemetry to gauge congestion levels like in traditional use cases of network telemetry, this telemetry is cleverly repurposed to signal the correct weighted fair share at the bottleneck.

With \oursystemsoze, every flow can be associated with a weight regardless of the number of flows or the size and topology of the network (scalability, generality), each flow's weight can be fine grain adjusted instantaneously at the host (granularity, flexibility), and the weighted bandwidth allocation is realized in a matter of several RTTs (agility). These capabilities provide a building block for supporting various application scenarios such as critical path prioritization, coflow straggler mitigation, intelligent bandwidth sharing across jobs, altruistic bandwidth sharing, shortest flow prioritization, etc.

Our contributions can be summarized as follows:

\begin{itemize}[noitemsep, topsep=0pt]
    \item \oursystemsoze shows that one network telemetry is enough to reflect the network sharing status and serve as a channel to coordinate multiple flow senders, thus offering an effective control knob for weighted bandwidth allocation. 

    \item \oursystemsoze provides a novel decentralized framework to enforce the weighted max-min fair allocation for different flows across the network with high precision and agility, without per-flow, topology, or routing information.


    \item We show that \oursystemsoze can be seamlessly incorporated into existing transport layer mechanisms such as TCP and eRPC. In the evaluation, we show that \oursystemsoze reduces the job completion time for the TPC-H benchmark up to $0.59\times$ and $0.79\times$ on average. 

\end{itemize}

This paper is organized as follows. 
    \secref{sec:motivation} motivates weighted bandwidth allocation and discusses challenges in designing an ideal weighted bandwidth allocation system in a large-scale cloud. 
    \secref{sec:design} introduces our proposal \oursystemsoze with thorough design details and proves that leveraging the network telemetry feature can achieve highly accurate weighted bandwidth allocation.  
    \secref{sec:implementation} and \secref{sec:evaluation} provide implementation details and extensive experiments to demonstrate the effectiveness of \oursystemsoze. \secref{sec:relatedworks} discusses related work and we conclude in \secref{sec:conclusion}.

\section{Motivation}
\label{sec:motivation}
In this section, 
    we firstly introduce the benefits of weighted bandwidth allocation for the network (\secref{subsec:motivation1}); 
    Next, we list the essential requirements for an efficient and powerful weighted allocation solution (\secref{subsec:motivation2}); 
    Lastly, we motivate that the in-network telemetry (INT) can be a low-cost and efficient solution to achieve weighted bandwidth allocation (\secref{subsec:motivation4}).

\subsection{Weighted Bandwidth Allocation}
\label{subsec:motivation1}

Instead of a rigid bandwidth allocation strategy, weighted bandwidth allocation can give modern applications extra flexibility to adapt the underlying bandwidth allocation intentionally for better performance, which benefits many applications, like ML training, Spark/Hadoop, databases, and RPCs.
    With the increasing demand for giant applications~\cite{achiam2023gpt, touvron2023llama} and low-cost function services~\cite{villamizar2016infrastructure, gos2020comparison}, modern cloud applications tend to rely on the data center network for efficient data exchange under distributed execution. 
        Take distributed training with data parallelism as an example, the tensors at different layers can be assigned different priorities for accessing the bandwidth resource, in order to overlap the communication and computation process and accelerate the overall training process \cite{jayarajan2019priority,hashemi2019tictac,bytescheduler}. 
        For map-reduce applications such as Spark and Hadoop, prioritizing the straggler flow over others during the shuffle operation is a common technique to shrink the job competition time \cite{chowdhury2014efficient, chowdhury2015efficient}. 


\subsection{Challenges}
\label{subsec:motivation2}

As the key infrastructure for interconnecting hardware in the cloud, one would expect the network to provide efficient and fine-grained weighted bandwidth allocation services to meet the various requirements from the applications. 
    However, networks in the production environment only provide coarse-grained differentiation, such as traffic aggregate classes with bandwidth reservations or strict priorities.
    The challenges in providing efficient and fine-grained weighted bandwidth allocation services for the network come from multiple aspects:

\textbf{Bottleneck resource recognition.}
    Unlike many other computational and storage resource, the data center network has multiple layers and each layer contains multiple alternative connections, where the routing of each flow is determined with random seeds and is difficult to predict. 
    For the flow that utilizes such networks, its path will travel multiple hops where every hop has a bandwidth capacity. On different hops, the flow will contend and share the bandwidth resources with an unpredictable group of other flows, but only one of the hops will become the bottleneck and limit its sending rate.
    Recognizing the bottleneck resource is therefore crucial for determining the weighted allocation of bandwidth resource but the bottleneck resource is highly unpredictable.

\textbf{High scalability with large network size and high concurrency.}
    Besides recognizing bottleneck resources, the intended system size adds another dimension of challenge: 
        1) Massive information: the amount of information we can collect from the network is massive, including the network topology, link bandwidth, and each flow's sender/receiver/routing. How to filter and pick the most useful information and reduce the input to the algorithm can be challenging.
        2) Global optimality: since the weighted bandwidth allocation needs to be enforced at the bottleneck resource to be effective, how to reach this global optimal state with low cost and consistently fast solving time is also challenging.

\textbf{Fine-grained weighted allocation.}
To support fine-grained policy changes, the weighted bandwidth allocation system should accommodate flow weight with fine granularity and enforce the desired allocation with high precision. 
    However, due to the network's scale, fine granularity and high accuracy are hard to obtain cost-effectively through high-precision computations in switch ASICs. Fine-grained weighted bandwidth enforcement must require careful algorithm design instead of relying on switch hardware capability.

\textbf{High agility in updates and changes.}
A weighted bandwidth allocation policy needs to be realized quickly, but this is challenging because the network environment dynamics can affect the allocations, and the solution must react quickly.
    Firstly, the network condition may change due to updates, like new flow arrivals, existing flow finishes, link failures, and routing changes.
    Secondly, the application may update the task weight based on the runtime information or user input. 
\subsection{Use Telemetry for Coordination}
\label{subsec:motivation4}

In-network telemetry (INT) is a common feature in switch ASICs today~\cite{INTbare, INTtomo, INTtri} that enables better network visibility by inserting the fine-grained switch-local information (e.g., queueing delay, timestamp, transmitted bytes, etc.) into the packet headers.
    The typical usage for the INT data is network monitoring, informing the network operator
    about the current network status, such as the queue depth or the link utilization. INT data is quantitative so that it can describe the status of the network accurately in a timely manner. 
In this paper, we want to use the INT feature to help achieve weighted bandwidth allocation. To start, we introduce the INT with a realistic example and demonstrate why the INT function can be used for information exchange and coordination.
    


Typical INT data include queueing level, link utilization, etc. As a concrete example, let us use queueing delay for demonstration:
    1) Multiple senders create an incast scenario on one link, and this link collects the queueing delay as the INT and informs all senders through packet headers. 
    2) The queueing delay may change if any sender changes its sending rate, but the actual change of queueing depends on the behavior of all senders together. 
    3) All flow senders will have the same observation about the queueing delay, such as the queueing increases if every sender increases its rate. 
2) and 3) imply that if flow senders change their behavior, the changes may be reflected in the queueing delay, and all the senders will sense the change at the same time. In this way, an information exchange channel can be constructed using queueing delay. 
    Thus, we argue for a distinctly different usage for INT: using passive INT data to coordinate different hosts.

Such an information exchange channel through INT data has many unique benefits. 
    First, the in-network telemetry feature on switches is simple and low-cost compared to other types of information exchange, which preserves deployability and scalability. 
    Second, the in-network telemetry data is inherently quantitative, which can be used to achieve fine-grained goals with precision. 
    Thirdly, the in-network telemetry feature can operate on a per packet granularity at full line rate, therefore timely information exchange can be achieved at very high speed. 
    %
All of the above properties of in-network telemetry are favorable for designing a low-cost scalable system for bandwidth allocation, but we still need to design an algorithm to make the best use of the INT data.

\section{Design}
\label{sec:design}

Driven by the observations in~\secref{sec:motivation}, we propose \oursystemsoze, a decentralized weighted bandwidth allocation system for large-scale and highly dynamic cloud environments. Given the flow weights, \oursystemsoze can identify the bottleneck and enforce the weighted max-min fair bandwidth allocation with only one network telemetry.
In this section, 
    we firstly show how \oursystemsoze achieves the weighted allocation on a single switch, where the bottleneck is always the switch egress (\secref{subsec:design1}); 
    Then, we show that \oursystemsoze can be applied to arbitrary data center networks, where the bottleneck can be any hop (\secref{subsec:design2});
    Lastly, we summarize our system design (\secref{subsec:design3}) and discussions (\secref{subsec:design4}).

\subsection{A Single Switch Scenario}
\label{subsec:design1}

In this subsection, we show step-by-step how we derive a decentralized resource allocation algorithm from a straw-man algorithm. For the goal proposed in \secref{subsubsec:design11}, we first give a simple decentralized algorithm where a lot of information needs to be collected and exchanged in \secref{subsubsec:design12}; then we try to reduce the amount of information to be very tiny in \secref{subsubsec:design13}; finally, we pick only one piece of information and transform it into a different format, to make telemetry easy and practical on the switches in \secref{subsubsec:design14}.

\subsubsection{Goal: Decentralized Weighted Allocation}
\label{subsubsec:design11}

Denote the flows on one link are $\{f_0, f_1, ..., f_n\}$, their specified weights are $\{w_0, w_1, ..., w_n\}$, and the hop bandwidth is $B$. After converging to weighted bandwidth allocation, the weighted fair-share should be $\frac{B}{w_0 + w_1 + ... + w_n}$, and flow $f_k$'s rate, $r_k$, should be:

\begin{equation}
    r_k= \frac{w_k}{w_0 + w_1 + ... + w_n} \cdot B
\label{equ:goal}
\end{equation}

However, calculating this weighted allocation requires obtaining the sum of all flows' weights, where a centralized controller is usually required. 
    No matter when a new flow joins or an existing flow wants to change the weight, the centralized controller needs to be notified. 
    Thus, this could become the bottleneck of the whole system and prevent the weighted service from being agile, accurate, and scalable.

\subsubsection{Straw-man: Massive Information Exchange}
\label{subsubsec:design12}

A straw-man decentralized solution can be given as follows: \textbf{each flow $f_i$ could send its weight specification $w_i$ to every other flow}, then every flow sender is able to calculate its rate with~\eref{equ:goal} directly. 


However, the problem with this straw-man solution is that the communication overhead is too high for data center networks. If there are $n$ numbers of flows on the link, the total amount of information that needs to be exchanged is $O(n)$ for each flow sender.

\subsubsection{Reduce the Information Exchange}
\label{subsubsec:design13}

To reduce the information exchange for achieving the weighted resource allocation, \oursystemsoze splits \eref{equ:goal} into two equations. And those two equations are achieved if and only if \eref{equ:goal} is achieved (proved in \secref{appendix:goal-transform}):

\begin{equation}
\begin{dcases*}
    r_0 + r_1 + ... + r_n = B \\
    \frac{r_0}{w_0} = \frac{r_1}{w_1} = ... = \frac{r_n}{w_n}
\end{dcases*}
\label{equ:goal_v1}
\end{equation}

For the first equation in \eref{equ:goal_v1}, \oursystemsoze finds that the sum of all flows' rates is the arrival rate of the link; For the second equation in \eref{equ:goal_v1}, \oursystemsoze further reformats the equation with respect to $\max(\frac{r_i}{w_i})$ and $\min(\frac{r_i}{w_i})$. Interestingly, now we are able to rewrite \eref{equ:goal_v1} from any single flow $k$'s view in a decentralized manner:

\begin{equation}
\begin{dcases*}
    link\_arrival\_rate = \sum_{i=0}^n{r_i} = B \\
    \frac{r_k}{w_k} = \min_{i \in \left[0,1,...,n\right]}\left(\frac{r_i}{w_i}\right) = \max_{i \in \left[0,1,...,n\right]}\left(\frac{r_i}{w_i}\right)
\end{dcases*}
\label{equ:goal_v2}
\end{equation}

If and only if every flow individually observes that \eref{equ:goal_v2} has been achieved, the weighted allocation is achieved for all flows. Thus, potentially, \textbf{only the information about whether $\sum^n_{i=0}{r_i} = B$ and $\frac{r_k}{w_k} = \min(\frac{r_i}{w_i}) = \max(\frac{r_i}{w_i})$ are satisfied is required to be exchanged}. In this way, \oursystemsoze largely reduced the exchanged information from $O(n)$ to $O(1)$ for each flow sender. 

\subsubsection{Conduct Information Exchange with a Convergence Algorithm and In-network Telemetry}
\label{subsubsec:design14}

Although the information's content is determined, how to conduct the exchange remains a problem. 
In this subsection, \oursystemsoze leverages the in-network telemetry and designs a convergence algorithm for every flow to converge to the equilibrium in \eref{equ:goal_v2} with zero coordination between flows.

\circled{1} \textbf{Use the link itself for information exchange.} 
    Previous methods usually exchange information only among servers (flow senders), either in a master-slave architecture or an all-reduce architecture. However, the information exchange among servers is always off-path and leads to extra overhead. 
    Thus, \oursystemsoze uses an on-path device for information exchange: the link. 
    Since flows compete for the link bandwidth resource, they must travel this link, which makes it the best channel to deliver information among servers or flow senders.

A specific new technique for links: \textbf{in-network telemetry.} \oursystemsoze leverages a new and common feature of the commodity switches: In-network Telemetry (INT), which allows the switch to tag some information on the packet.
    Specifically, \oursystemsoze only uses one of the telemetry data --- \textbf{queueing delay}. The queueing delay is tagged in the data packet header on the forwarding path and reflected back to the sender with the ACK packet header on the reverse path. Only the queueing delay on the forwarding path is required and it is represented by a single 2-byte field in each packet header.
    This operation is simple and does not require the use of programmable switches.

\circled{2} \textbf{Queueing delay inherently implies link arrival rate.} 
The queueing delay's dynamic inherently reflects the link arrival rate, more specifically, the first derivative of the queueing delay is the difference between the link arrival rate and the link bandwidth. Denote the queueing delay as $D$, link arrival rate as $R$, and link bandwidth as $B$:

\begin{equation}
    \frac{dD}{dt} = \frac{R - B}{B}\text{, }\forall D > 0
\label{equ:queue_function_1}
\end{equation}

As shown in \eref{equ:queue_function_1}, as long as the queueing delay is stabilized around a non-zero value, the link arrival rate is equal to the bandwidth. Since each flow could individually observe queueing delay, the first equation in \eref{equ:goal_v2} can be verified in a decentralized manner.

\circled{3} \textbf{Design different queueing delays to represent different weighted fair-share.} 
Since the queueing delay can be stabilized around any value to indicate the link arrival rate, the actual value of the queueing delay can be used to represent another selected property.
In \oursystemsoze, different queueing delays represent different weighted fair-share with a unique one-to-one mapping, where the weighted fair-share $wfs$ is defined as $wfs=\frac{r_i}{w_i}$ from \eref{equ:goal_v1}. Specifically, the larger the weighted fair share, the smaller the queueing delay. In addition, we denote a flow's rate divided by its weight as its "rate-per-weight". For networks in weighted allocation status, each flow's rate-per-weight is equal to the $wfs$.

\begin{equation}
\begin{aligned}
    &target\_delay = T(\frac{r_i}{w_i}) \\
    T(x) < T(&y), \forall\text{ } x, y \text{ where } x > y
\end{aligned}
\end{equation}

With this relation between the queueing delay and the weighted fair share, each flow could compare its rate-per-weight with the $wfs$ value from the queueing delay and find out if it is smaller or larger than the indicated value. In this way, a possible convergence algorithm can be created to converge to the second equation in \eref{equ:goal_v2}.

\begin{figure}[t!]
     \centering
     \includegraphics[width=0.3\textwidth]{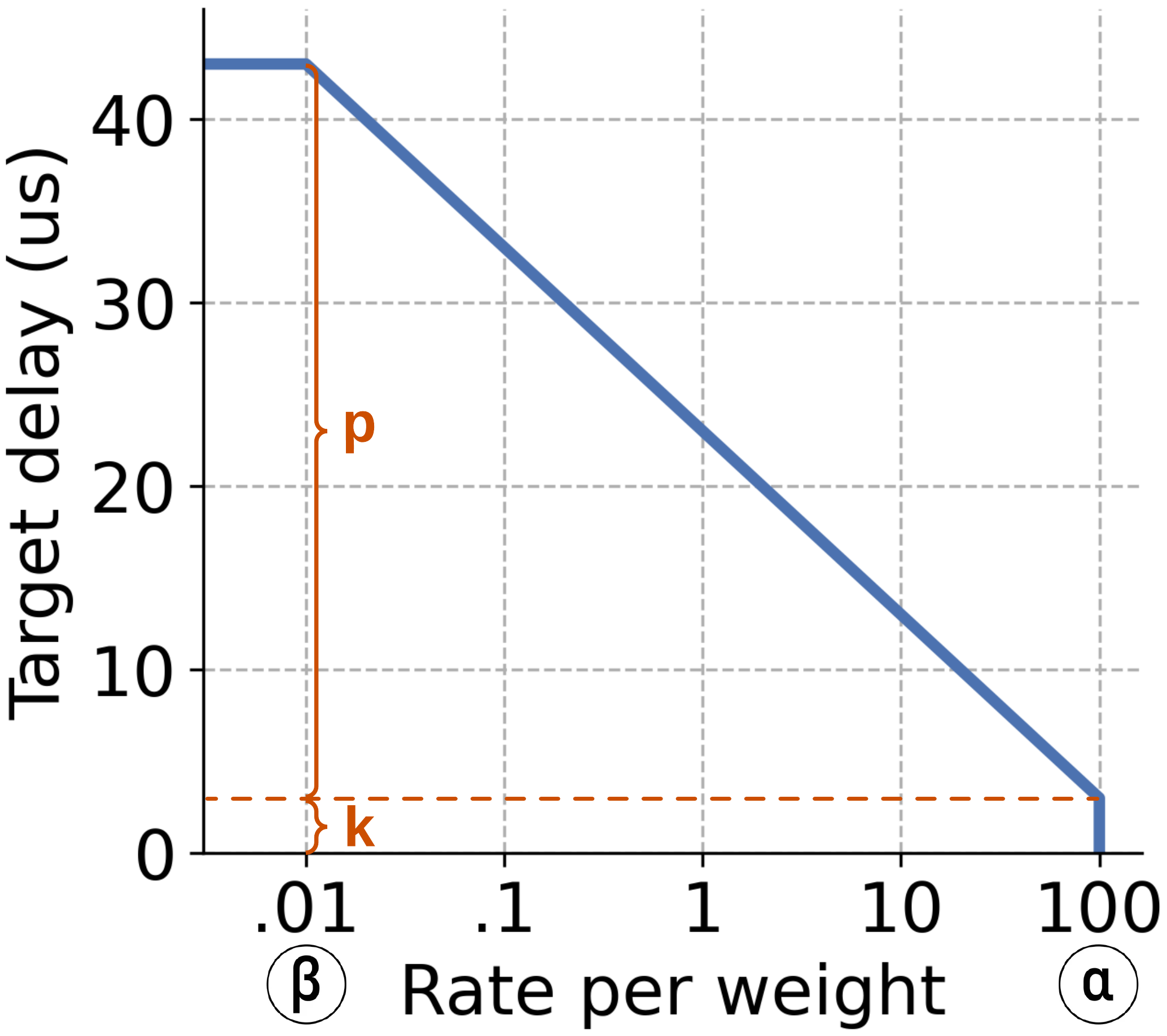}
\vspace{-2mm}
\caption{Target function must be monotonically decreasing. $p$ is the queueing delay scaling and $k$ is always $>0$ for full utilization; $\alpha$ and $\beta$ should be set to the highest and lowest rate-per-weight, depending on the scenarios.}
\label{fig:target}
\vspace{-6mm}
\end{figure}

\circled{4} \textbf{Provide a convergence algorithm that uses insights} \circled{2} \textbf{and} \circled{3}\textbf{ simultaneously.} 
The convergence algorithm will converge to a final steady state indicated by \eref{equ:goal}, where the weighted fair share is $\frac{B}{w_0+w_1+...+w_n}$, and the queueing delay on that link is a function of the weighted fair share.

\textbf{The target function} differentiates flows with larger $\frac{r}{w}$ from flows with smaller $\frac{r}{w}$, so that the function is monotonically-decreasing as in \fref{fig:target}. With this, each flow calculates a queueing delay according to their own $\frac{r}{w}$.

\begin{equation}
    T\left(\frac{r}{w}\right) = p \cdot \frac{ln(\alpha)-ln(\frac{r}{w})}{ln(\alpha)-ln(\beta)} + k
\label{equ:target}
\end{equation}

In \eref{equ:target}, $k$ is the minimal queueing delay when the link is saturated by one flow;
$p$ is the queueing delay scaling to differentiate between different flows. $\alpha$ and $\beta$ indicate the most frequent range of weighted fair share: $\alpha$ is the highest rate-per-weight, which is usually the link bandwidth divided by the smallest weight; $\beta$ is the lowest rate-per-weight, which can be determined with the traffic pattern. The update interval is referred as $\Delta t$, where $\Delta t = \frac{RTT}{CWND}$ for per-packet update.

\textbf{The update function} tries to change a flow's sending rate so that its target delay can match the observed queueing delay.

\begin{equation}
    U\left(\frac{r}{w}, D\right) = \left(\frac{T^{-1}(D)}{\frac{r}{w}}\right)^m
\label{equ:update}
\end{equation}

In \eref{equ:update}, $m$ is a smoother parameter smaller than 2, which is tunable to achieve either faster convergence or more stable final state.

In Algorithm 1, we give an incredibly simple decentralized algorithm for each flow sender to adjust their rate and converge to weighted allocation. 
    After receiving the queueing delay information from the link, each flow calculates a $ratio$ based on its current rate-per-weight and the received queueing delay, then updates the rate accordingly. 
    From any initial bandwidth allocation, the final allocation will always comply with the weighted fair share as in \eref{equ:goal}.

\begin{figure}[t!]
     \centering
     \hfill
     \begin{subfigure}[b]{0.23\textwidth}
         \centering
         \includegraphics[width=\textwidth]{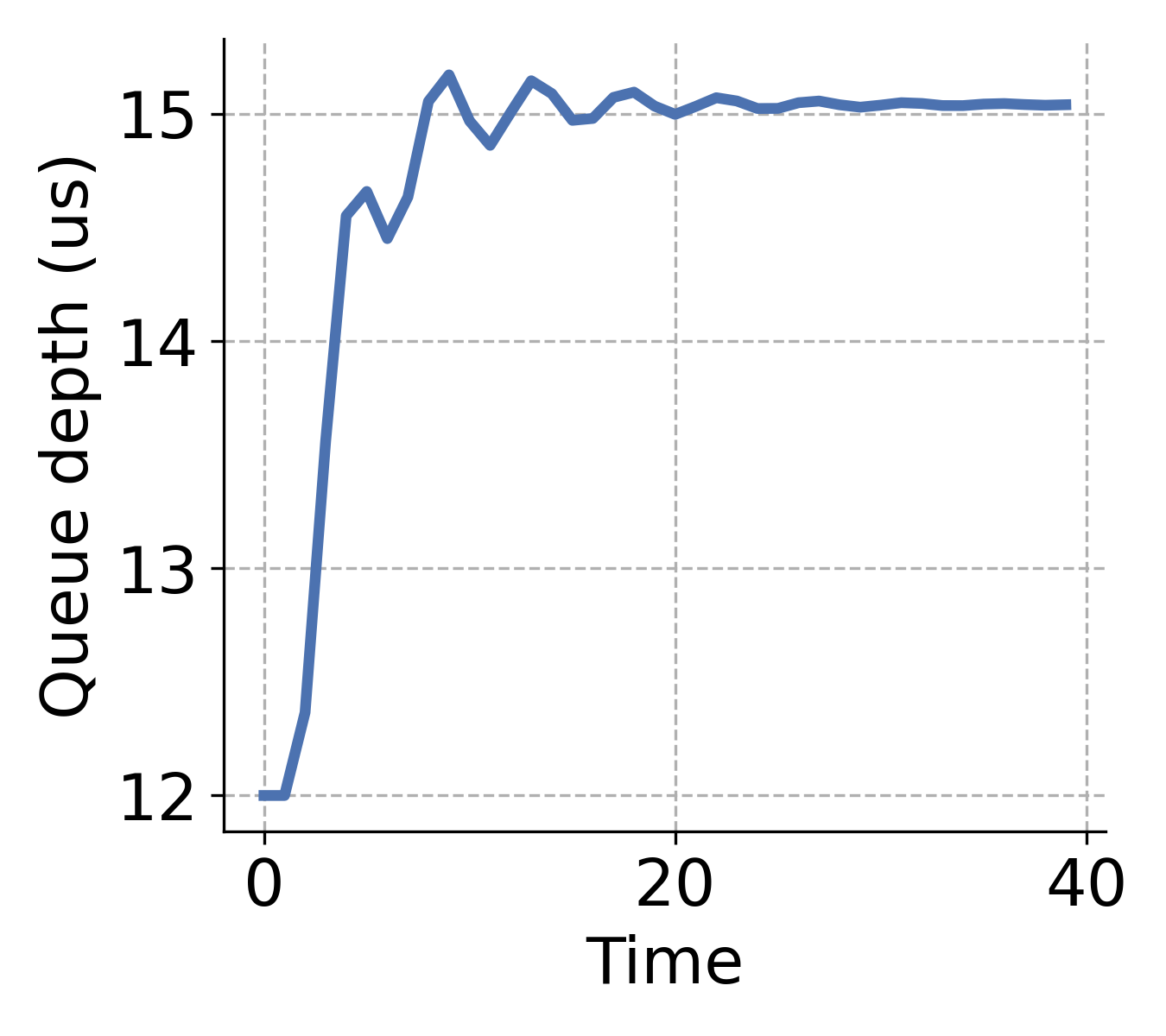}
         \vspace{-6mm}
         \caption{Convergence to the target queueing delay.}
         \label{fig:queue}
     \end{subfigure}
     \hfill
     \begin{subfigure}[b]{0.23\textwidth}
         \centering
         \includegraphics[width=\textwidth]{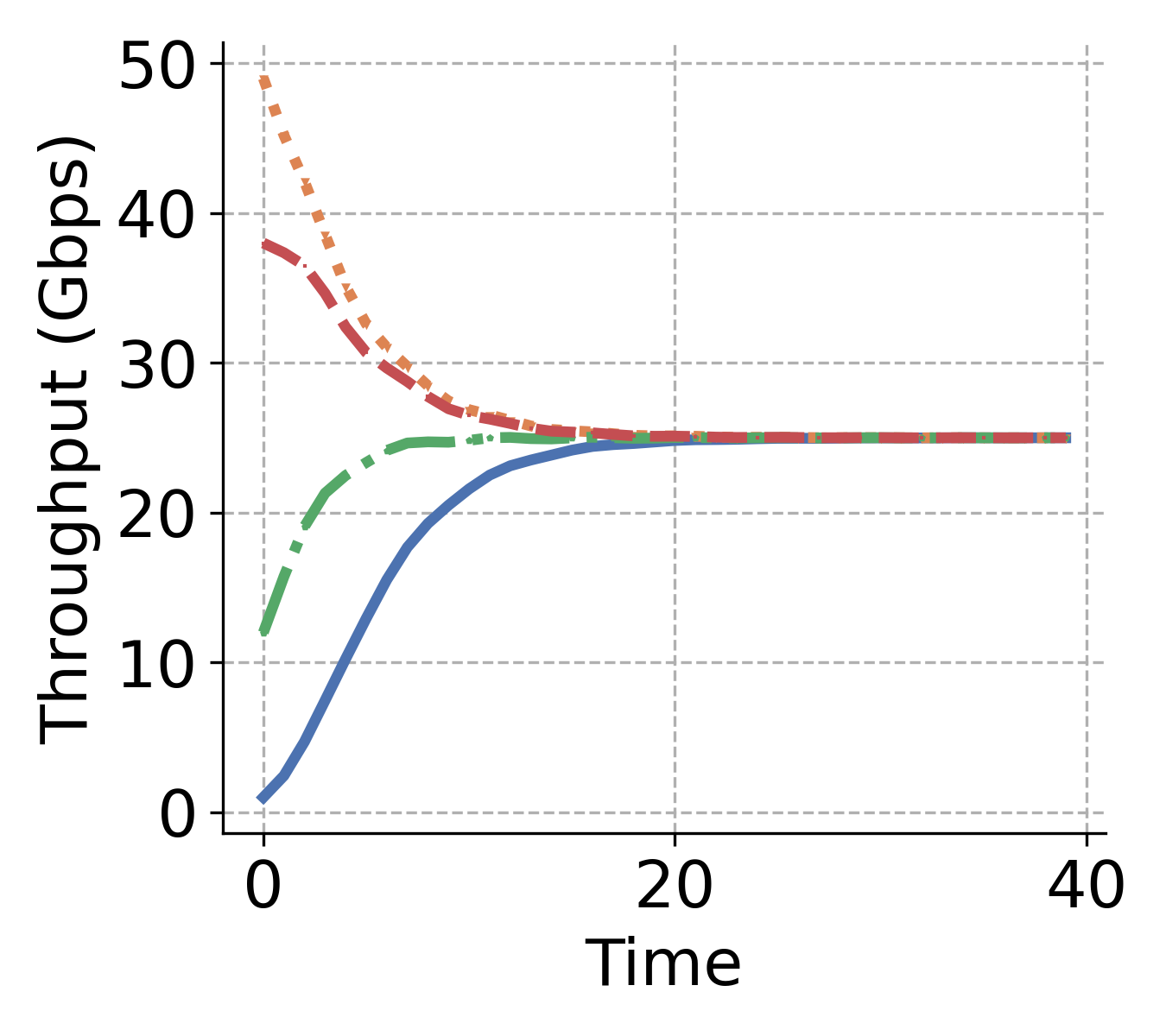}
         \vspace{-6mm}
         \caption{Convergence to weighted fair-share.}
         \label{fig:wfs}
     \end{subfigure}
     \hfill
\vspace{-2mm}
\caption{Convergence for 4 flows on a single switch.}
\label{fig:convergence}
\vspace{-6mm}
\end{figure}

\subsubsection{Proof of Convergence to Weighted Allocation}
\label{subsubsec:design15}

Although the above decentralized algorithm has no explicit information exchange, the queueing delay that contributed by all the flows is also observed by all the flows, which conducts an efficient information exchange towards the weighted bandwidth allocation. With the above design, we are able to have the following two lemmas:

\begin{algorithm}[t!]
    \caption{\oursystemsoze's Rate Adjustment Algorithm}
    \SetKwInOut{Parameter}{Parameter}
    \KwIn{$w_k, r_k$: weight and rate for flow $f_k$;}
    \SetKwFunction{Fmain}{MainFunctionForFlowK}
    \SetKwFunction{Fsend}{Send}
    \SetKwFunction{Freceive}{RecvPktWithINT}
    \SetKwProg{Fn}{Function}{:}{}
    \SetKwFunction{Ft}{T}
    \SetKwFunction{Fu}{U}
    \Fn{\Fmain{}}
    {   
        $signal$ = \Freceive{} \\
        \If{$now - last\_update > rtt$}{
            $ratio = \Fu\left(\frac{r_k}{w_k}, signal\right)$ \\
            $r_k = r_k \cdot ratio$
        }
    }
\label{alg:soze}
\end{algorithm}

\begin{lem}[Convergence to Weighted Fairness]{lem:conv-to-fair}
    \oursystemsoze converges to the weighted fairness on a link if and only if $0 < m < 2$ in the update function.
\end{lem}

\vspace{-2mm}
\begin{lem}[Convergence to Target Queueing]{lem:conv-to-queue}
    \oursystemsoze converges to the target queueing delay level, if and only if $p>\frac{\Delta t}{2}\cdot\left[\ln(\alpha) - \ln(\beta)\right]$
\end{lem}

The idea for proving the above two lemmas is that all the flows together contribute to the queueing and maintain the queueing at the target delay level. 
    The observed delay is the same across all flows, which provides a guidance for achieving fairness; 
    The observed delay is maintained at a certain level, which provides the guarantee of achieving full utilization.
The complete proof is included in Appendix~\ref{appendix:fairness-single} and Appendix~\ref{appendix:util-single}.





\subsection{Arbitrary Network Scenario}
\label{subsec:design2}

Based on the weighted allocation in a single-switch scenario (\secref{subsec:design1}), \oursystemsoze also achieves weighted bandwidth allocation in the arbitrary network scenario, such as the widely used multi-layer network architecture in production data centers. 
    In this section, we firstly introduce the goal of the weighted resource allocation in \secref{subsubsec:design21}; 
    Then, to accommodate the complex scenario, we explore the properties of the flow bottleneck in \secref{subsubsec:design22};
    Lastly, we extend the collected INT signal to be the maximum queueing delay from all links on the path in \secref{subsubsec:design23}. 

\subsubsection{Goal: Weighted Max-min Allocation}
\label{subsubsec:design21}

For the network-wide scenario, a flow may travel multiple links along the path and contend with different sets of other flows. However, for any flow traveling multiple hops, there will be at least one bottleneck link, which determines the flow rate. Thus, the weighted allocation should happen on the bottleneck link for any flow, which is also the goal of weighted max-min fair allocation.
According to the previous papers, the weighted max-min fair is defined as follows:

\begin{defn}[Weighted Max-min Fair~\cite{marbach2002priority, allalouf2008centralized}]{def:1} 
For all flows $\{f1, ..., fn\}$ in the network, denote their weight to be $\{w_{f1}, ..., w_{fn}\}$. A rate allocation $\{r_{f1}, ..., r_{fn}\}$ is weighted max-min fair when for each flow $f$, any increase in $r_f$ would cause a decrease in the transmission rate for some flow $f'$ satisfying $\frac{r_{f'}}{w_{f'}} \le \frac{r_f}{w_f}$.
\end{defn}

\begin{figure}
    \centering
    \includegraphics[width=0.48\textwidth]{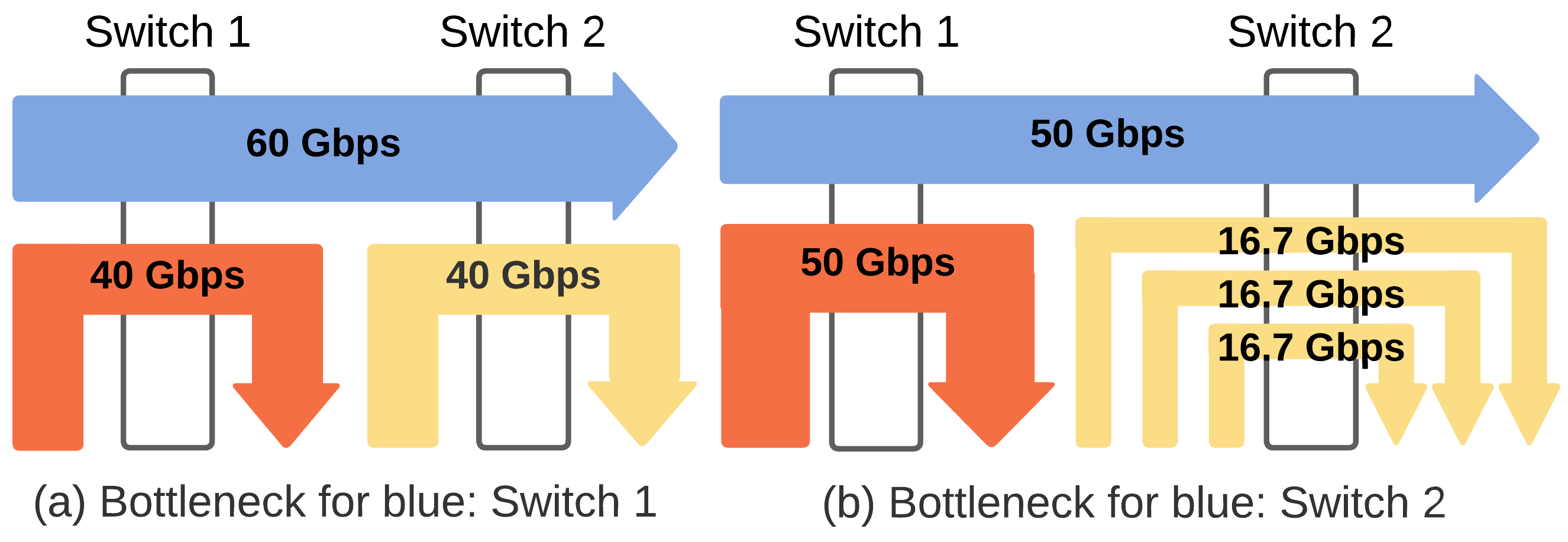}
    \vspace{-6mm}
    \caption{Bottleneck changes in weighted max-min fair: blue flow weight: 3, red flow weight: 2, yellow flow weight: 1.}
    \vspace{-4mm}
    \label{fig:wmmf}
\end{figure}

In the weighted max-min fair, each hop divides the bandwidth to let each bottleneck flow have the same "rate-per-weight". 
    As \fref{fig:wmmf} shows, the "rate-per-weight" is the same for the blue flow and the red flow, and the rate allocation is equivalent to having "weight" number of individual flows.

With the above analysis, we realize that recognizing the bottleneck becomes the new challenge when achieving the weighted max-min fair in arbitrary networks. In the next subsection, we will show how to simply recognize the bottleneck hop and obtain the INT signal from the bottleneck.






\subsubsection{The Bottleneck Hop Properties}
\label{subsubsec:design22}

In \secref{subsec:design1}, \oursystemsoze already achieves the weighted bandwidth allocation on a single switch in a decentralized manner. Thus, in order to achieve the weighted max-min fair, the major challenge is to recognize the bottleneck hop for each flow, then achieve weighted allocation on the bottleneck hop. According to the definition, achieving fair allocation on the bottleneck hop for every flow is just achieving weighted max-min fair.

From the definition, we could derive a lemma that could help us identify each flow's bottleneck hop, namely, the hop that prevents the flow from increasing the rate further. 

\begin{lem}[Bottleneck Hop Properties]{lemma:1}
    When achieving weighted max-min fair, each flow will have the largest rate-per-weight among all flows on its bottleneck hop and not on any other saturated hop. 
\end{lem}

Intuitively, a flow must have the largest rate-per-weight on its bottleneck hop, otherwise, it could take more bandwidth from the flows on this hop who has a higher rate-per-weight; a flow cannot have the largest rate-per-weight on a non-bottleneck saturated hop, since there must be some other flow on that hop could steal more bandwidth from it.
    The full proof of the above lemma is included in the Appendix~\ref{appendix:bottleneck-signal}.

With Lemma~\ref{lemma:1}, we could easily know that if a flow has the largest rate-per-weight only on its bottleneck hop, then among all hops that this flow travels, the bottleneck hop must have the highest queuing delay. Because on non-bottleneck hops, there must be a flow with larger rate-per-weight, and And our target function is monotonically decreasing. Thus, the queueing delay on any other hops is lower than the bottleneck queueing delay. The proof is included in Appendix~\ref{appendix:bottleneck-signal}.

\subsubsection{Recognize Bottleneck with $maxQD$ Signal}
\label{subsubsec:design23}


With the above properties, we could know that a flow does not have the largest rate-per-weight on hops other than the bottleneck hop. Thus, the bottleneck hop always has the highest queueing delay for that flow. In the single-switch scenario, we use the queueing delay ($QD$) signal from one hop to achieve the weighted allocation on that hop. Similarly, for a flow that travels multiple hops in an arbitrary network, we can use the maximum per-hop queueing delay ($maxQD$) signal to achieve weighted allocation on the bottleneck hop of that flow. 

The algorithm does not need to know on which hop the $maxQD$ signal was collected, switches simply compare and replace the packet header to keep the $maxQD$ signal is enough to achieve weighted max-min fair on the bottleneck hop. With this solution, $maxQD$ becomes the signal in Algorithm 1.

With this new INT signal, we could derive our theorem on achieving weighted max-min fairness with \oursystemsoze. The full proof is given in Appendix~\ref{appendix:bottleneck-signal}.

\begin{theo}[Weighted Max-min Fairness]{theo:wmmf}
    For every flow in an arbitrary network, \oursystemsoze converges to a weighted max-min fair allocation, if and only if $0 < m < 2$ in the update function and $p>\frac{\Delta t}{2}\cdot\left[\ln(\alpha) - \ln(\beta)\right]$ in the target function.
\end{theo}
\subsection{System Design Summary}
\label{subsec:design3}

With all the above analysis on how to achieve weighted max-min fairness in arbitrary networks, we are able to give our system design in this section.

\begin{figure}
    \centering
    \includegraphics[width=1.0\linewidth]{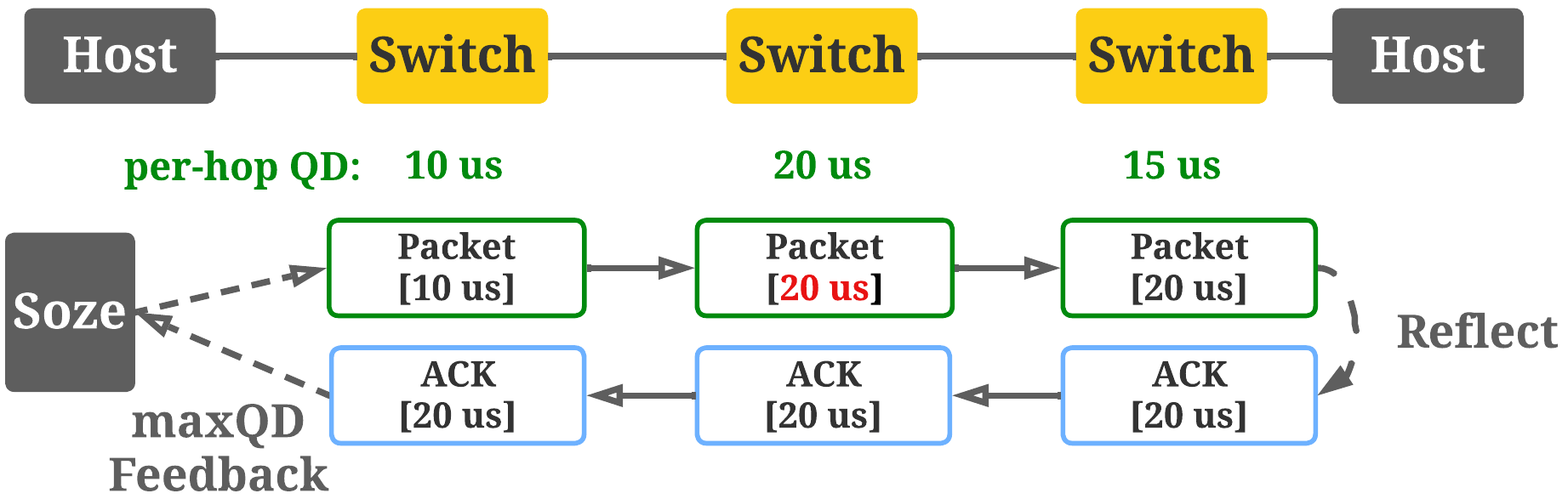}
    \vspace{-6mm}
    \caption{System design of \oursystemsoze.}
    \vspace{-6mm}
    \label{fig:system}
\end{figure}

\vspace{1mm}
\noindent\textbf{Collect $maxQD$ in data packet and reflect in ACK packet.}
As shown in \fref{fig:system}, each packet contains a fixed-length packet header to store the $maxQD$ INT data, which usually takes 2 bytes in the field. 
    When the data packet travels the forwarding path, it compares and keeps the highest per-hop queueing delay on each hop. 
    After reaching the receiver side, the receiver host will attach the $maxQD$ information to the ACK packet and send it back to the sender.
Thus, the $maxQD$ signal is collected at the sender side for every ACK packet.

\vspace{1mm}
\noindent\textbf{Host control with the INT signal.}
The rate control is conducted at the sender host.
    Firstly, after receiving the INT signal from the packet, the host will parse the $maxQD$ data from the packet header and use it to determine the multiplicative rate update ratio.
    Then, the rate can be controlled with either a traditional $CWND$ specification or with a rate-based specification, such as the pacing in some CC algorithms~\cite{li2019hpcc}.
        Note that, when using the window-based protocol, we are still using the "rate" rather than the congestion window directly. The rate of the flow needs to be calculated with $CWND$ and $RTT$ by $rate = \frac{CWND\cdot pktsize}{RTT}$.
    Lastly, \oursystemsoze serves both as congestion control and as a weighted resource allocation protocol. Thus, \oursystemsoze should be on the networking stack, either in the kernel or running on the NIC hardware. The applications should give the weight specification through an API.

\vspace{1mm}
\noindent\textbf{Conduct rate update in per-packet manner.}
Typical CC algorithm conducts rate update for each RTT, because they are using AIMD-based algorithm for convergence. 
In \oursystemsoze, 
    the convergence to weighted fairness is performed with a new adaptive-update algorithm in Algorithm 1. 
    Moreover, each individual $maxQD$ signal can reflect the network condition and calculate the rate update.
Thus, achieving a per-packet rate update becomes beneficial to weight enforcement, making the convergence faster and stabilizing the rate enforcement. 

\subsection{Discussion}
\label{subsec:design4}

\vspace{1mm}
\noindent\textbf{Weight distribution policy.}
With \oursystemsoze, each application could individually determine the weight of its flows and obtain the corresponding bandwidth from the network. However, if each application increases its weight to maximum blindly, the weighted fairness will not benefit any application. For practical usage, there should be guidelines for weight determination and restrictions on the application's behavior.

The network operator could define any usage policy based on their specific workloads and service-level objectives, which, we emphasize, is not the focus of this paper. We use one simple idea for a discussion as follows:
    1) Each application has a total weight that it can distribute across different hosts and flows, but the application can pay the service provider to get more total weight.
    2) An application can decrease the weight of any flow when possible, namely, not fully utilize all the weights.
    3) When an application wants to increase the weight of flows, some other flows in the same application need to decrease their weight so that the sum of the weights in an application is generally the same.
This simple policy gives applications the ability to be altruistic to others, but when the application wants to benefit themselves, some penalties will be added to regularize their behavior.


\vspace{1mm}
\noindent\textbf{Policy enforcement monitor.}
For applications inside the cloud, \oursystemsoze expects them to follow the policy for weight distribution from the cloud provider. However, there is still the possibility that an application does not follow the policy. To monitor whether all the applications' behavior follows the policy, a logging system could be used.
    1) Whenever the networking stack receives a weight specification from the application through the API, \oursystemsoze will log the timestamp, flow ID (e.g., 5-tuple), and the specified weight. 
    2) Periodically, each application's weight update logs will be gathered to a policy checker, where the policy will be verified through the logs from multiple machines.

\vspace{1mm}
\noindent\textbf{Queuing as INT signal.}
Although in this paper we use queuing delay as the feedback signal, the queuing level is controlled by the target function in \eref{equ:target}. Thus, the queue can be maintained at a relatively low level, as shown in \fref{fig:stepinout-erpc-rtt}. Besides queuing delay, the design of \oursystemsoze also applies to other INT signals, such as link utilization or ECN marking ratio. For different signals, the properties of the system also vary, which will be discussed in future studies.

\section{Implementation}
\label{sec:implementation}

We implemented \oursystemsoze with both the Linux kernel module to replace Linux's default Cubic transport and kernel-bypassing network transport --- eRPC. We also provide simulator implementation on the NS-3 simulator for large-scale experiments.

\vspace{1mm}
\noindent\textbf{Switch implementation.}
We implement the queueing delay signal on Tofino switches~\cite{tofino} with 9 lines of code. 
    When every packet reaches the exit point of the switch, the switch uses the low-pass filter $LPF()$ function to collect the queueing signal in the "sampling mode". By controlling the sampling interval, we can adjust the signal to be each individual packet's queueing delay or smoothed queueing delay over multiple packets. In \oursystemsoze, the default setting is to output the smoothed queueing delay over 10 $\mu$s.

\vspace{-3mm}
\begin{minted}
[
frame=lines,
framesep=2mm,
% baselinestretch=1.2,
% bgcolor=LightGray,
fontsize=\footnotesize,
linenos
]
{python}
Lpf<bit<32>, bit<10>>(size=1) lpf_queue;  # type="SAMPLE"
queue_input = (bit<32>) eg_intr_md.deq_timedelta;
queue_output = lpf_queue.execute(queue_input, 0);
\end{minted}
\vspace{-2mm}

\vspace{1mm}
\noindent\textbf{Host implementation 1: kernel module.}
We implemented a prototype of \oursystemsoze as a Linux kernel module with 241 lines of code, which can be installed on Linux without recompiling the kernel.
    By replacing the kernel module for congestion control, we could receive the maximum per-hop queueing signal in a TCP option field. For the functions $exp()$ and $log()$ in \oursystemsoze, we implemented efficient approximation functions without using the STL library. 
    For application integration, we added one TCP socket option and used that field to configure the weight for this specific TCP socket. 
    To deliver the weight specification, we allow the application to modify the value of the TCP socket option field as the weight value.
    For \oursystemsoze, the parameters are set to: $p=20 \mu s$, $k=3 \mu s$, $m=0.25$. As a baseline, the default algorithm for Linux is TCP-Cubic. The switch port bandwidth is set to 25 Gbps.


\vspace{1mm}
\noindent\textbf{Host implementation 2: eRPC.} 
Alternatively, we also implemented \oursystemsoze in an RPC library --- eRPC~\cite{kalia2019datacenter}, an open source RPC library that supports Ethernet, InfiniBand, and RoCE. 
    We add an additional field in the erpc header to carry the max queueing delay information, and use the same Tofino switches to attach those signals to each data packet. 
    We replace the Timely CC algorithm in erpc with \oursystemsoze and added supporting features, such as packet header changes and queueing signal processing, in 1972 lines of code. 
    For application integration, eRPC uses userspace networking with polling, so applications can directly communicate with eRPC through the API and reconfigure the weight parameter for eRPC connections.
        The parameters for \oursystemsoze are set to: $p=20 \mu s$, $k=3 \mu s$, $m=0.25$. The bandwidth is 25 Gbps.

\vspace{1mm}
\noindent\textbf{NS-3 simulator implementation.}
We also implement \oursystemsoze in NS-3~\cite{ns3}.
The switches in the simulator have been customized to provide the maximum per-hop queueing delay. The packet header with "maxQD" will be updated on every hop to keep the maximum queueing delay so far. This $maxQD$ information is sent back to the sender through the ACK packets.
For \oursystemsoze, unless otherwise noted, the default parameters in the target function are set as $p=20 \mu s$, $k=3 \mu s$, $m=0.25$. For DCQCN, we use all the parameters suggested in \cite{zhu2015congestion, zhu2016ecn}. For HPCC, all the original settings from \cite{li2019hpcc} are kept unless noted: $W_{AI}=80$ Bytes, $maxstage=5$, and $\theta=95\%$.

\vspace{1mm}
\noindent\textbf{Principles for choosing parameters.}
The choice of the parameters is based on the following principles: $p$ controls the granularity of the INT signal, a higher $p$ value leads to higher queuing delay but less rate oscillation, so we tend to choose the minimal $p$ value that gives a relatively low rate enforcement oscillation. $k$ controls the base queuing delay, which may cause the utilization to be less than 100\% if $k$ is too small. Thus, we choose the minimal $k$ that provides full link utilization. $m$ controls the convergence speed, higher $m$ leads to faster convergence but worse rate oscillation. So, there is a wide spectrum of $m$ that may suit different workload patterns.
\section{Evaluation}
\label{sec:evaluation}

We will evaluate \oursystemsoze on both the testbed and the large-scale simulator.
    Firstly, we demonstrate application use cases for \oursystemsoze on the eRPC testbed in \secref{subsec:evaluation1};
    Secondly, we also evaluate the efficiency of \oursystemsoze as the congestion control protocol alone and compared with existing industry solutions in \secref{subsec:evaluation2};
    Lastly, in \secref{subsec:evaluation3}, we conduct micro-benchmark experiments in the simulator to show that \oursystemsoze can achieve fine-grained weighted max-min fair allocation rapidly on a large scale.

\vspace{1mm}
\noindent\textbf{eRPC testbed setup.}
For the eRPC testbed, we connect four servers with DPDK-capable CX-4 NIC to a Tofino-1 programmable switches in a star topology. The flow is the RPC write request sent from one host to another, where the request size is 7 MB and the response size is 32 bytes. When multiple hosts send write requests to the same host, the switch egress port to that destination host will become the bottleneck. In the eRPC testbed, we compare our scheme with Timely~\cite{mittal2015timely}, the default CC used in eRPC.

\vspace{1mm}
\noindent\textbf{NS-3 simulator setup.}
In the NS-3 simulator, we built a network with 1024 servers and 320 switches in a fat-tree topology. The simulator uses 100 Gbps links with 1 $\mu$s link delay, 32 MB buffer size, and 1000 bytes packet payload size. In the simulator, we compare \oursystemsoze with DCQCN and HPCC. All the parameter settings are adopted from their original paper.


\vspace{-1mm}
\subsection{Application Case Study}
\label{subsec:evaluation1}

Firstly, we evaluate the application benefits from \oursystemsoze in several scenarios on the eRPC testbed. 
The scenarios below are a small subset of the benefits for demonstration. 
We then evaluate \oursystemsoze on the TPC-H jobs in the NS-3 simulator.


\vspace{-1mm}
\subsubsection{Prioritize Critical Path inside a Job}

\begin{figure}[t!]
    \centering
    \includegraphics[width=0.35\textwidth]{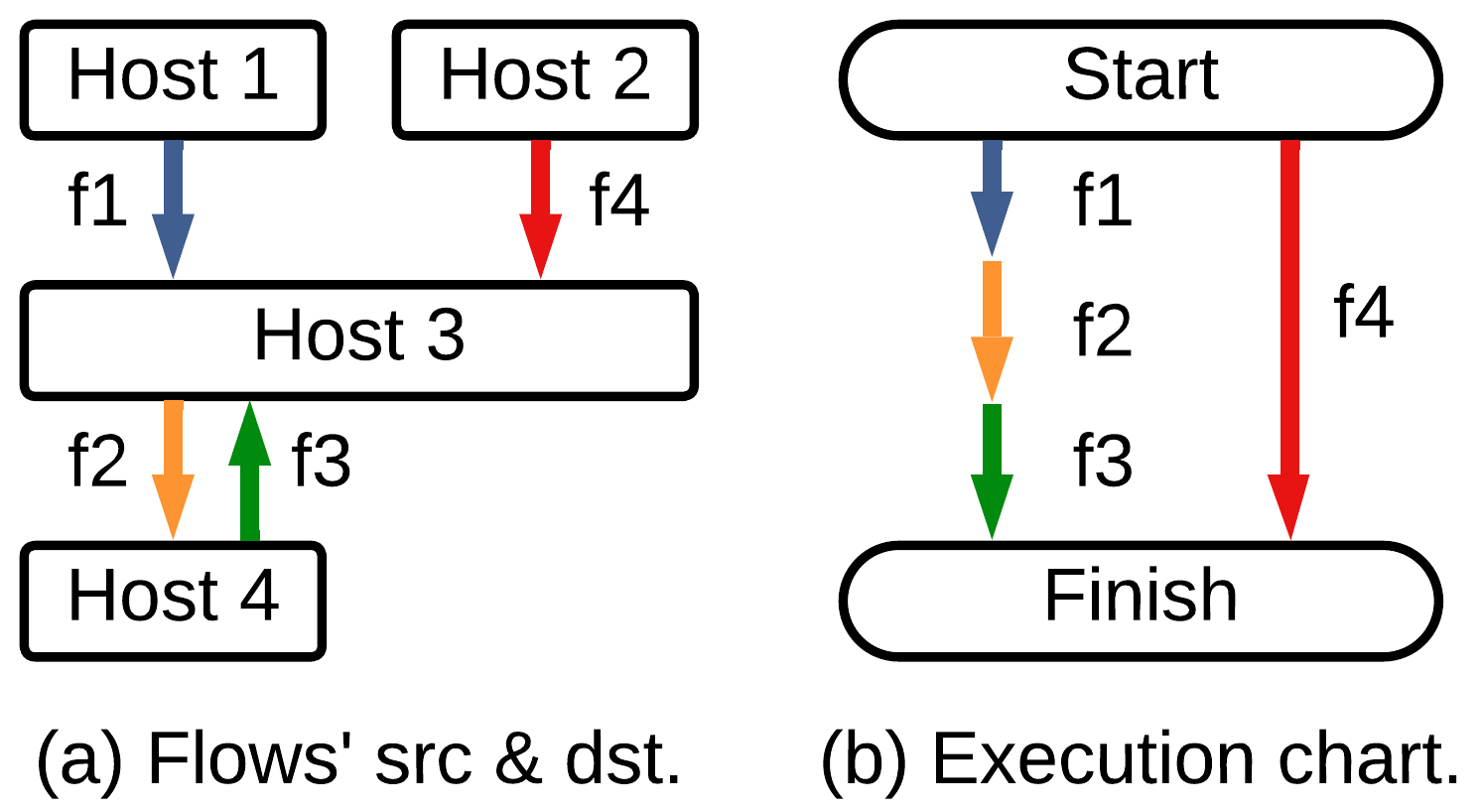}
\vspace{-2mm}
\caption{Scenario for critical path acceleration.}
\label{fig:jct_a}
\vspace{-4mm}
\end{figure}

\begin{figure}[t!]
     \centering
     \hfill
     \begin{subfigure}[b]{0.235\textwidth}
         \centering
         \includegraphics[width=\textwidth]{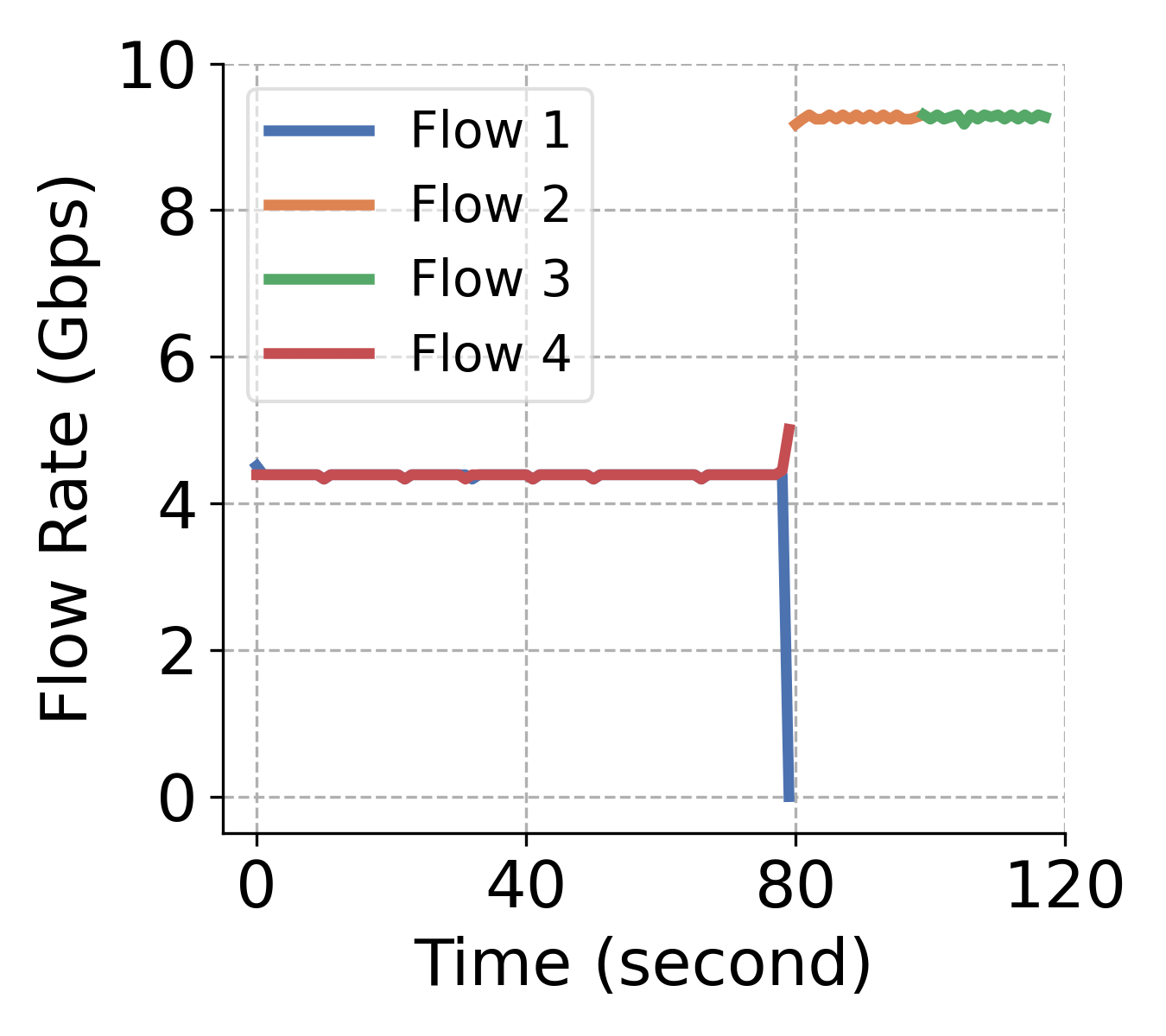}
         \vspace{-5mm}
         \caption{Fair allocation.}
         \label{fig:jct1}
     \end{subfigure}
     \hfill
     \begin{subfigure}[b]{0.235\textwidth}
         \centering
         \includegraphics[width=\textwidth]{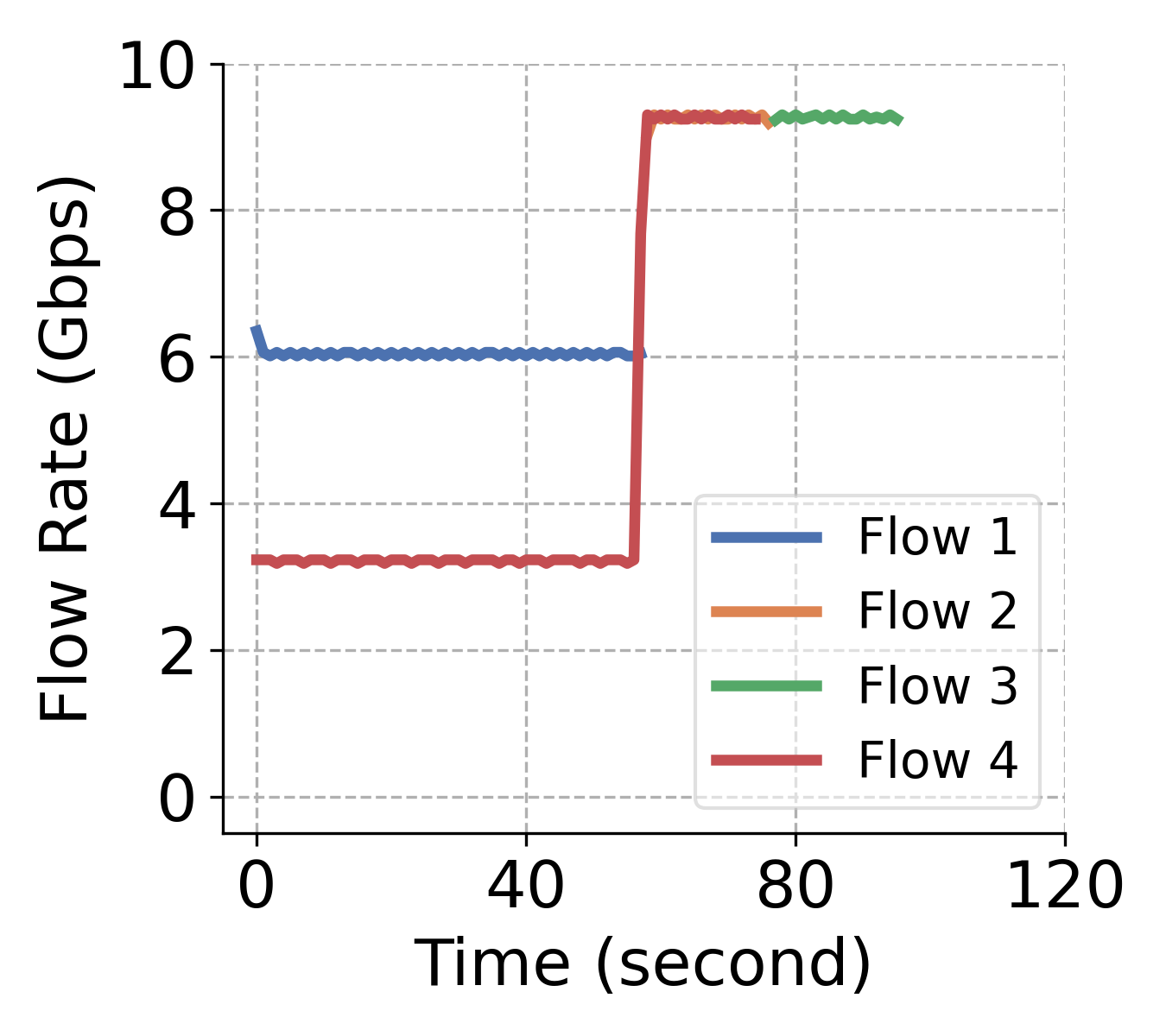}
         \vspace{-5mm}
         \caption{Weighted: f1:f4=2:1.}
         \label{fig:jct2}
     \end{subfigure}
     \hfill
\vspace{-6mm}
\caption{Reduce completion time by prioritizing critical path.}
\label{fig:jct-result}
\vspace{-5mm}
\end{figure}

\begin{figure*}[t!]
     \centering
     \hfill
     \begin{subfigure}[b]{0.235\textwidth}
         \centering
         \includegraphics[width=\textwidth]{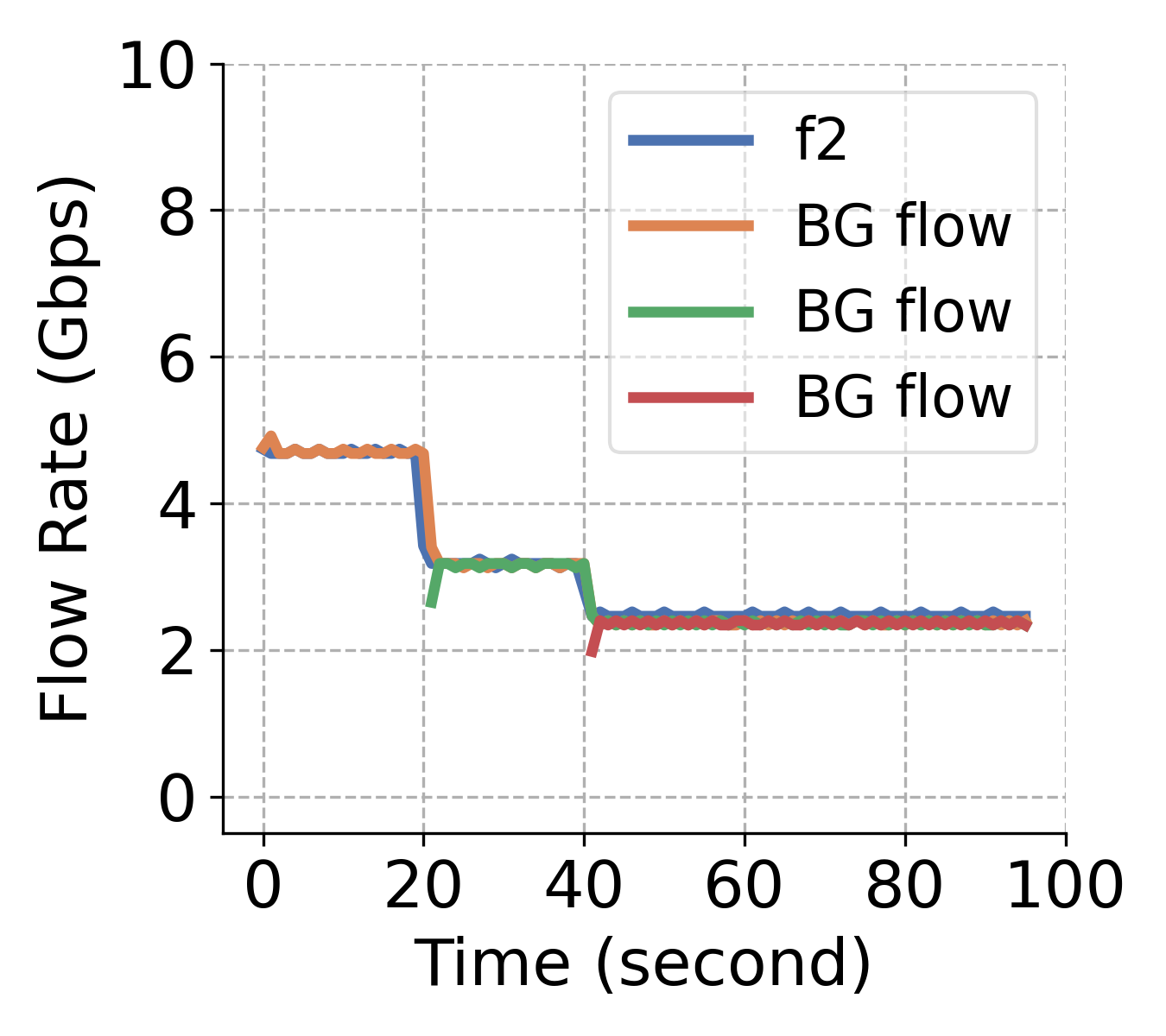}
         \vspace{-6mm}
         \caption{Fair allocation for f2.}
         \label{fig:straggler1}
     \end{subfigure}
     \hfill
     \begin{subfigure}[b]{0.235\textwidth}
         \centering
         \includegraphics[width=\textwidth]{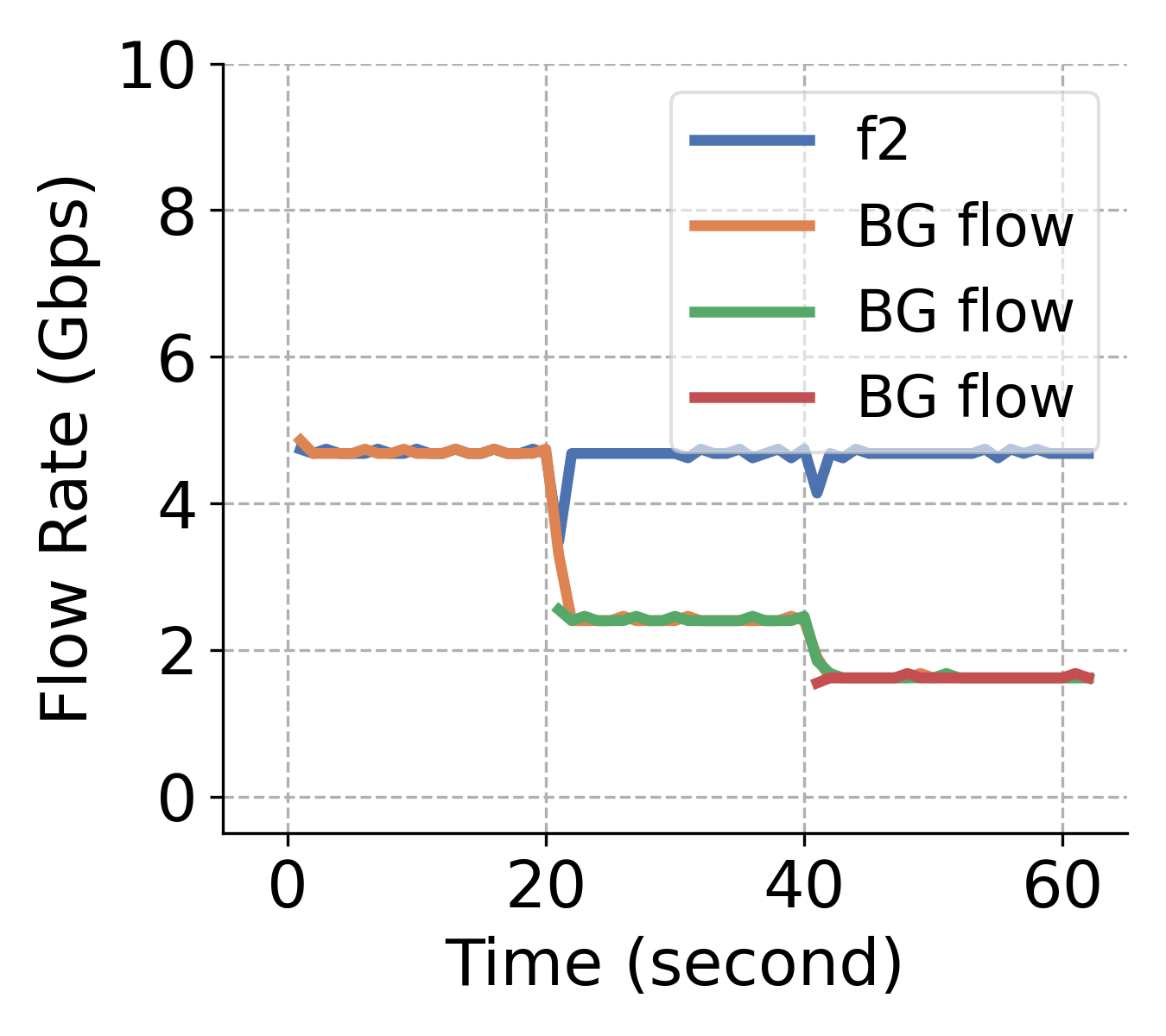}
         \vspace{-6mm}
         \caption{Straggler mitigation for f2.}
         \label{fig:straggler2}
     \end{subfigure}
     \hfill
     \begin{subfigure}[b]{0.235\textwidth}
         \centering
         \includegraphics[width=\textwidth]{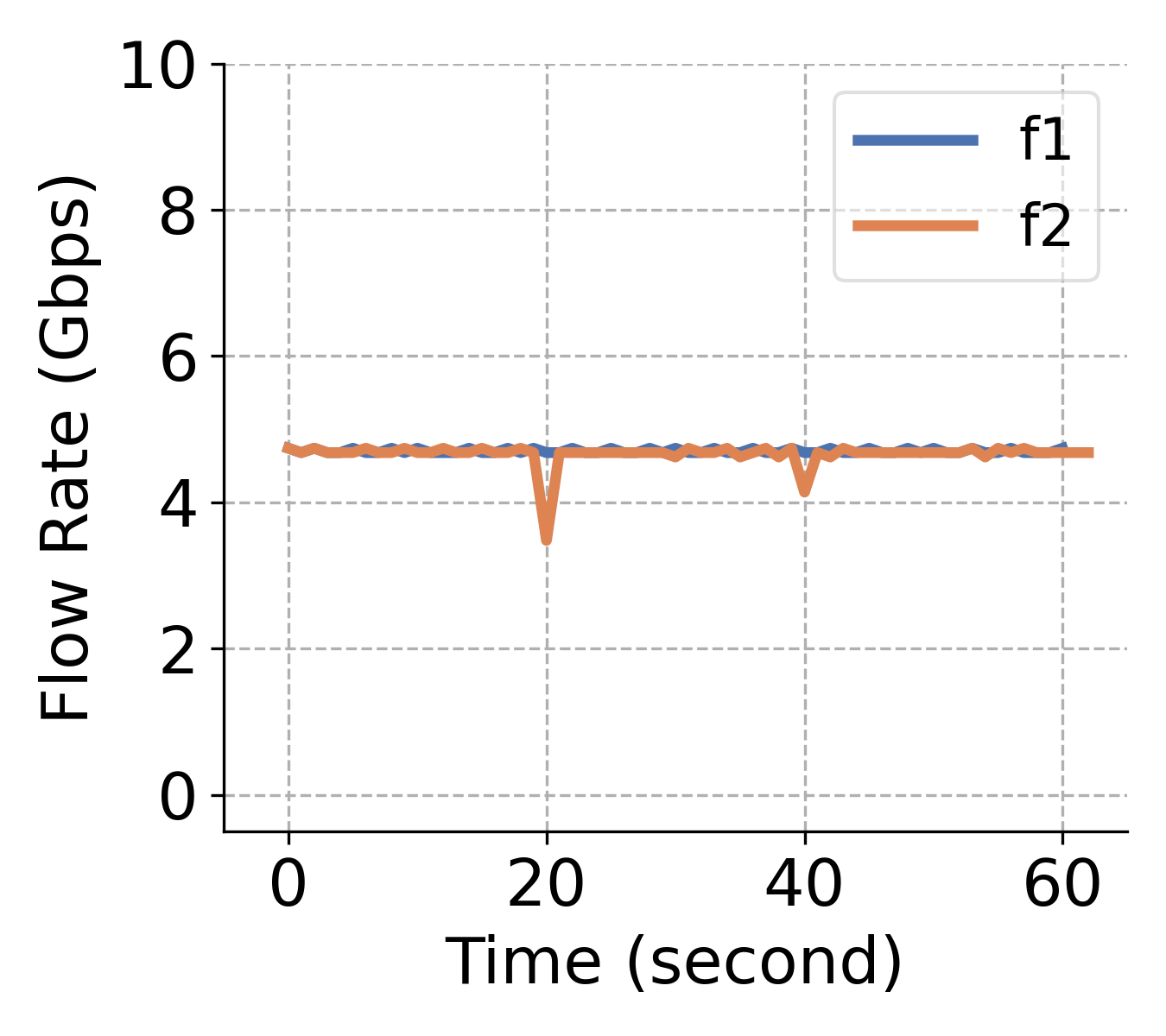}
         \vspace{-6mm}
         \caption{Rates after mitigation.}
         \label{fig:straggler3}
     \end{subfigure}
     \hfill
     \begin{subfigure}[b]{0.235\textwidth}
         \centering
         \includegraphics[width=\textwidth]{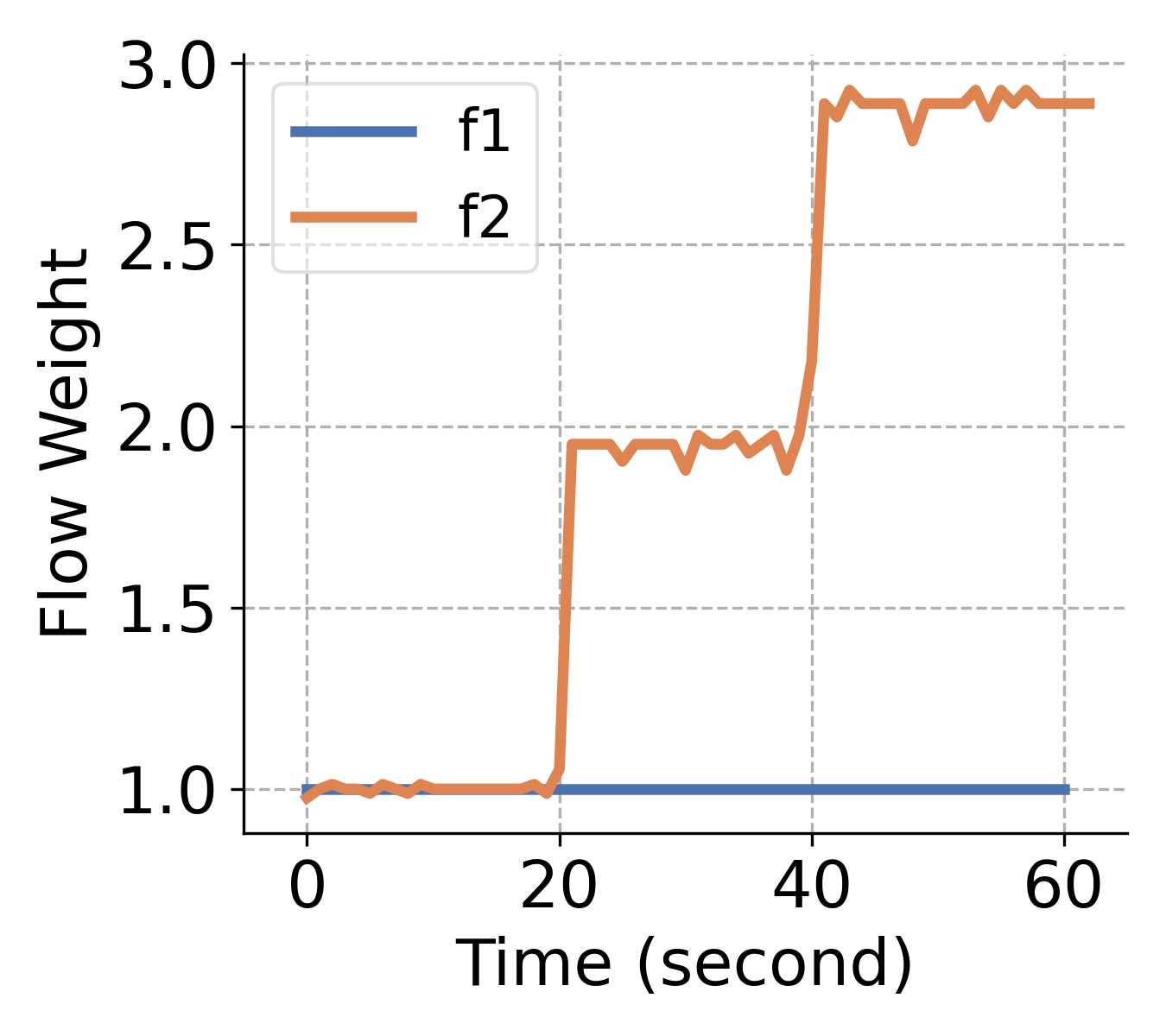}
         \vspace{-6mm}
         \caption{Weights after mitigation.}
         \label{fig:straggler4}
     \end{subfigure}
     \hfill
\vspace{-2mm}
\caption{Mitigate straggler.}
\label{fig:jct}
\vspace{-4mm}
\end{figure*}

\begin{figure*}[t!]
     \centering
     \hfill
     \begin{subfigure}[b]{0.235\textwidth}
         \centering
         \raisebox{3mm}{\includegraphics[width=\textwidth]{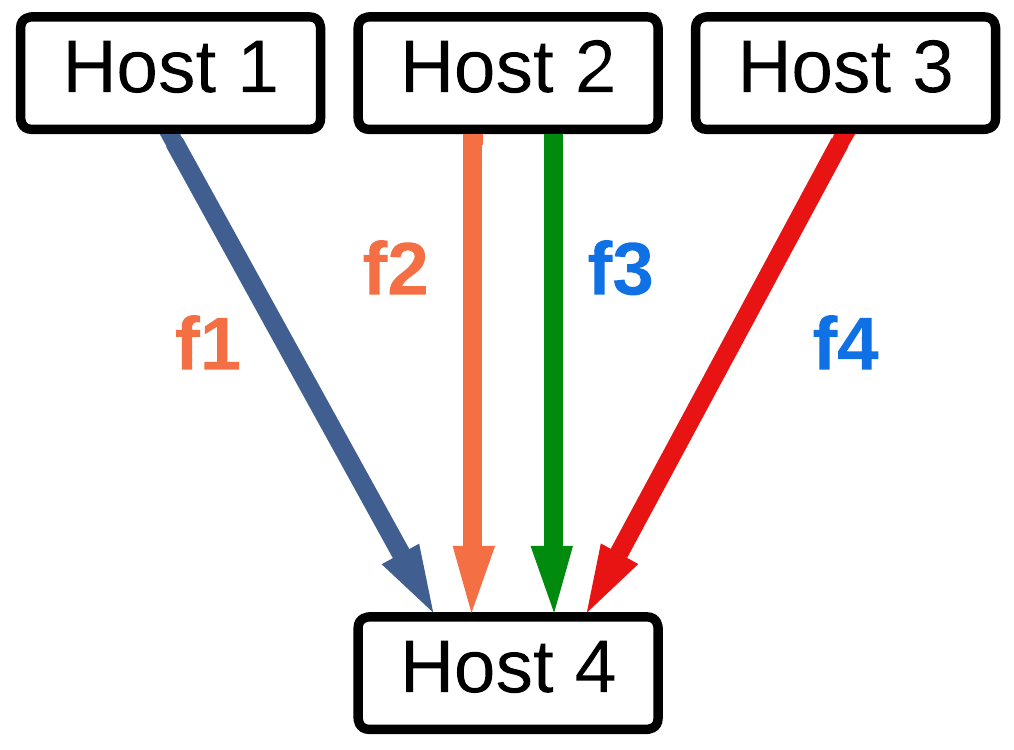}}
         \vspace{-6mm}
         \caption{Experiment setup.}
     \end{subfigure}
     \hfill
     \begin{subfigure}[b]{0.235\textwidth}
         \centering
         \includegraphics[width=\textwidth]{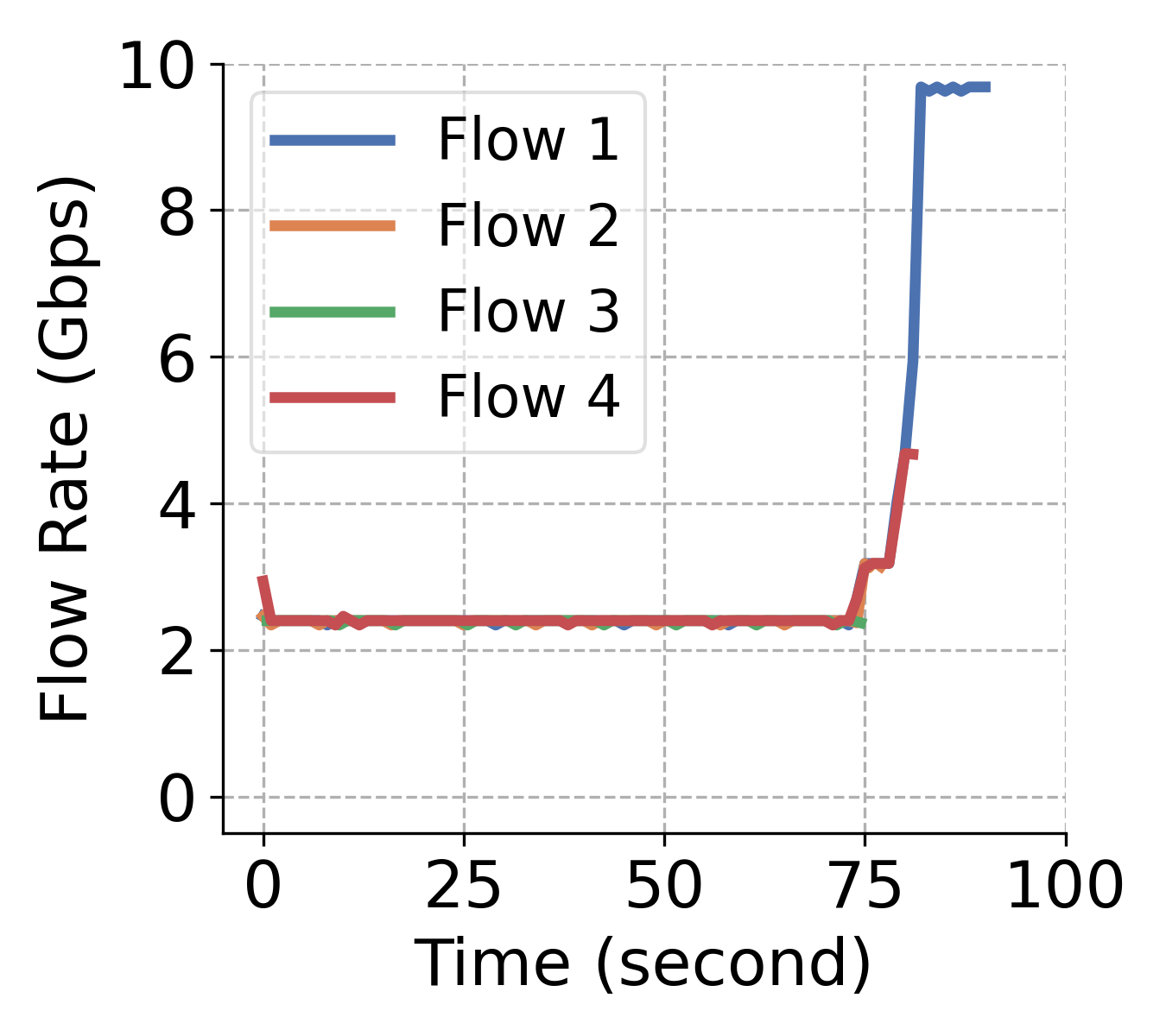}
         \vspace{-6mm}
         \caption{Fair allocation.}
     \end{subfigure}
     \hfill
     \begin{subfigure}[b]{0.235\textwidth}
         \centering
         \includegraphics[width=\textwidth]{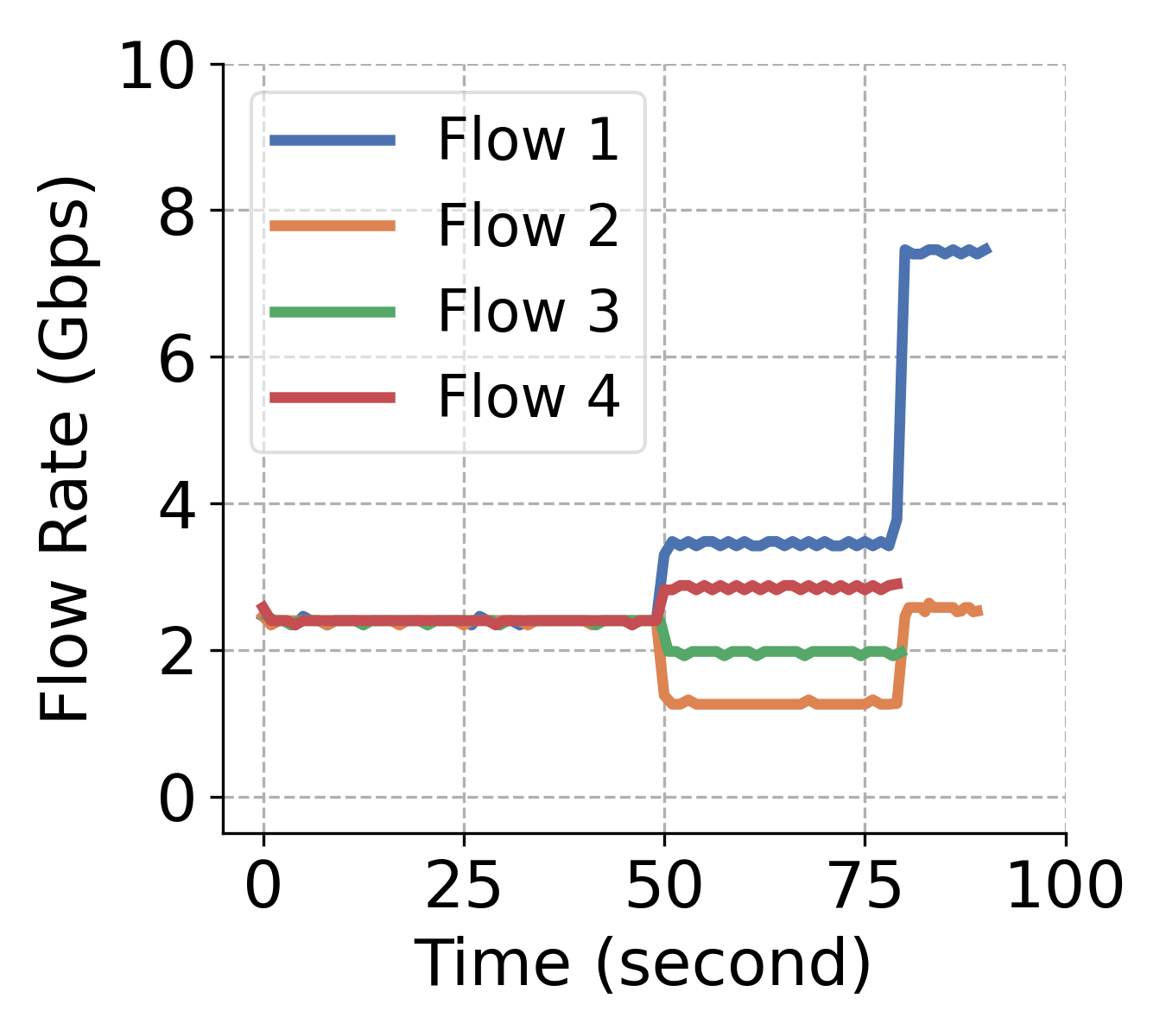}
         \vspace{-6mm}
         \caption{Weighted allocation.}
     \end{subfigure}
     \hfill
     \begin{subfigure}[b]{0.235\textwidth}
         \centering
         \includegraphics[width=\textwidth]{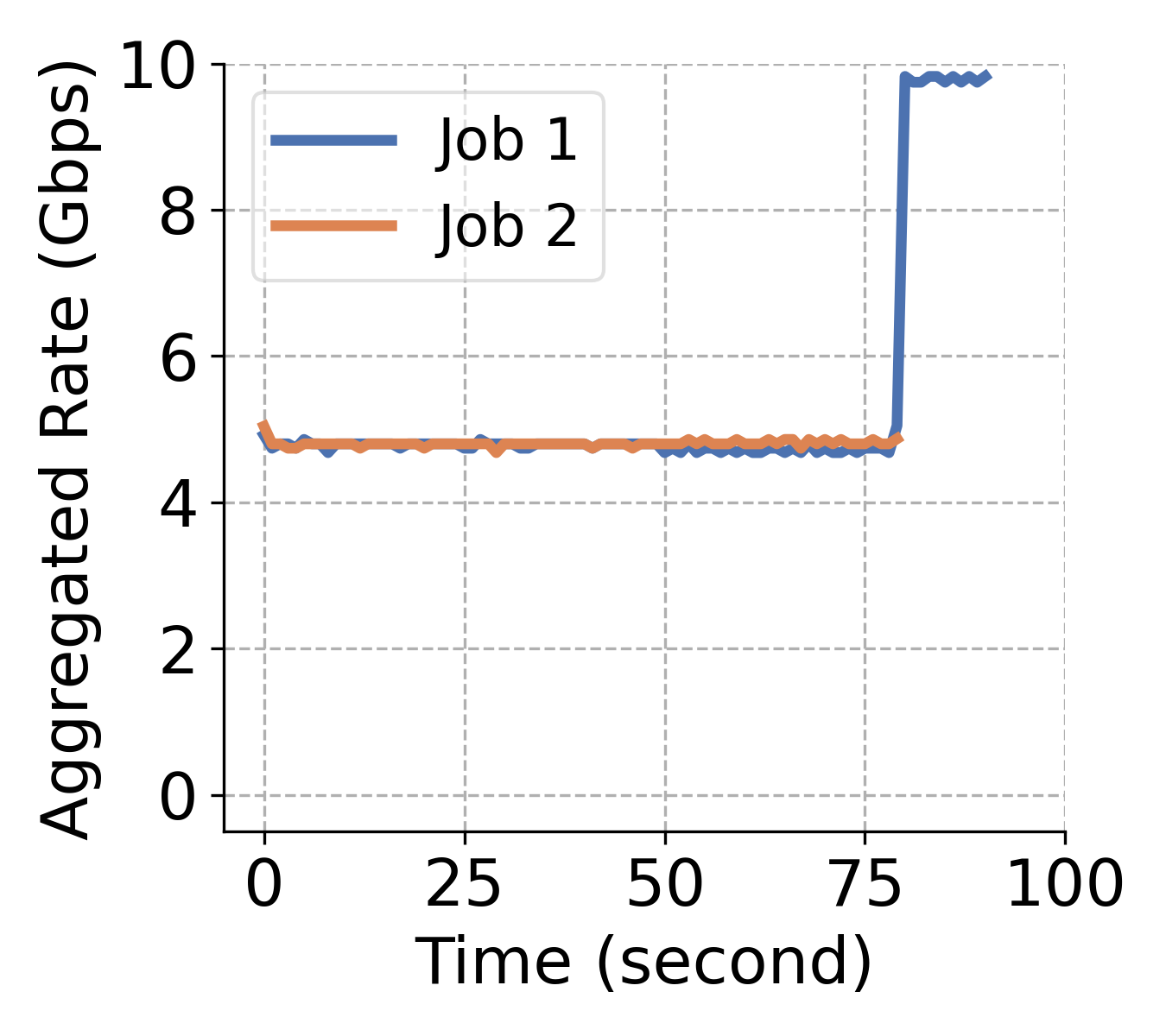}
         \vspace{-6mm}
         \caption{Aggregated rates.}
         \label{fig:appfair_d}
     \end{subfigure}
     \hfill
\vspace{-2mm}
\caption{a) The flows f1 and f2 belong to job 1, and the flows f3 and f4 belong to job 2; b) The fair allocation equally allocates the bandwidth among all four flows, and rebalances when any flow finishes. Under the fair allocation, job 2 finishes at 82 seconds. c) The weighted allocation will update the weights at time 50 s, letting f3 and f4 finish at the same time and reducing the completion time for job 2 to 79 seconds; d) For the weighted allocation, each job always has the same aggregated bandwidth as before the weight changes, so that the bandwidth rebalancing only happens within a job.}
\label{fig:appfair}
\vspace{-4mm}
\end{figure*}

Consider stage-based applications such as map-reduce, distributed matrix multiplication, DNN training, etc., where there are several computation and communication phases. The end-to-end application performance depends on the performance of the critical path. 
    Enabled with modern application-level abstractions and critical-path analysis~\cite{liang2021joint,wang2021mxdag,pan2022efficient}, relative priorities between flows of different paths can be determined prior to execution, and \oursystemsoze can allocate bandwidths between flows according to those priorities. 

As shown in \fref{fig:jct_a}(a), a job consists of 4 flows: flow f1 has 20 GB of data, flow f2 has 10 GB, flow f3 has 10 GB, while flow f4 has 20 GB. As shown in \fref{fig:jct_a}(b), f2 can only start when flow f1 finishes, and f3 can only start when f2 finishes. And flow f1 and flow f4 share the same bottleneck. 

When fairly sharing the resource between flow 1 and flow 4 as in \fref{fig:jct1}, the start of flow 2 will be delayed, and the job completion time is 117 seconds. Once we recognize the critical path, we can prioritize flow 1 over flow 4 by setting their weight to be $\frac{4}{3}$ and $\frac{2}{3}$, so that flow 1 gets twice bandwidth than flow 4 and the sum of their weights still add to $2$. As shown in \fref{fig:jct2}, prioritized allocation reduces the job completion time to 96 seconds.

\vspace{-1mm}
\subsubsection{Identify and Mitigate Stragglers inside a Coflow}

A common technique for mitigating stragglers in a coflow is to dynamically change the inter-flow priority. Such a scenario can be an interesting use case for \oursystemsoze. Due to the zero-coordination and fast-convergence properties, \oursystemsoze can implement the weighted bandwidth allocation with arbitrary granularity among the flows of a coflow to mitigate stragglers, and thus minimize the coflow-completion time. 

This experiment tries to identify the straggler for an all-shuffle job. The figure focuses on two flows inside one coflow, where flow f2 is twice as large as flow f1. When the resources are fairly shared by f2 and background flows, flow f2 will decrease the rate as shown in \fref{fig:straggler1}; To mitigate the straggler and make flow f2 finish around the same time as flow f1, we monitor the progress of flow f2, if the progress cannot catch up with f1 under current sending rate, the weight of flow f2 will increase as in \fref{fig:straggler4}, and flow f2 will take more bandwidth from the background traffic and maintain the same rate as flow f1 as shown in \fref{fig:straggler2} and \fref{fig:straggler3}.

\begin{figure*}[t!]
     \centering
     \hfill
     \begin{subfigure}[b]{0.235\textwidth}
         \centering
         \raisebox{3mm}{\includegraphics[width=\textwidth]{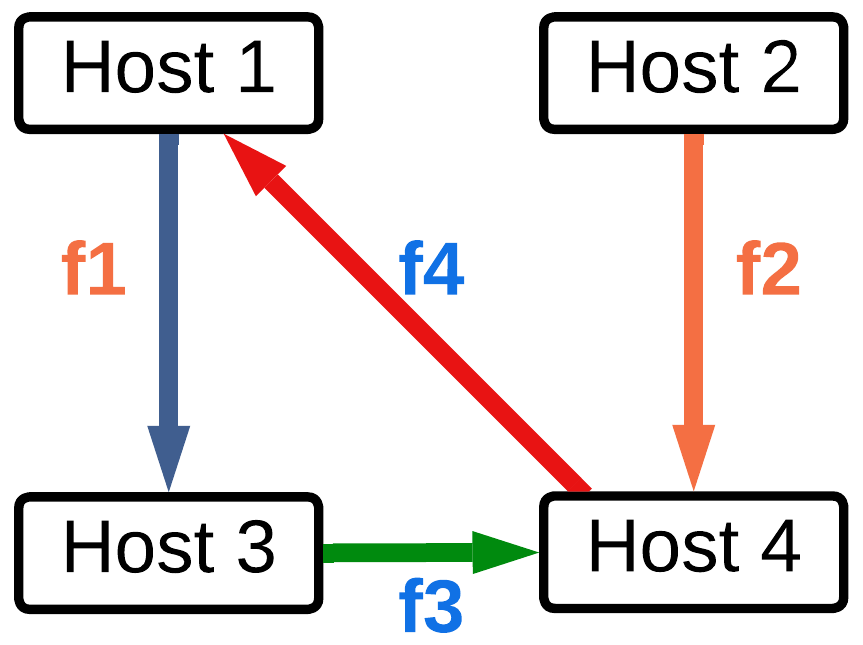}}
         \vspace{-6mm}
         \caption{Experiment setup.}
         \label{fig:altruistic-1}
     \end{subfigure}
     \hfill
     \begin{subfigure}[b]{0.235\textwidth}
         \centering
         \includegraphics[width=\textwidth]{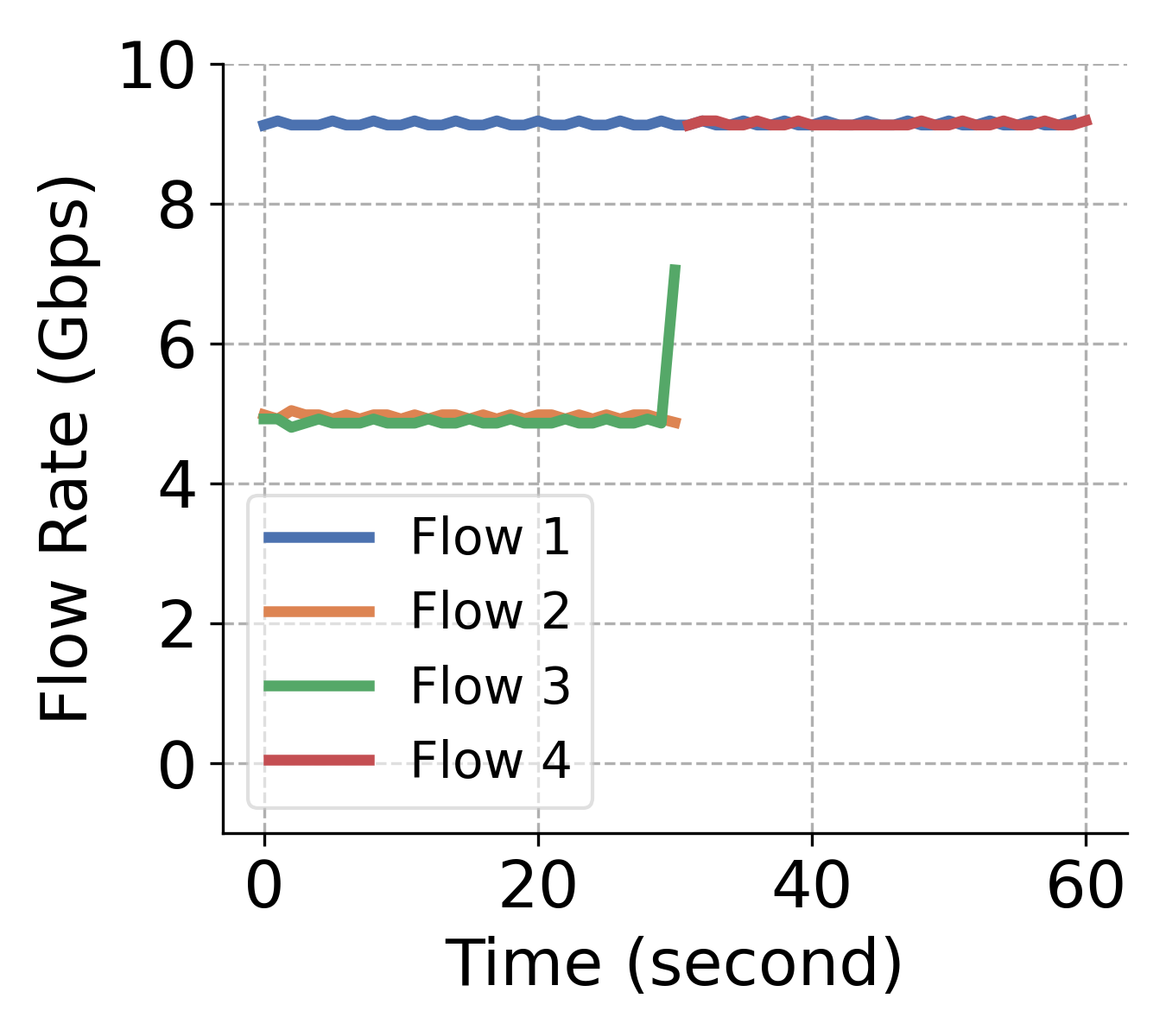}
         \vspace{-6mm}
         \caption{Fair allocation.}
         \label{fig:altruistic-2}
     \end{subfigure}
     \hfill
     \begin{subfigure}[b]{0.235\textwidth}
         \centering
         \includegraphics[width=\textwidth]{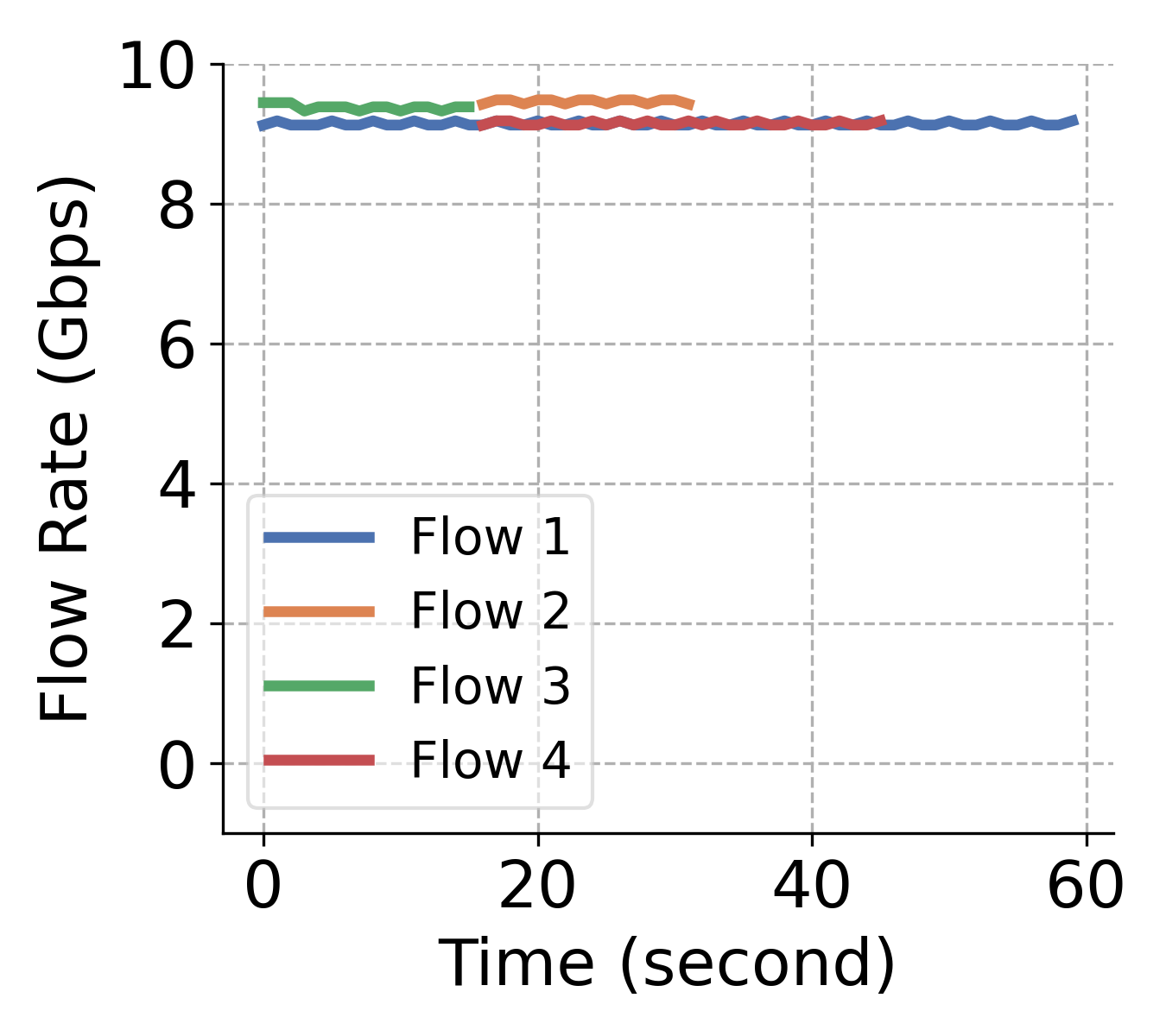}
         \vspace{-6mm}
         \caption{Optimal allocation.}
         \label{fig:altruistic-3}
     \end{subfigure}
     \hfill
     \begin{subfigure}[b]{0.235\textwidth}
         \centering
         \includegraphics[width=\textwidth]{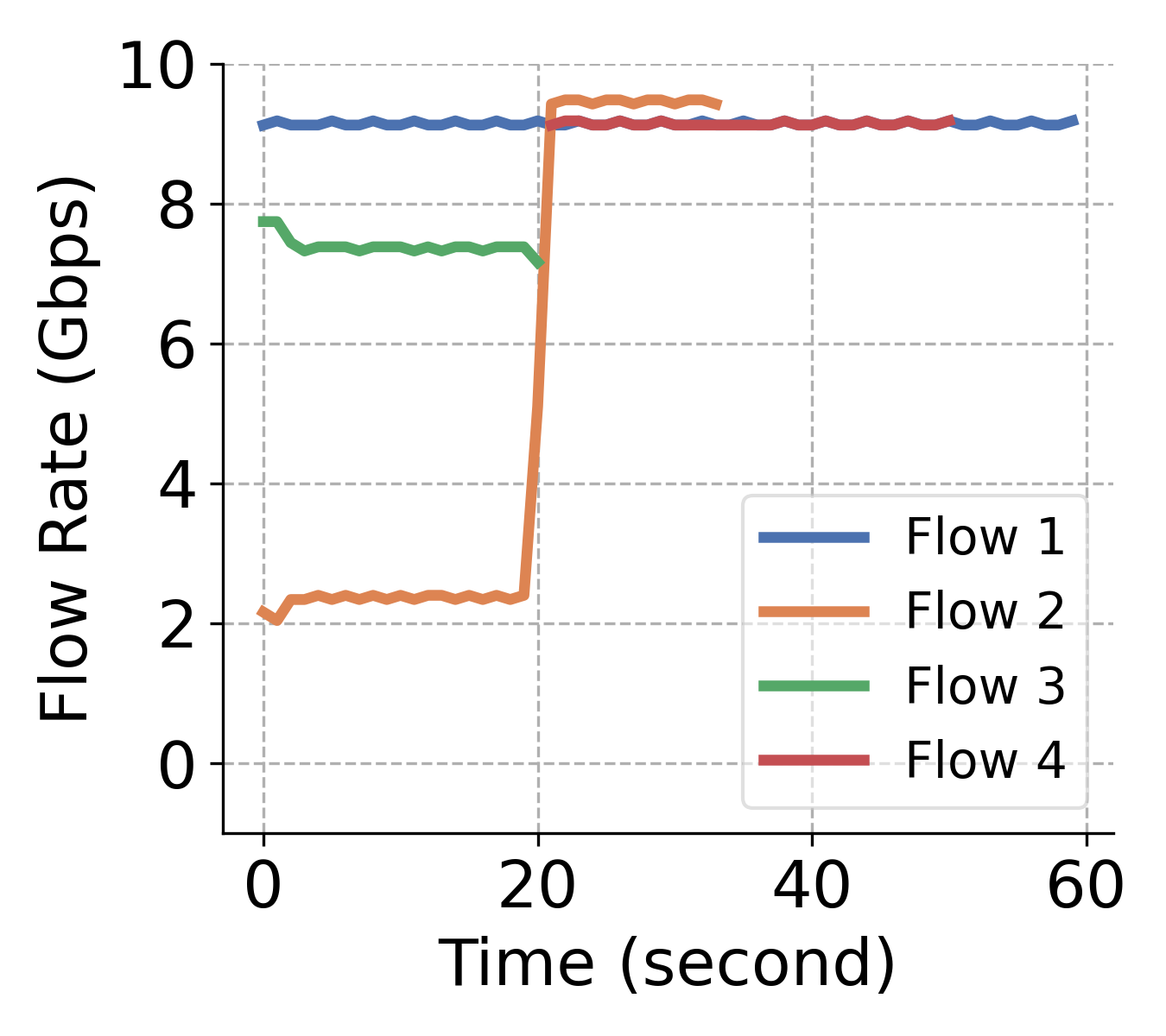}
         \vspace{-6mm}
         \caption{Altruistic allocation.}
         \label{fig:altruistic-4}
     \end{subfigure}
     \hfill
\vspace{-2mm}
\caption{a) The flows f1 (28 GB) and f2 (7 GB) belong to job 1, where f1 f2 start simultaneously. The flow f3 (7 GB) and f4 (14 GB) belongs to job 2, where f4 can only started when f3 finishes; b) With the fair allocation, f2 and f3 fairly share the bandwidth and job 2 finishes at time 60 s; c) Under the optimal allocation, f2 will initially give all bandwidth to f3. Thus, job 2 finishes at time 45 s; d) With the altruistic allocation, f2 shares the bandwidth with the guarantee that it can still finish on time with its bandwidth. The job 2 finishes at around time 50 s.}
\label{fig:altruistic}
\vspace{-4mm}
\end{figure*}

\vspace{-1mm}
\subsubsection{Share Resource on Common Bottleneck}

Another great property of \oursystemsoze is that we can reallocate the resource among flows within the same application, while the application generally keeps the same total bandwidth utilization and minimizes the effect on other applications' behavior. This property can be helpful, especially when some jobs share the same bottleneck in the network, such as the firewalls or load balancers. When applications share the same bottleneck, the resource allocation for each application can be perfectly isolated if the sum of weights remains the same. 

In this experiment, we have two jobs in total. As shown in \fref{fig:appfair}, one job has two parallel flows f1 and f2, while the other job has two other parallel flows f3 and f4. Both jobs need data from every parallel flow for further computation, so we want them to finish at the same time to minimize waiting time. Because all the flows have different sizes, we change the weight of every flow at time 100 seconds. 
    However, to keep the same job-level resource allocation, we choose the weight carefully so that the sum of all the flows' weights within one application remains the same.
    As you can see in Figure \fref{fig:appfair_d}, no matter before or after the weight changes, the bandwidth that each job occupied in total is always around 5 Gbps, while each individual flow's bandwidth has been updated depending on the size. 
In summary, by restricting the sum of all flows' weights insides one job to be a constant, each job reduce the interference with other jobs' resource; Only when different jobs shares the same bottleneck, there is no interference among different jobs.

\vspace{-1mm}
\subsubsection{Altruistic Scheduling among Multiple Jobs}

Although the weighted allocation could help us prioritize the critical execution path, sometimes the critical path cannot be accelerated due to resource limitations and job characteristics. In such cases, other non-critical execution paths could be altruistic by giving up some resources to other jobs, as long as the non-critical execution paths can finish at the same time as the critical path, the job completion time is not harmed. However, the bandwidth altruistically given to other jobs may benefit their completion time.

In the example of \fref{fig:altruistic}, we have two jobs: one job has two parallel flows, f1 and f2, while the other job has two sequential flows f3 and f4. 
In the fair allocation case, flows f2 and f3 fairly share the inbound bandwidth at host 4, and both finish at around 30 seconds. Then, flow f4 starts and finishes at 60 seconds. However, because flow f1 is much larger than flow f2, we could let flow f2 give up some bandwidth without hurting the job completion time. 
In the optimal allocation scenario, flow f2 can give all the bandwidth to flow f3 and achieve the minimum job completion time. However, this is not safe because one job does not know the size of the other job's flow.
Thus, in the altruistic allocation case, we let f2 get only 25\% of the bandwidth, because this is the minimum bandwidth to guarantee completion at the same time with flow f1. With this altruistic behavior, flow f3 finished much earlier at 20 seconds, so that its job also finished at time 60 seconds, which is 10 seconds earlier than the fair allocation.

\vspace{-1mm}
\subsubsection{Shortest-flow Prioritization}

In this experiment, we create an in-cast scenario where three hosts send RPC write requests to one host. 
Note that this experiment is not under the assumption that all flows come from the same job and need to follow the weight distribution policy. 
The size of the RPC write varies based on the workload, from around 700 MB to 7000 MB. Because the three senders may send RPCs of different sizes at the same time, prioritizing the RPCs with the smallest size could approximate shortest job first scheduling and reduce the overall average flow completion time. Thus, based on the flow size, we calculate a specific weight for flow with the equation $weight = \frac{max\_flow\_size}{flow\_size}$.

\begin{figure}[t!]
    \vspace{-4mm}
     \centering
     \hfill
     \begin{subfigure}[b]{0.235\textwidth}
         \centering
         \raisebox{5mm}{\includegraphics[width=\textwidth]{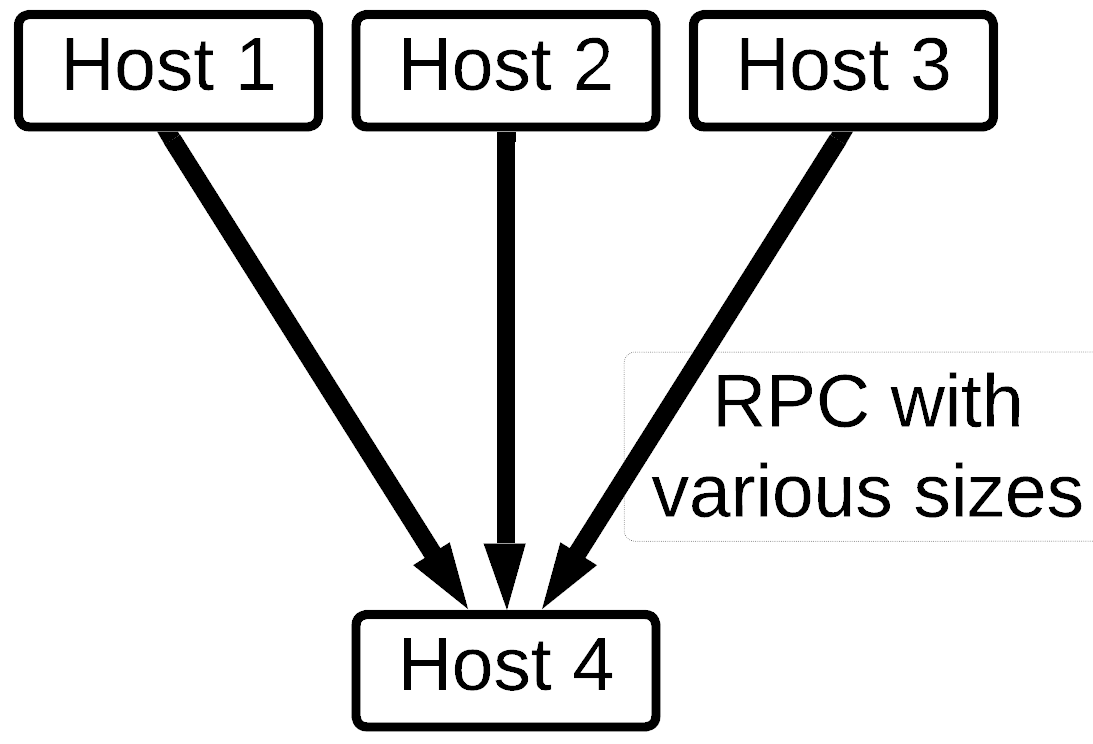}}
         \vspace{-7mm}
         \caption{Experiment setup.}
     \end{subfigure}
     \hfill
     \begin{subfigure}[b]{0.235\textwidth}
         \centering
         \includegraphics[width=\textwidth]{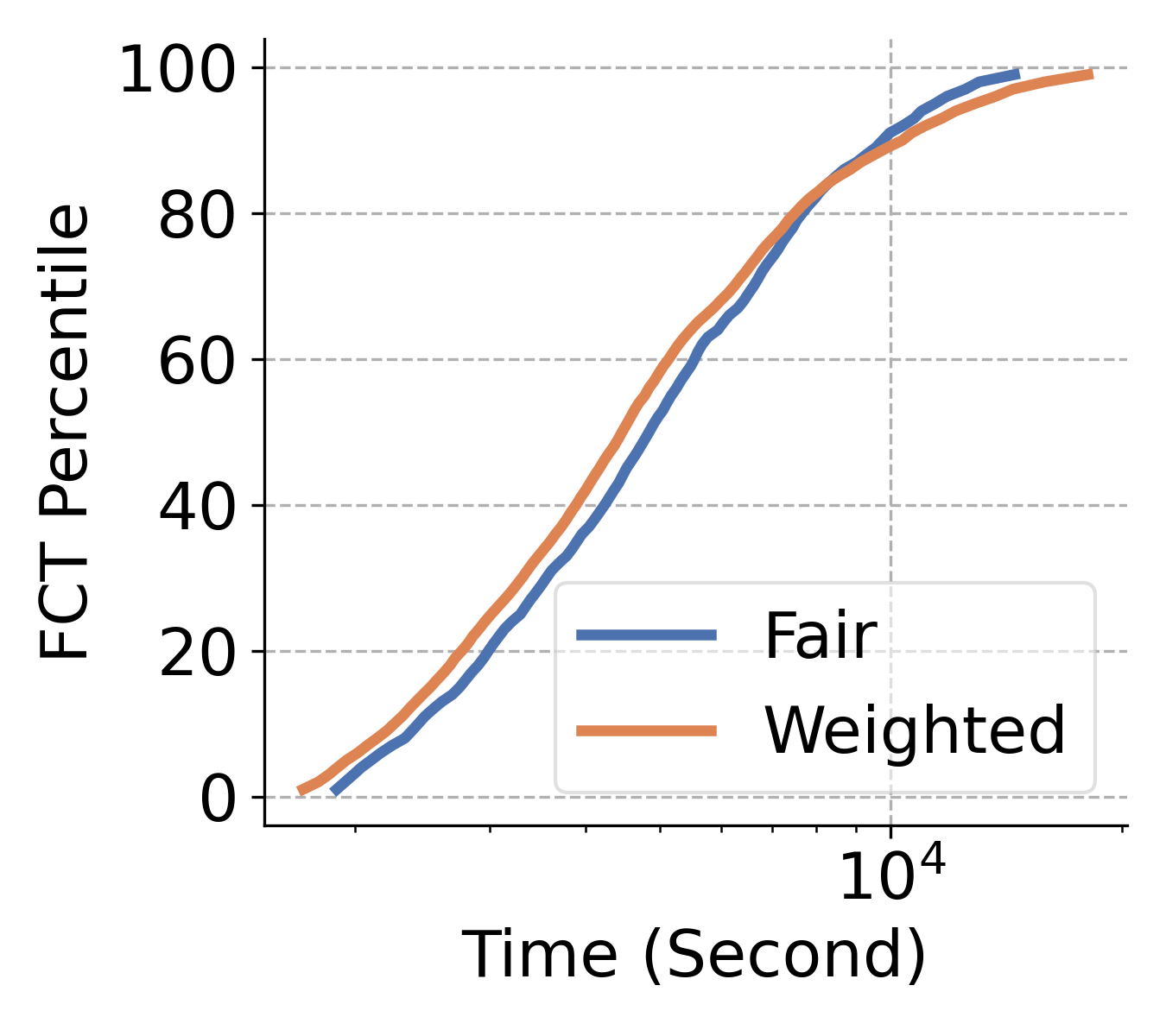}
         \vspace{-7mm}
         \caption{Flow completion time.}
     \end{subfigure}
     \hfill
\vspace{-6mm}
\caption{Shortest flow prioritization.}
\label{fig:sff}
\vspace{-6mm}
\end{figure}

As we can see in \fref{fig:sff}, when we update the weight according to the flow size, smaller flows finish faster than the fair allocation case. As a trade-off, only a small portion of large flows suffer from longer completion time than fair allocation, while more than 80\% of the flows benefit from the approximately shortest-flow first allocation. Moreover, unlike a strict shortest-flow first policy, like the preemption-based schedulers, \oursystemsoze never starves the large flows.

\vspace{-1mm}
\subsubsection{TPC-H Benchmark Acceleration}

\begin{figure}
    \centering
    \includegraphics[width=0.95\linewidth]{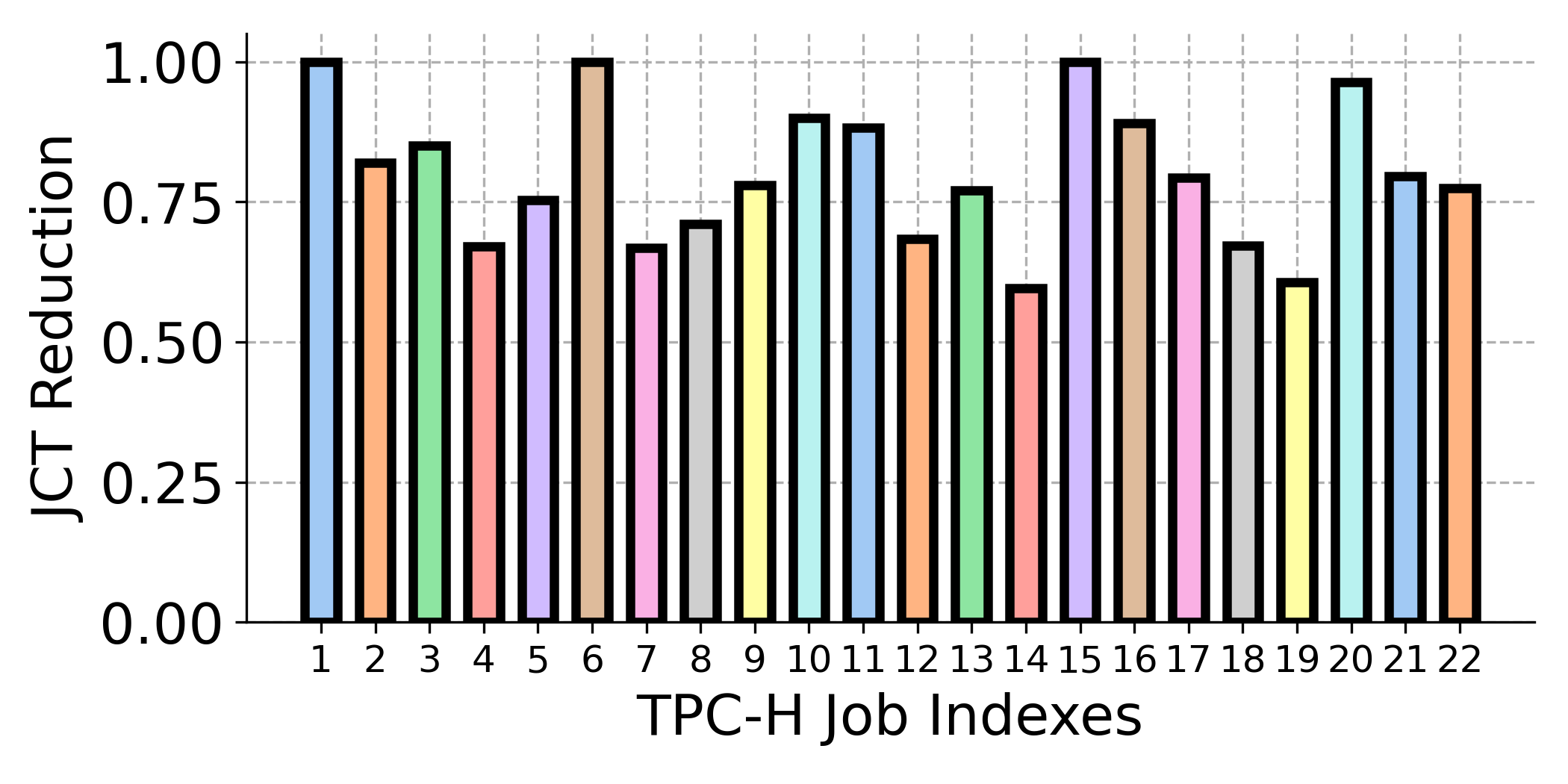}
    \vspace{-4mm}
    \caption{TPC-H jobs acceleration with weighted allocation.}
    \vspace{-4mm}
    \label{fig:tpch}
\end{figure}

Furthermore, we also test the 22 jobs from the TPC-H benchmarks~\cite{TPCH} in our NS-3 simulator, in which each job is an execution DAG with tasks. Each task is randomly placed on a host within a fat-tree topology. For every flow in the DAG, we calculate the longest distance from this flow to the end of the DAG and use this distance to determine the flow's weight. This simple policy is designed based on the insight that the longer execution path needs more resources. In \fref{fig:tpch}, with \oursystemsoze providing the weighted allocation, the average job completion time is reduced by $0.79\times$, and the maximum reduction is $0.59\times$. Only jobs that have only one execution path do not receive benefits from \oursystemsoze.

\vspace{-1mm}
\subsection{Efficient Congestion Control}
\label{subsec:evaluation2}

\vspace{-1mm}
\subsubsection{Step-in \& Step-out Experiment in Testbed}

To show that \oursystemsoze can serve as an efficient congestion control algorithm, we conduct a step-in and step-out experiment on the eRPC testbed with three hosts sending traffic to one host. Every 50 seconds, a new flow was added to the incast host from a different sender until time 100 seconds, then every 50 seconds, one flow will terminate and give bandwidth back. 

\begin{figure}[t!]
     \centering
     \hfill
     \begin{subfigure}[b]{0.235\textwidth}
         \centering
         \includegraphics[width=\textwidth]{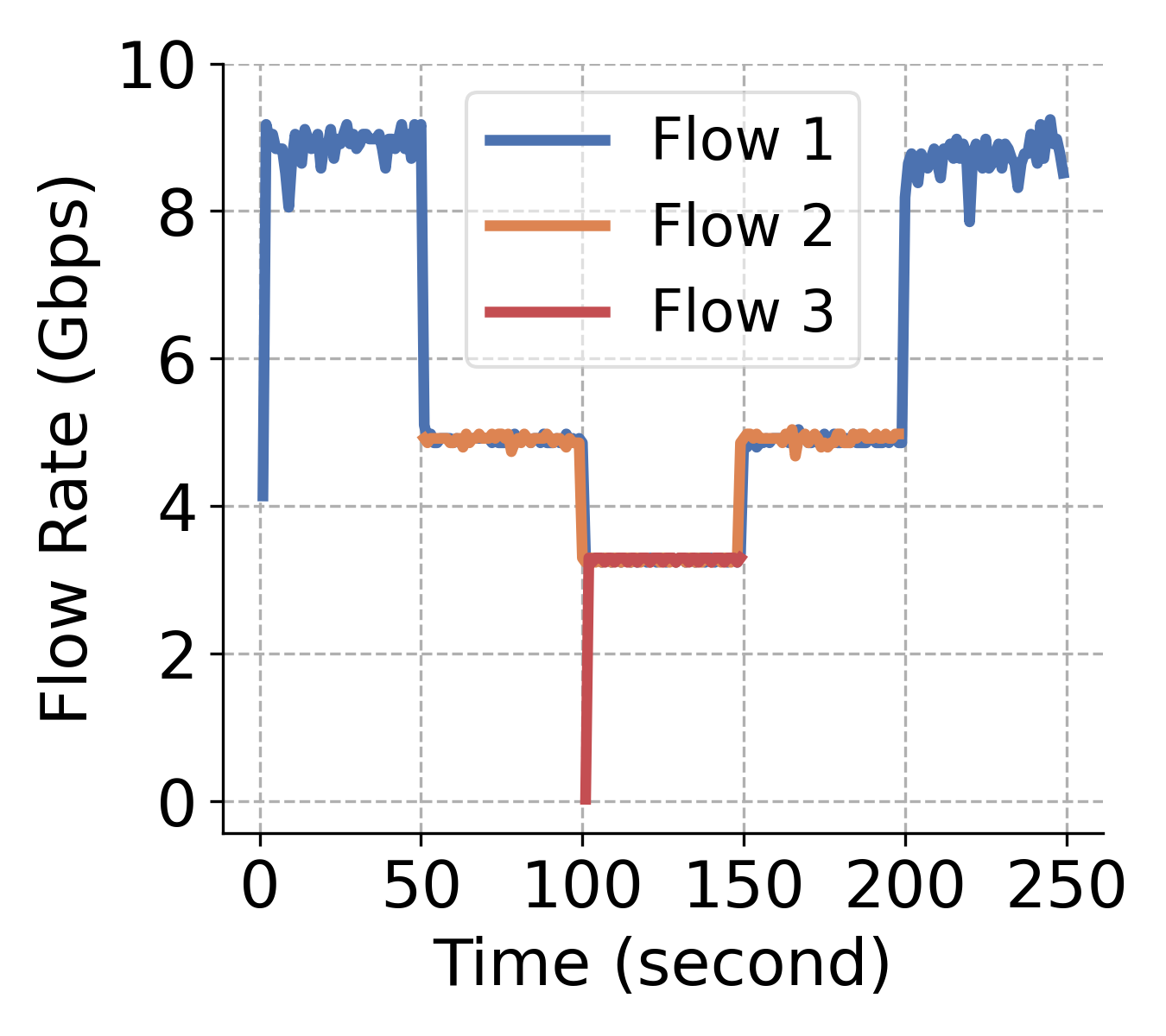}
         \vspace{-6mm}
         \caption{Timely.}
         \label{fig:stepinout-erpc-1}
     \end{subfigure}
     \hfill
     \begin{subfigure}[b]{0.235\textwidth}
         \centering
         \includegraphics[width=\textwidth]{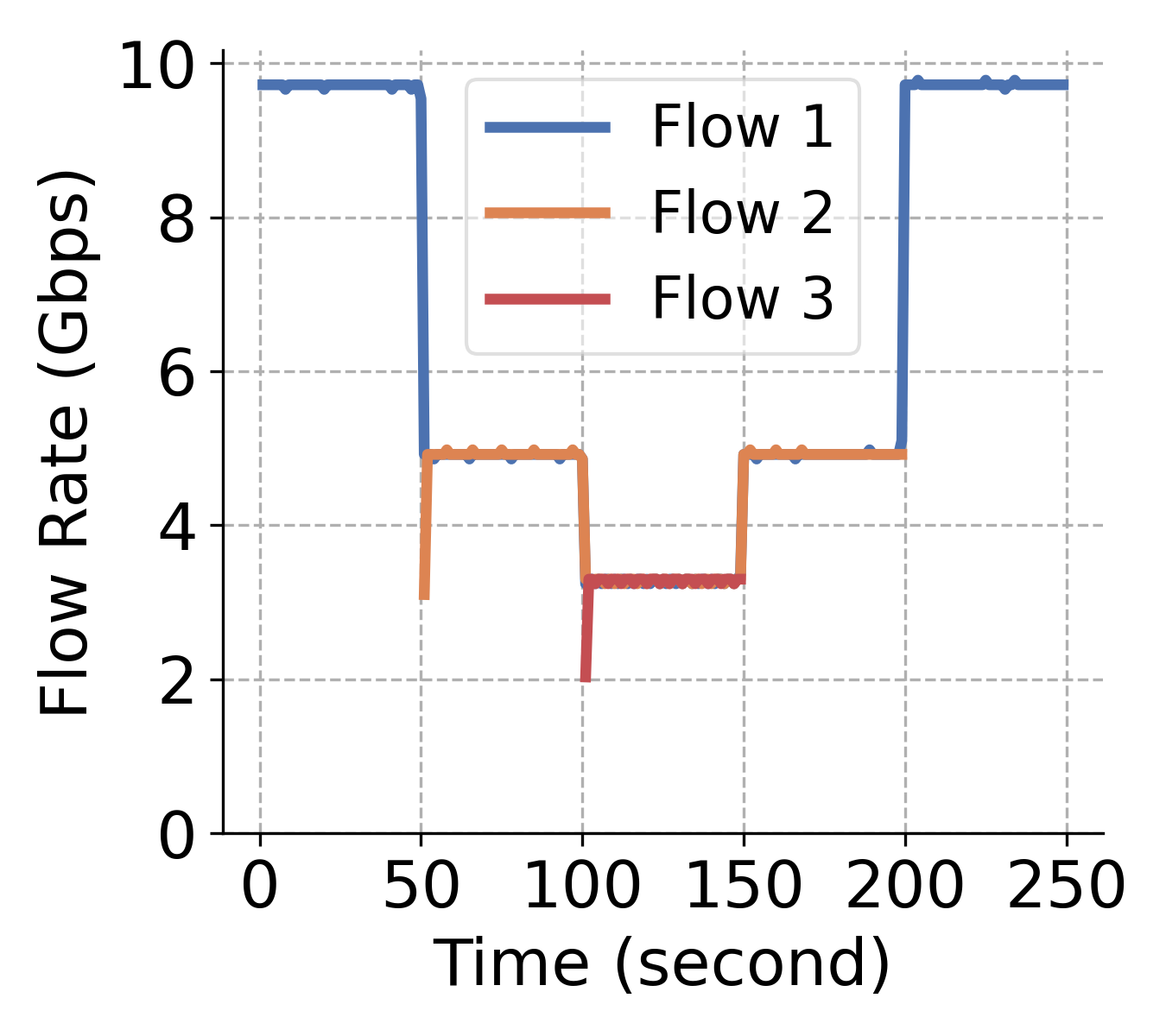}
         \vspace{-6mm}
         \caption{\oursystemsoze.}
         \label{fig:stepinout-erpc-2}
     \end{subfigure}
     \hfill
\vspace{-6mm}
\caption{Step-in \& step-out experiment on eRPC testbed. \oursystemsoze achieves higher utilization and faster convergence to the optimal state than Timely.}
\label{fig:stepinout-erpc}
\vspace{-5mm}
\end{figure}

As shown in \fref{fig:stepinout-erpc}, \oursystemsoze outperforms Timely, the default congestion control algorithm in eRPC.
    When there is only one flow, \oursystemsoze achieves higher utilization than Timely; 
    and when a new flow arrives or an existing flow completes, \oursystemsoze converges to the new optimal allocation faster than Timely because of the proposed adaptive MIMD algorithm.

\begin{figure}[t!]
     \centering
     \hfill
     \begin{subfigure}[b]{0.235\textwidth}
         \centering
         \includegraphics[width=\textwidth]{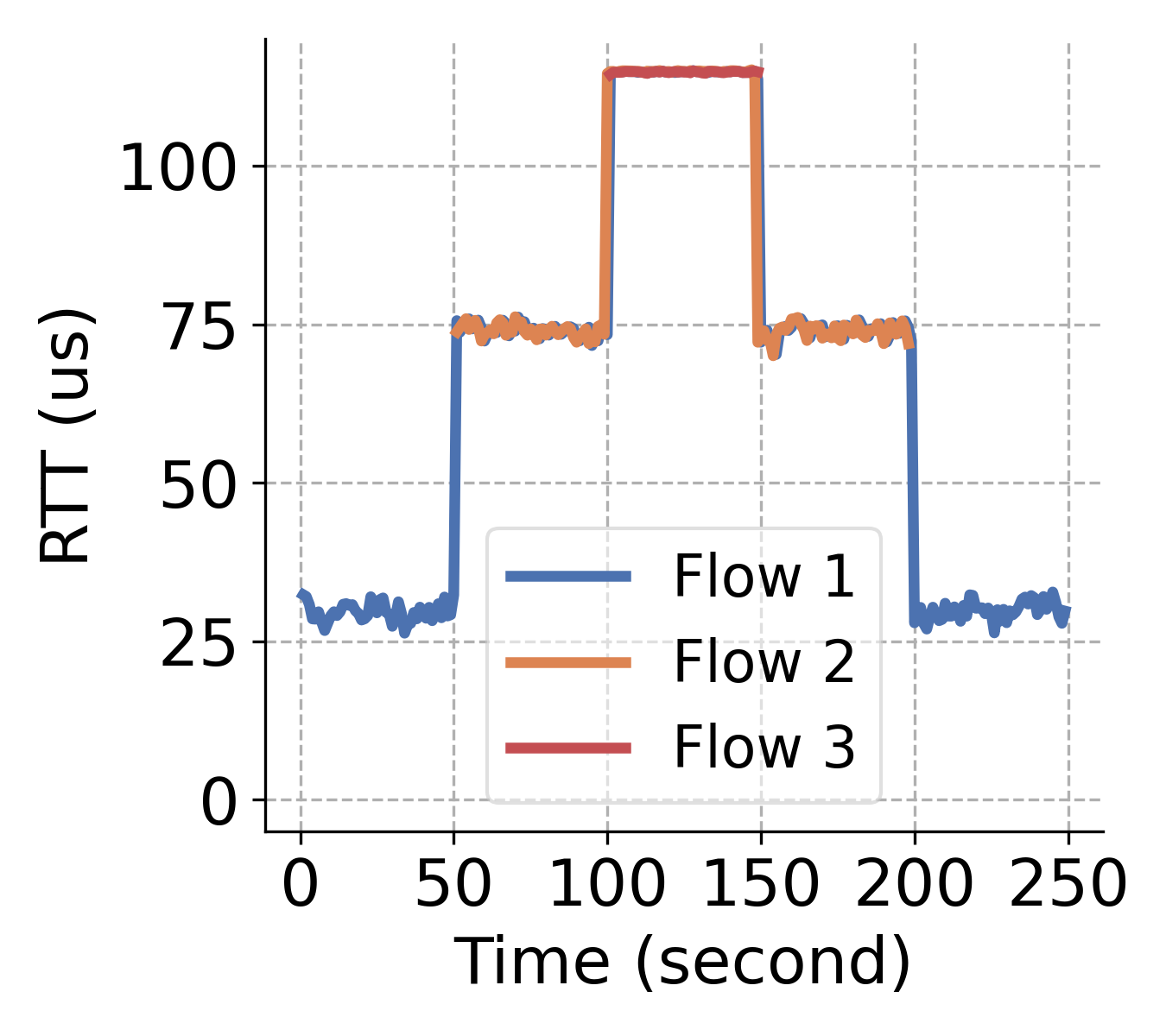}
         \vspace{-6mm}
         \caption{Timely.}
         \label{fig:stepinout-erpc-rtt-1}
     \end{subfigure}
     \hfill
     \begin{subfigure}[b]{0.235\textwidth}
         \centering
         \includegraphics[width=\textwidth]{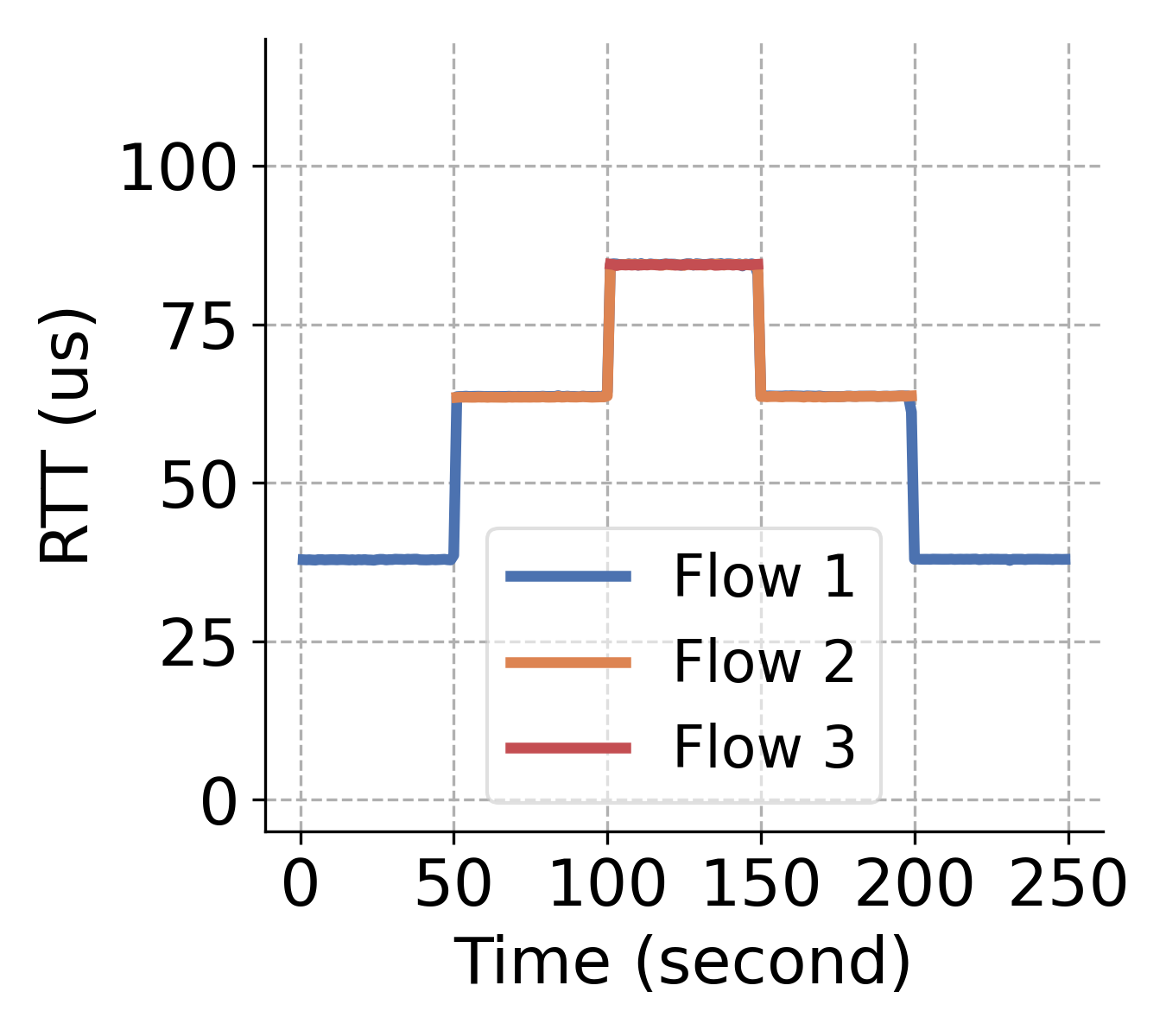}
         \vspace{-6mm}
         \caption{\oursystemsoze.}
         \label{fig:stepinout-erpc-rtt-2}
     \end{subfigure}
     \hfill
\vspace{-6mm}
\caption{\oursystemsoze achieves lower RTT than Timely by using queueing level as the signal to indicate congestion.}
\label{fig:stepinout-erpc-rtt}
\vspace{-5mm}
\end{figure}

In \fref{fig:stepinout-erpc-rtt}, we also show the round-trip time (RTT) for both Timely and \oursystemsoze. The RTT of \oursystemsoze increases with more number of flows competing on the link, because our target function is monotonically decreasing as the fair-share rate of the link increases. Nevertheless, \oursystemsoze still achieves lower RTT than Timely, because we use a range of queueing delay to indicate different levels of congestion, and the maximum queueing delay in \eref{equ:target} bounds the RTT of \oursystemsoze.

\vspace{-1mm}
\subsubsection{Step-in \& Step-out Experiment in NS-3}

\begin{figure}[t!]
     \centering
     \hfill
     \begin{subfigure}[b]{0.235\textwidth}
         \centering
         \includegraphics[width=\textwidth]{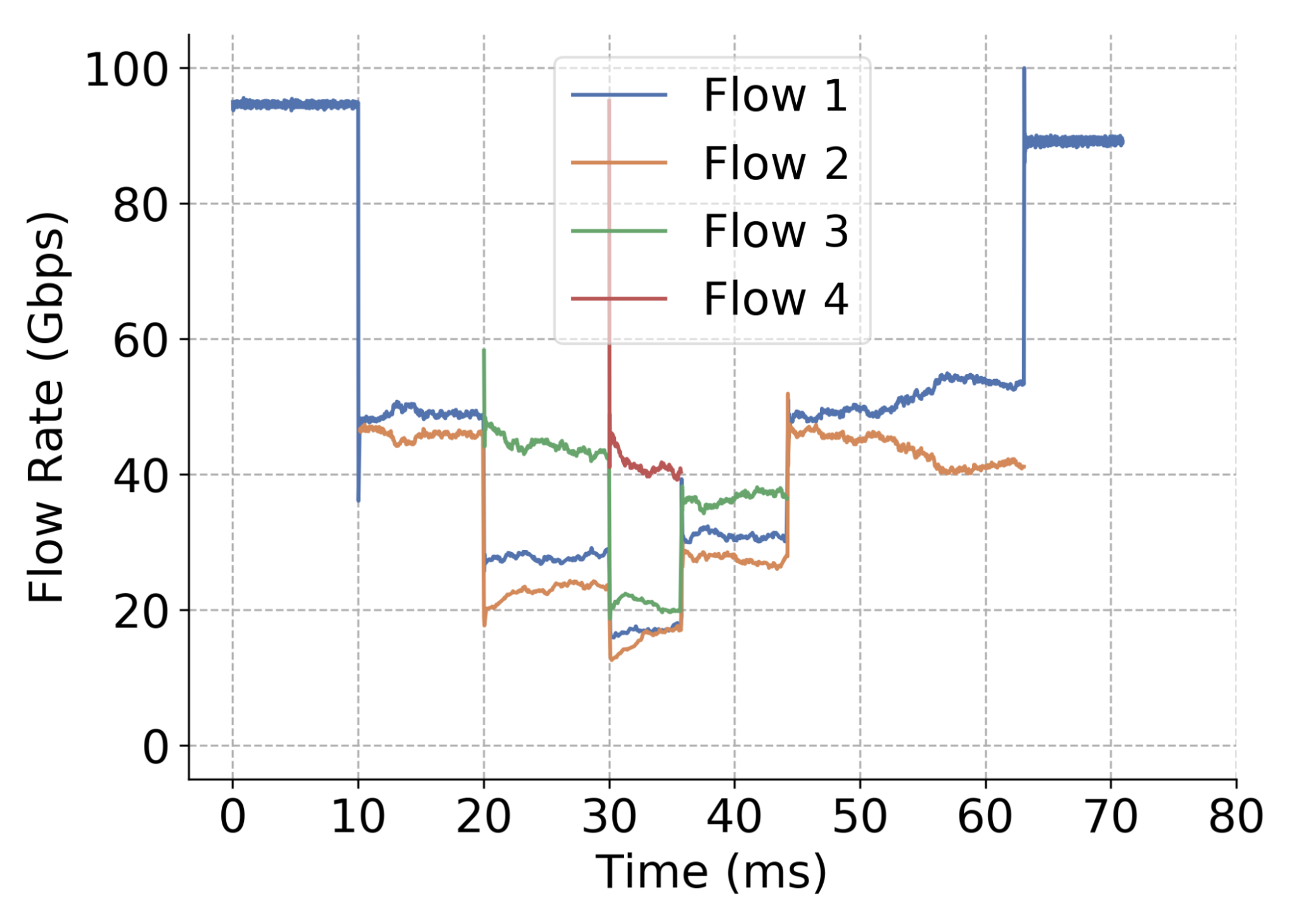}
         \vspace{-6mm}
         \caption{HPCC.}
         \label{fig:stepinout-ns3-1}
     \end{subfigure}
     \hfill
     \begin{subfigure}[b]{0.235\textwidth}
         \centering
         \includegraphics[width=\textwidth]{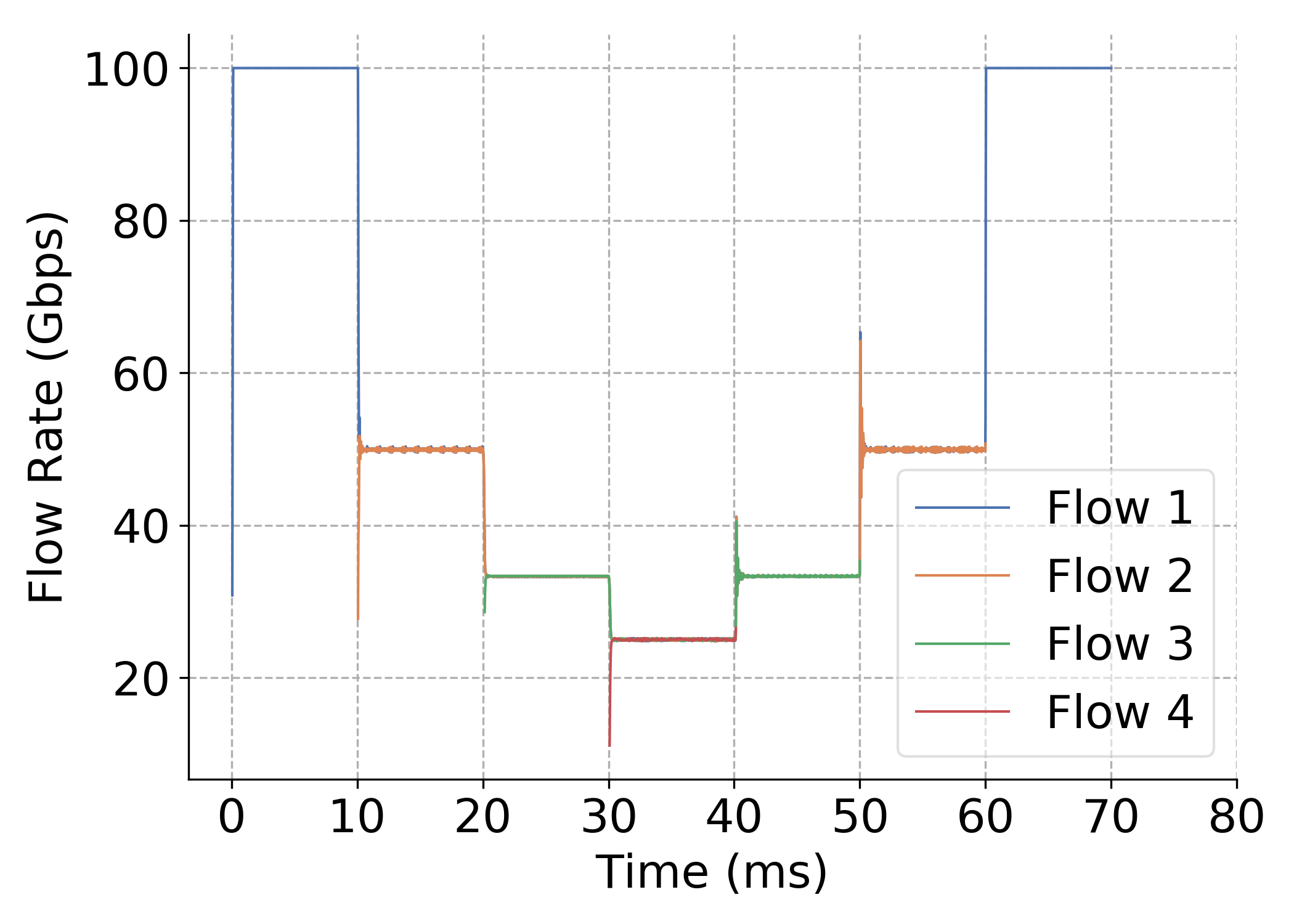}
         \vspace{-6mm}
         \caption{\oursystemsoze.}
         \label{fig:stepinout-ns3-2}
     \end{subfigure}
     \hfill
\vspace{-6mm}
\caption{Step-in \& step-out experiment in NS-3 simulator. Compared with HPCC, \oursystemsoze achieves a more stable and accurate rate allocation.}
\label{fig:stepinout-ns3}
\vspace{-5mm}
\end{figure}

In the NS-3 simulator, we also conducted a step-in and step-out experiment with hosts under different ToRs sending traffic to the same destination. We compared \oursystemsoze with one recent data center congestion control algorithm using in-network telemetry --- HPCC.
In this set of experiment, 4 flows are added every 10 milliseconds, and each flow completes every 10 milliseconds in sequence.

Compared to HPCC, the bandwidth allocation in \oursystemsoze is very accurate and stable in \fref{fig:stepinout-ns3}. Because \oursystemsoze uses the queueing delay as the signal to indicate the level of congestion, more specifically, the fair-share rate of the link. While for HPCC, the in-network signal is only used to indicate whether the link is congested or not, so that the bandwidth allocation is not accurate nor stable for HPCC due to the binary status provided by its in-network signal.

\vspace{-1mm}
\subsubsection{FCT Slowdown}

\begin{figure}[t!]
     \centering
     \hfill
     \begin{subfigure}[b]{0.235\textwidth}
         \centering
         \includegraphics[width=\textwidth]{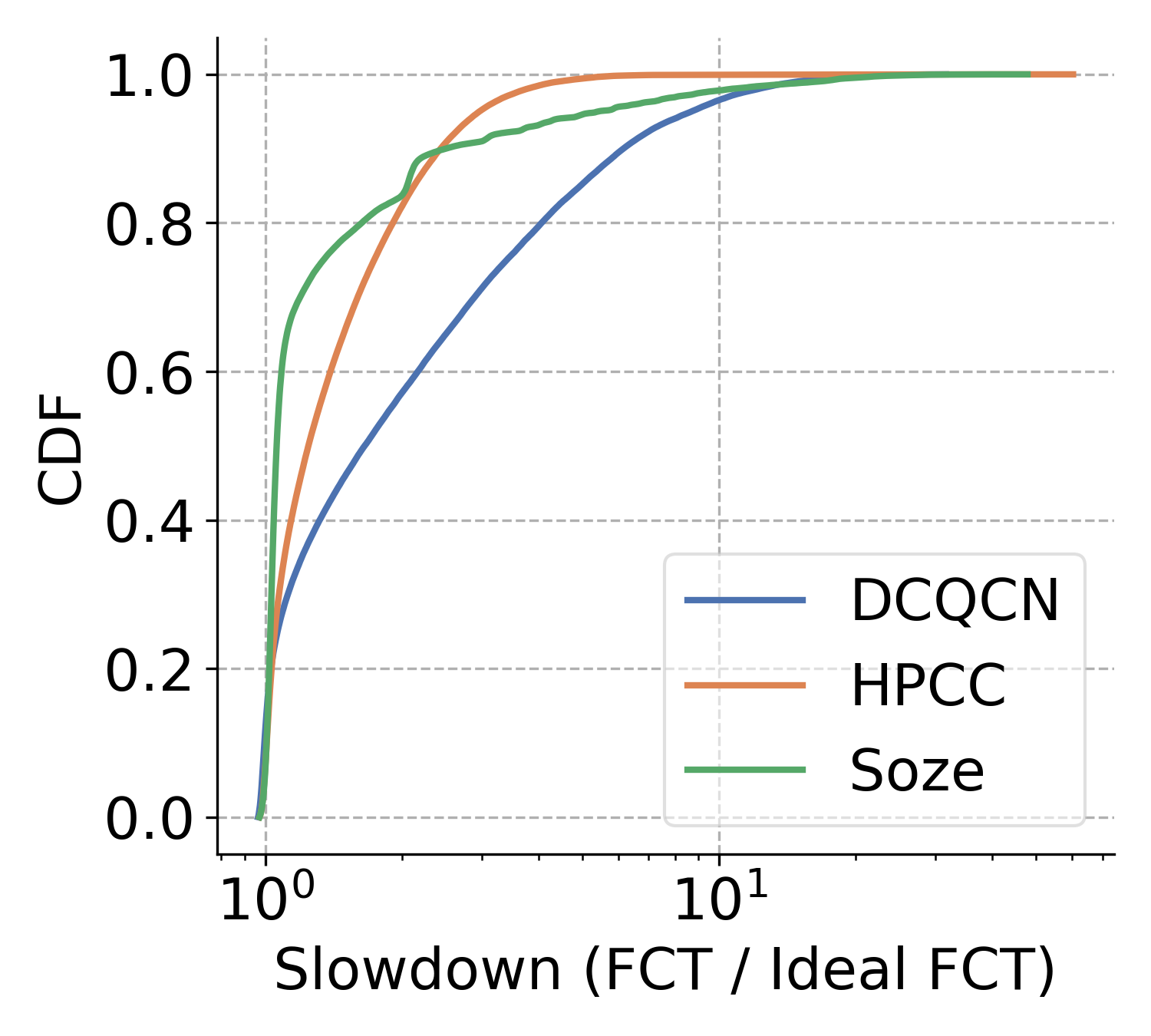}
         \vspace{-6mm}
         \caption{40\% load.}
         \label{fig:fct-slowdown-1}
     \end{subfigure}
     \hfill
     \begin{subfigure}[b]{0.235\textwidth}
         \centering
         \includegraphics[width=\textwidth]{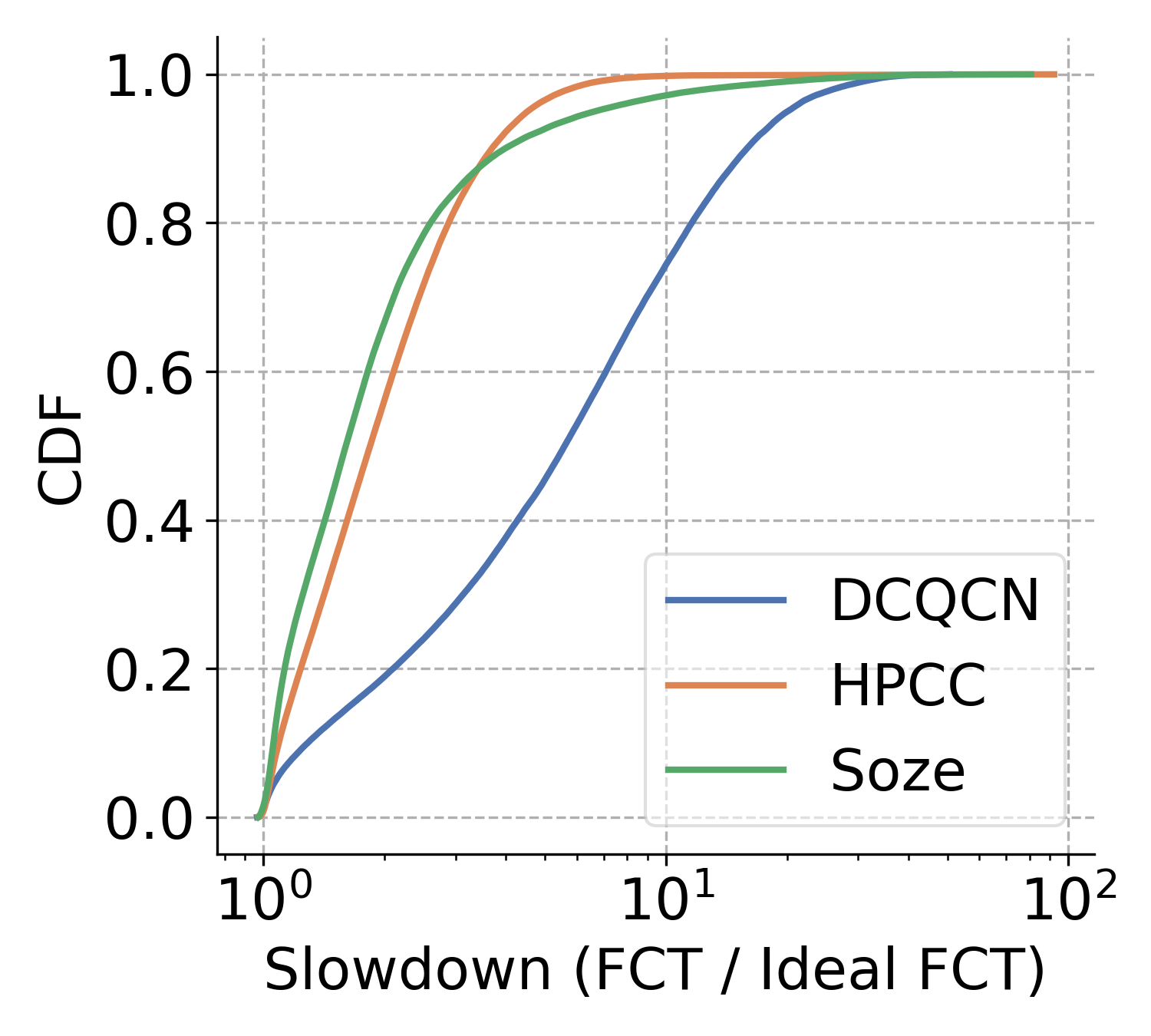}
         \vspace{-6mm}
         \caption{80\% load.}
         \label{fig:fct-slowdown-2}
     \end{subfigure}
     \hfill
\vspace{-6mm}
\caption{FCT slowdown under different network loads.}
\label{fig:fct-slowdown}
\vspace{-4mm}
\end{figure}

In order to evaluate the performance of \oursystemsoze under high network load with a large amount of flows, we run the Google RPC workload~\cite{sigelman2010dapper} with randomly selected sender and receiver. And we compare with two datacenter congestion control solutions --- HPCC and DCQCN. 

\begin{figure}[t!]
     \centering
     \hfill
     \begin{subfigure}[b]{0.235\textwidth}
         \centering
         \includegraphics[width=\textwidth]{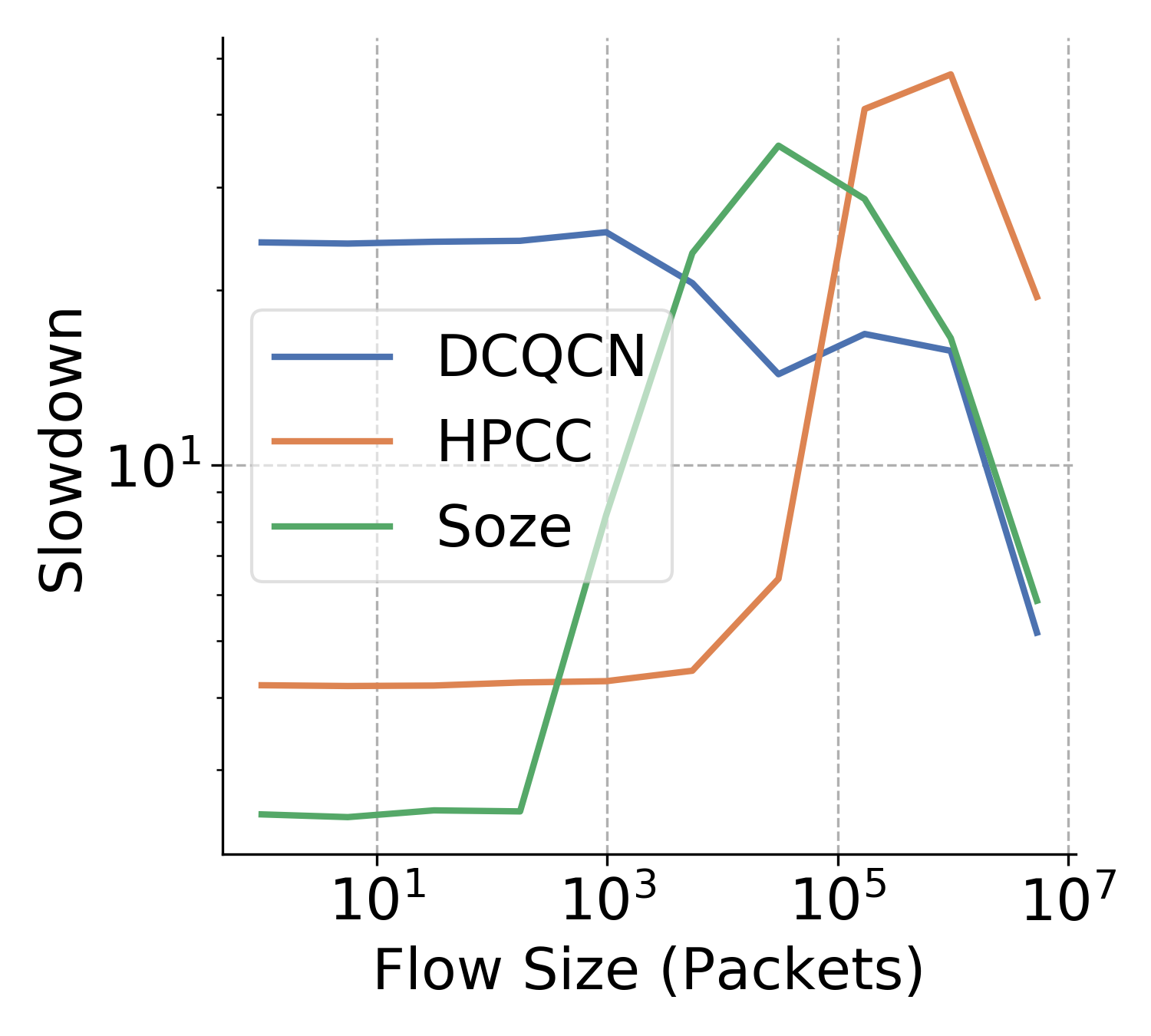}
         \vspace{-6mm}
         \caption{40\% load.}
         \label{fig:fct-slowdown-f-1}
     \end{subfigure}
     \hfill
     \begin{subfigure}[b]{0.235\textwidth}
         \centering
         \includegraphics[width=\textwidth]{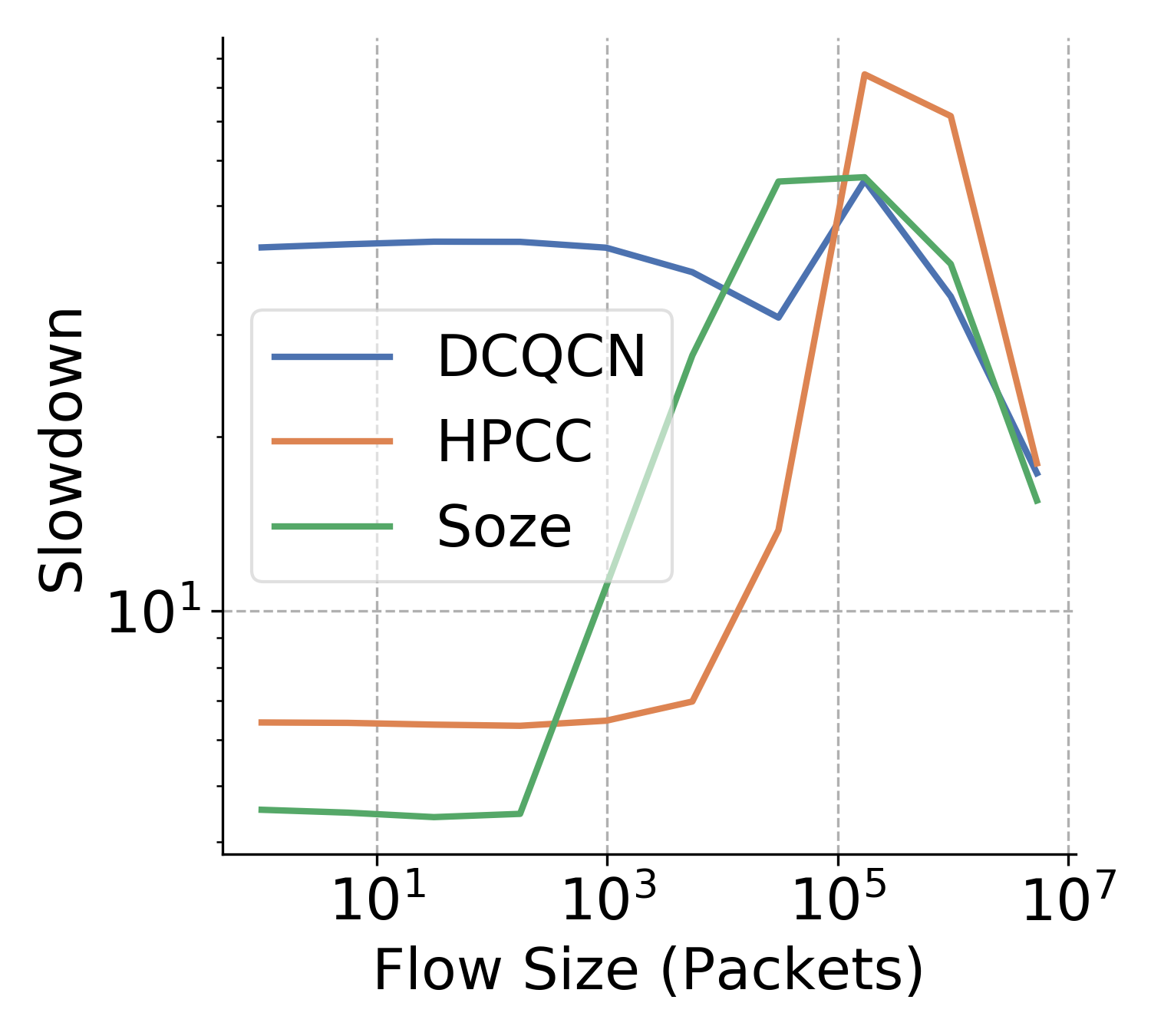}
         \vspace{-6mm}
         \caption{80\% load.}
         \label{fig:fct-slowdown-f-2}
     \end{subfigure}
     \hfill
\vspace{-6mm}
\caption{FCT slowdown for different flow size under different network loads.
}
\label{fig:fct-slowdown-f}
\vspace{-4mm}
\end{figure}

In \fref{fig:fct-slowdown}, under both 40\% and 80\% load, \oursystemsoze achieves lower FCT slowdown than HPCC and DCQCN, especially for the short flows. Because \oursystemsoze could grab bandwidth rapidly with fast convergence. 
As shown in \fref{fig:fct-slowdown-f}, \oursystemsoze provides more benefits for flows with less than 1000 packets. While for flows with larger data size (> 100k packets), \oursystemsoze also provides comparable performance with DCQCN and outperforms HPCC, leading to shorter tail FCT slowdown in \fref{fig:fct-slowdown}. 
\vspace{-1mm}
\subsection{Micro-benchmark}
\label{subsec:evaluation3}

In this subsection, we compare \oursystemsoze with alternative solutions with a set of micro-benchmark experiments on the NS-3 simulator to demonstrate \oursystemsoze's weighted max-min fairness, high agility, fine granularity, and scalability.

\vspace{-1mm}
\subsubsection{Weighted Max-min Fairness}

In the experiment for weighted max-min fairness, we create a scenario where flow 1 only travels switch 1; flow 2, 3, and 4 travel switch 1 and switch 2; flow 4 and 5 only travel switch 2. 
    The link bandwidth is 100 Gbps and the default weight for every flow is 1, so that flow 1 get 40 Gbps at the beginning. 
    Every 10 milliseconds, the weight of flow 1 will be increased by 1. 
        When the weight is 2, flow 1 still gets 40 Gbps because the bottleneck hop for flow 2, 3, and 4 is still switch 2. 
        Only when the weight increases to be higher than 3, the bottleneck hop becomes switch 1 and flow 1 will take more bandwidth from other flows on switch 1.

\begin{figure*}[t!]
     \centering
     \hfill
     \begin{subfigure}[b]{0.235\textwidth}
         \centering
         \raisebox{6mm}{\includegraphics[width=\textwidth]{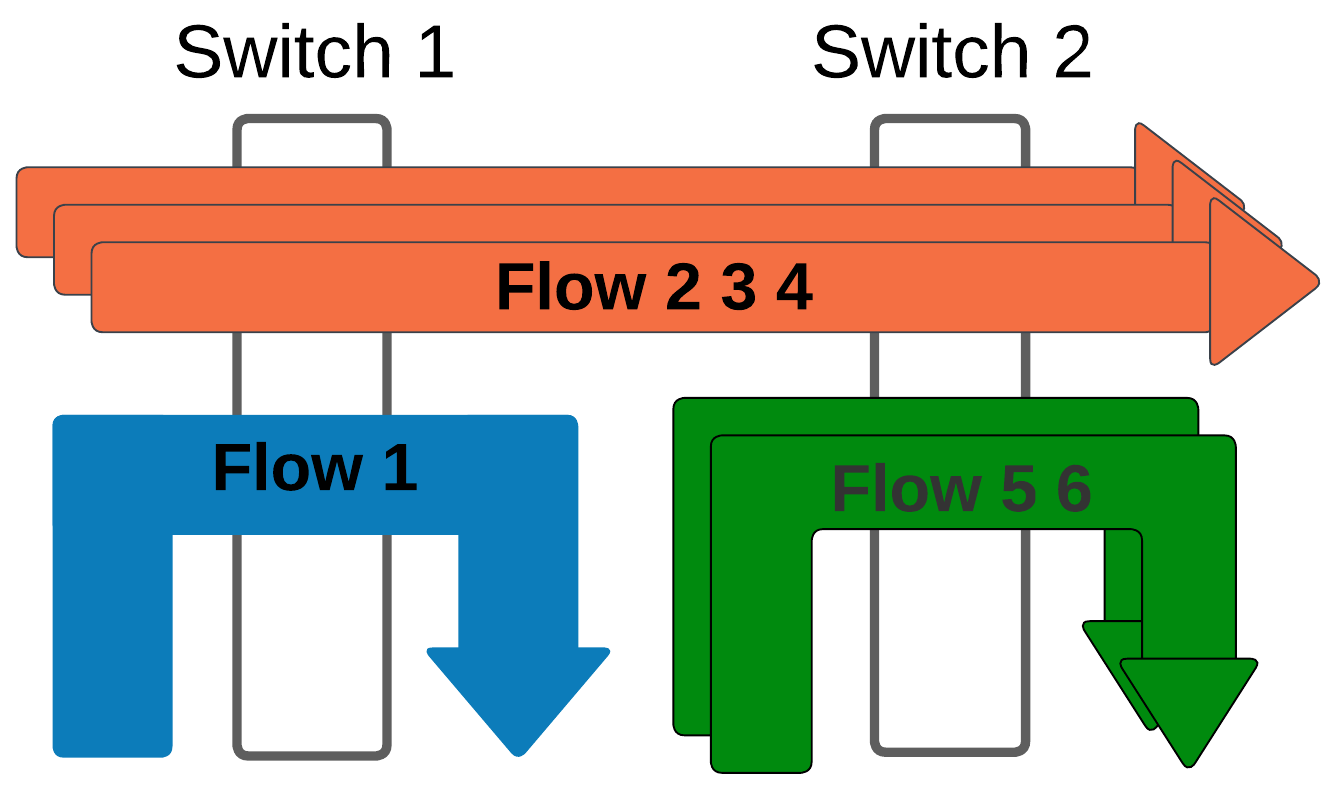}}
         \vspace{-7mm}
         \caption{Experiment scenario.}
         \label{fig:maxmin1}
     \end{subfigure}
     \hfill
     \begin{subfigure}[b]{0.235\textwidth}
         \centering
         \includegraphics[width=\textwidth]{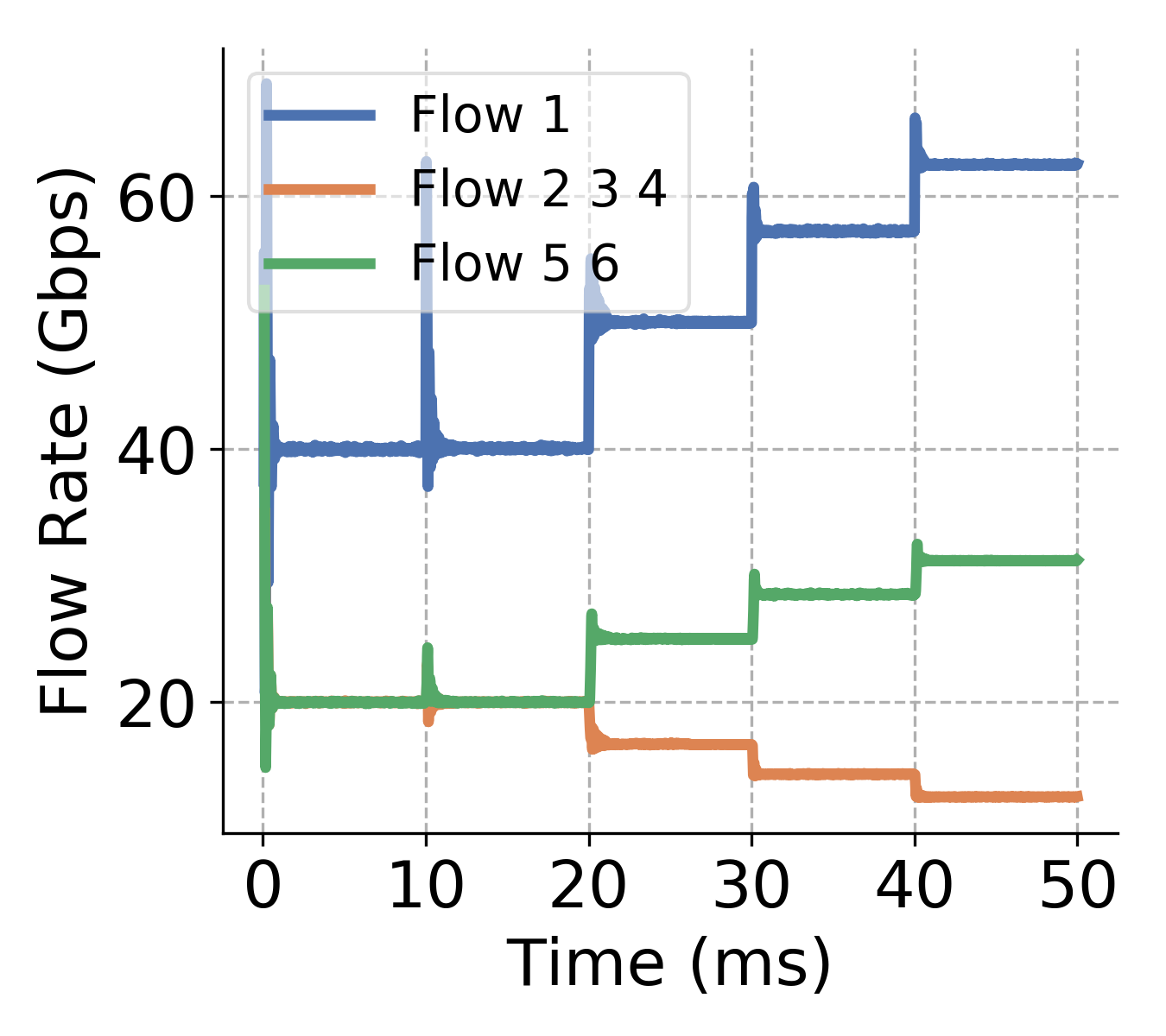}
         \vspace{-7mm}
         \caption{\oursystemsoze with increasing weight.}
         \label{fig:maxmin2}
     \end{subfigure}
     \hfill
     \begin{subfigure}[b]{0.235\textwidth}
         \centering
         \includegraphics[width=\textwidth]{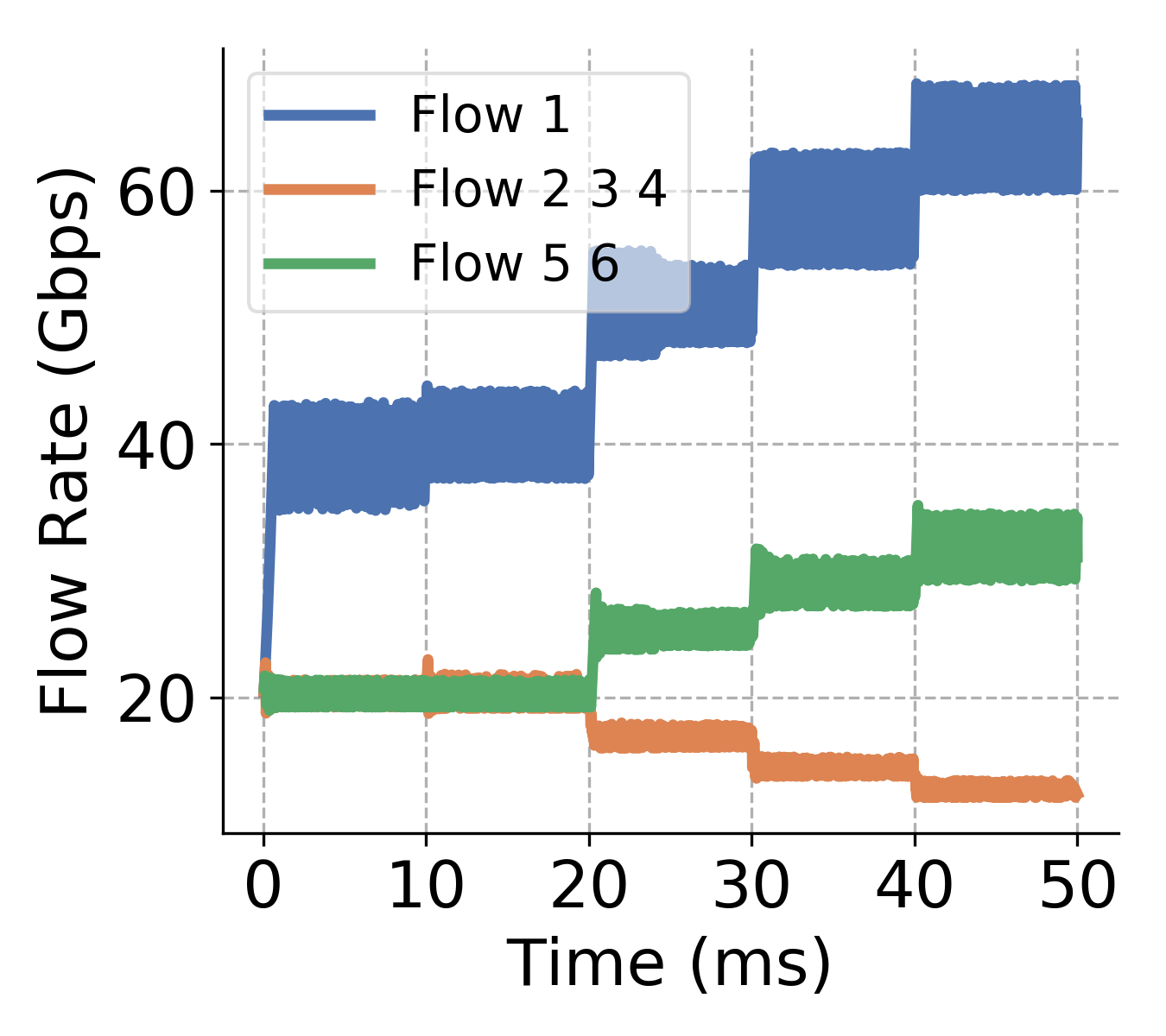}
         \vspace{-7mm}
         \caption{WRR with AIMD.}
         \label{fig:maxmin3}
     \end{subfigure}
     \hfill
\vspace{-3mm}
\caption{a) Flow 1 and flow 2, 3, 4 compete on switch 1; flow 5 and 6 compete with flows on switch 2. During time 0 ms to time 50 ms, we increase the weight of flow 1 from 1 to 5 every 10 ms. b) \oursystemsoze achieves accurate weighted sharing. When the weight of flow 1 is smaller than 2, flow 1 remains at 40 Gbps since switch 2 is the most congested hop; but when the weight is higher than 2, flow 1's rate increases because the congested hop has become switch 1. c) For comparison, we repeat the experiment using weighted round-robin (WRR) scheduling in switches to achieve flow 1's increasing weight; AIMD rate control is used to adapt to the weight changes.}
\label{fig:max-min}
\vspace{-2mm}
\end{figure*}

As we described in \secref{subsec:design2}, the weighted max-min fairness for every flow is determined by its bottleneck hop. In \fref{fig:max-min}, we show that the bottleneck hop for flow 2, 3, and 4 changes with the increase of flow 1's weight.
With this micro-benchmark experiment, we demonstrate that \oursystemsoze can recognize the bottleneck hop for every flow and achieve weighted max-min fairness rapidly after the weight is changed for any flow in the network. Moreover, the bottleneck hop changes at time 20 ms, and \oursystemsoze handles the changing bottleneck rapidly within 10 RTTs.
For comparison, we test how weighted round-robin (WRR) scheduling in switches combined with AIMD rate control would perform in this scenario. WRR enforces weighted sharing of bandwidth at the packet scheduling level but senders must still discover the achievable bandwidth at the end-to-end level. 
The AIMD control uses queuing delay as the explicit feedback signal: when the queuing delay is below 20 $\mu$s (this value is chosen to ensure good link utilization in the experiment), the CWND is increased by 1; when the queuing delay exceeds 20 $\mu$s, the CWND is reduced by $20\%$ (this value is chosen to limit oscillation). In the results in \fref{fig:maxmin3}, we can see that the rate for each flow is maintained correctly around the weighted max-min fair allocation by WRR scheduling. However, the link utilization is lower than with \oursystemsoze because the oscillation due to AIMD is larger.

\begin{figure*}[t!]
     \centering
     \hfill
     \begin{subfigure}[b]{0.235\textwidth}
         \centering
         \raisebox{3mm}{\includegraphics[width=\textwidth]{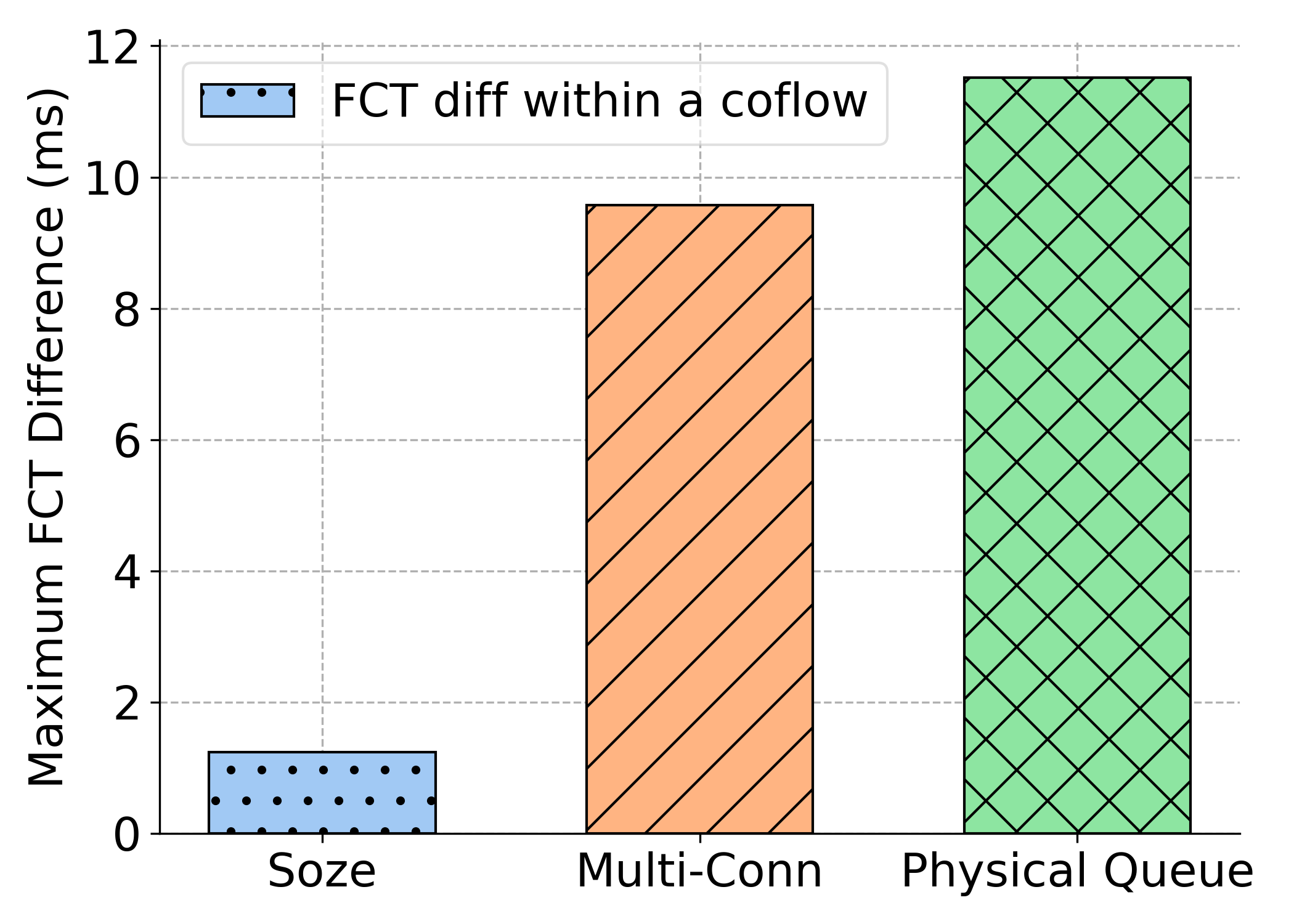}}
         \vspace{-6mm}
         \caption{Granularity comparison.}
         \label{fig:granularity4}
     \end{subfigure}
     \hfill
     \begin{subfigure}[b]{0.235\textwidth}
         \centering
         \includegraphics[width=\textwidth]{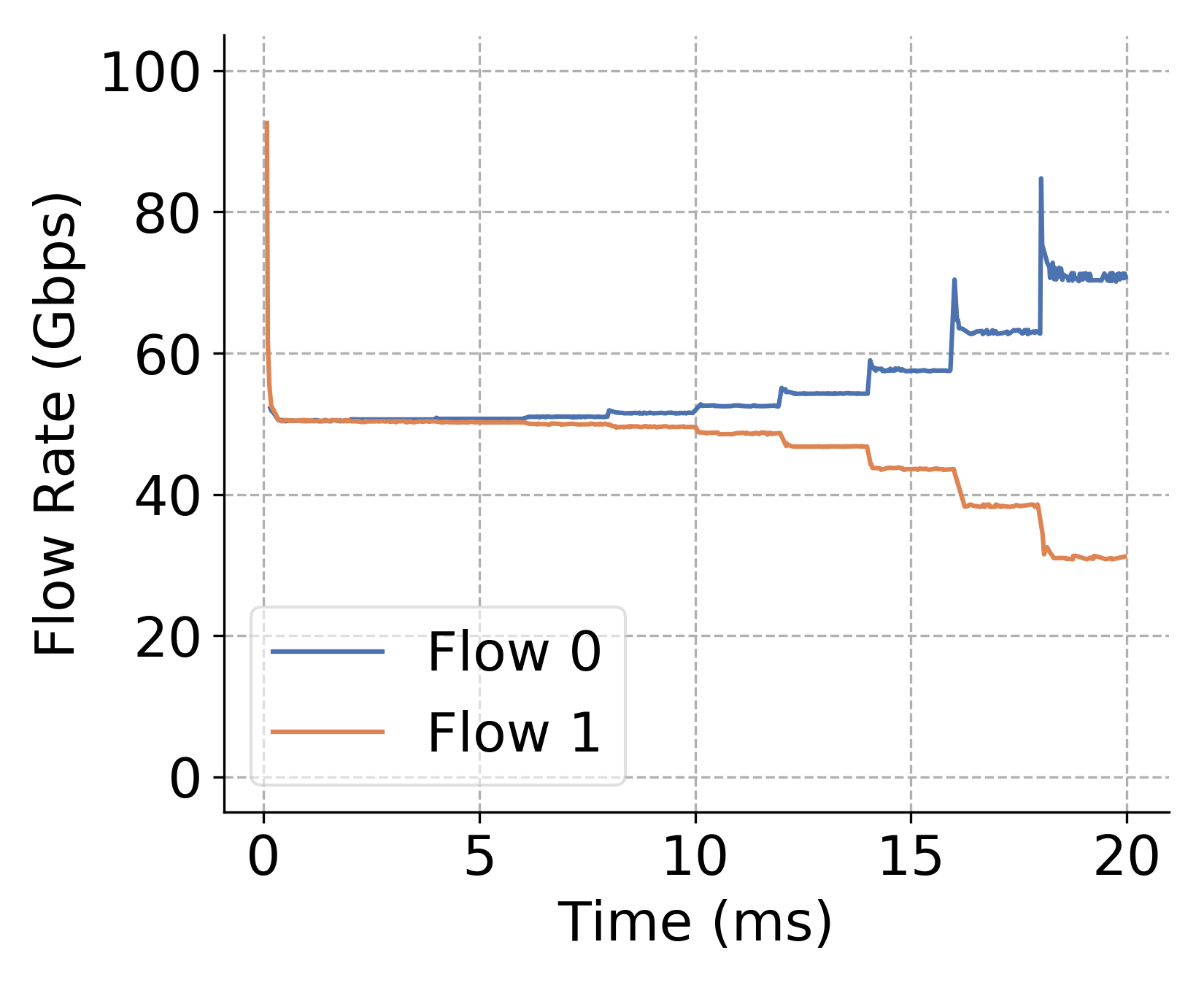}
         \vspace{-6mm}
         \caption{\oursystemsoze with weight updates.}
         \label{fig:granularity1}
     \end{subfigure}
     \hfill
     \begin{subfigure}[b]{0.235\textwidth}
         \centering
         \includegraphics[width=\textwidth]{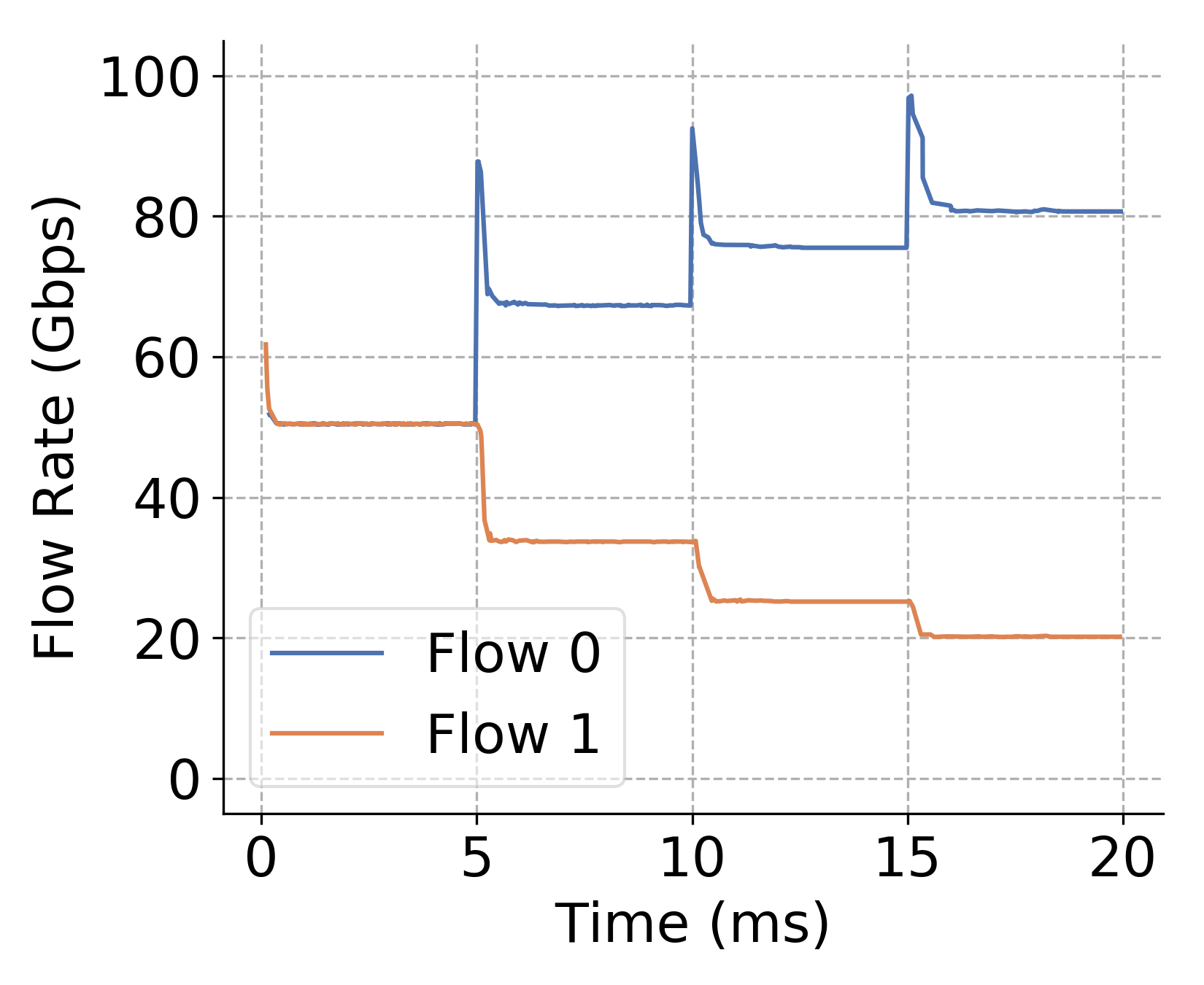}
         \vspace{-6mm}
         \caption{\oursystemsoze multi-connection.}
         \label{fig:granularity2}
     \end{subfigure}
    \hfill
    \begin{subfigure}[b]{0.235\textwidth}
        \centering
        \includegraphics[width=\textwidth]{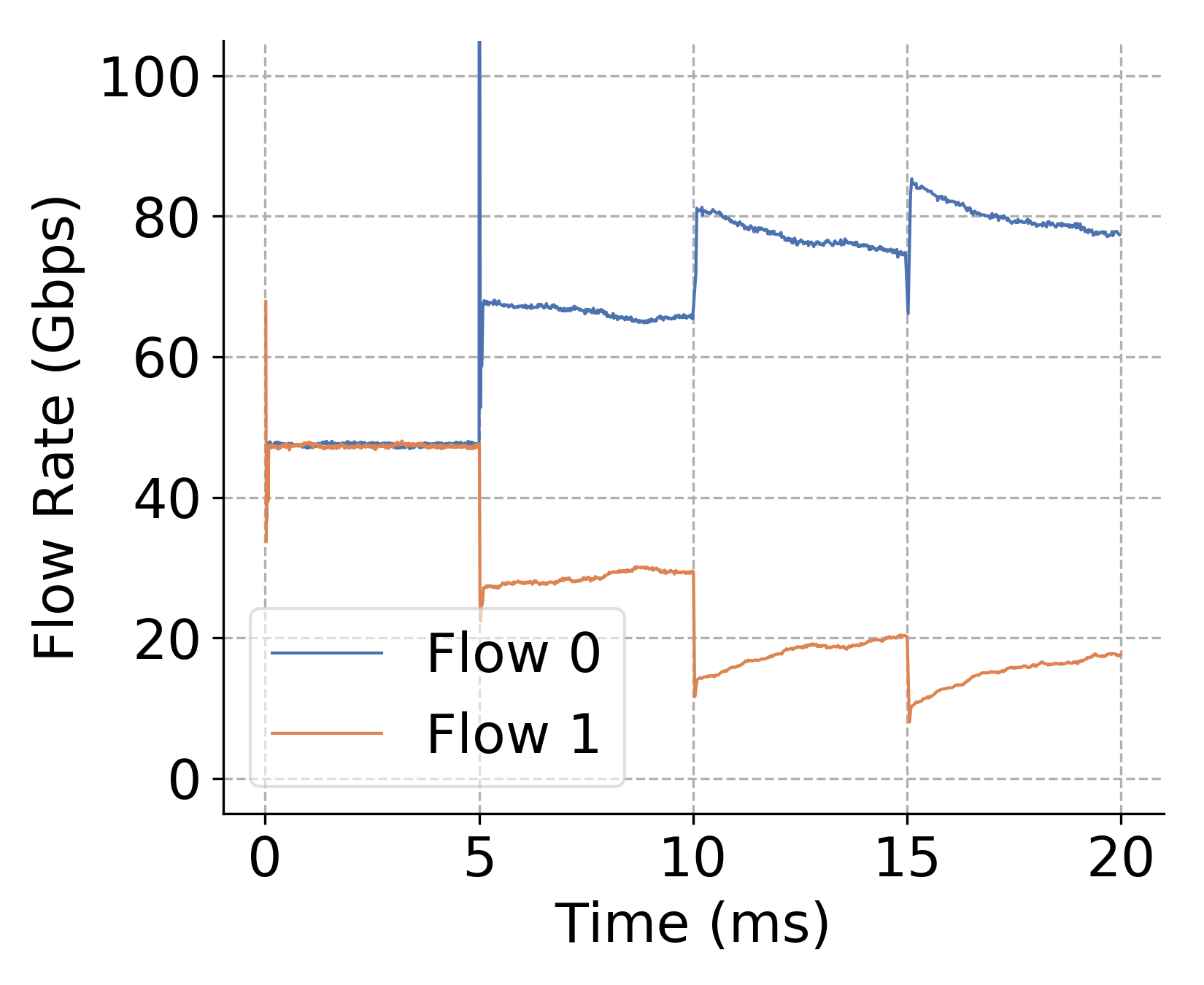}
        \vspace{-6mm}
        \caption{HPCC multi-connection.}
        \label{fig:granularity3}
    \end{subfigure}
    \hfill
\vspace{-2mm}
\caption{Granularity and agility comparison.}
\label{fig:granularity}
\vspace{-4mm}
\end{figure*}

\vspace{-1mm}
\subsubsection{Granularity Micro-benchmark}

To show the granularity of the weighted allocation, we make use of a coflow scenario, where each coflow is consist of 10 flows, and the flow size is randomly chosen from a range of [1 GB, 3 GB]. 
    In \oursystemsoze, according to the size of each flow, the flow will be assigned with a certain weight to complete at the same time with other flows: in \oursystemsoze's scenario, the weight is equal to the flow size. 
    For the multi-connection solution, we will choose the closest integer number with the flow size. For instance, there will be 2 connections for a flow with 2.4 GB data size and there will be 3 connections for a flow with 2.6 GB data size. 
    For the switch-based weighted round-robin solutions, we test on a scenario where each switch has 4 physical queues. We try to pack the 10 different flows into 4 physical queues and set a certain weight for each physical queue. The flows with similar data size will be packed into the same queue, and they will fairly share the bandwidth from that physical queue. The sum of the flow data size within a queue will be used to calculate the weight of each queue.

In \fref{fig:granularity4}, \oursystemsoze allows all the flows within a coflow to finish at nearly the same time, so that it has the minimum FCT difference among all the alternative solutions.
In contrast, the multi-connection solution ("Multi-Conn") only provides integer number of concurrent connections, which makes the weight enforcement rigid and cannot support non-integer weight; the switch-based weighted round-robin ("Physical Queue") relies on limited number of physical queues, and weight enforcement granularity will degrade when the number of flows exceeds the number of queues.

In addition, we also compare different schemes by changing weights or the number of connections. We create a simple in-cast scenario, where two flows are sent to the same destination host. 
    In \fref{fig:granularity1}, we increase the weight for flow 0 every 2 milliseconds by 5\textperthousand, 10\textperthousand, 20\textperthousand, etc. The figure shows that the rate allocation can be accurately stabilized around any level, even when the weight difference is relatively small. 
    In contrast, we also let both \oursystemsoze and HPCC have multiple connections to change the rate allocation. In \fref{fig:granularity2}, \oursystemsoze adds one connection for flow 0 every 5 milliseconds. The rate allocation changes accurately accordingly to the number of connections, but the allocation granularity is limited to weights with integer numbers. For HPCC with multi-connection in \fref{fig:granularity3}, the flow rates are also maintained around the weighted allocation, but the rate enforcement is less accurate and also suffers from coarse-granularity due to integer weights.

\vspace{-1mm}
\subsubsection{Agility Micro-benchmark}

With the same experiment in \ref{fig:granularity1} that changes the weights for \oursystemsoze, we also demonstrate the agility achieved by \oursystemsoze. 
    As shown in \fref{fig:granularity1}, on average, it only takes less than 10 RTTs for \oursystemsoze to converge to the new rate under the new weight.
        We can observe that there is a rate surge right after each weight change and before convergence. This is because we are increasing the weight of flow 0, thus the new converged queueing delay is higher than before, so the rate needs to exceed the line rate briefly to stack the switch queue.

\vspace{-1mm}
\subsubsection{Scalability Micro-benchmark}

Scalability is the key to supporting large-scale systems with consistent performance despite increasing scale. \oursystemsoze is a fully distributed system and can inherently scale to datacenters of any size; in contrast, the water-filling algorithm is a centralized solver, which needs to gather all the information to one master node and use its resources for computation.
In the following experiments, we test the convergence time of \oursystemsoze and the solving time of the water filling to demonstrate the capability to scale for those two solutions. 

\begin{figure}[t!]
     \centering
     \hfill
     \begin{subfigure}[b]{0.235\textwidth}
         \centering
         \includegraphics[width=\textwidth]{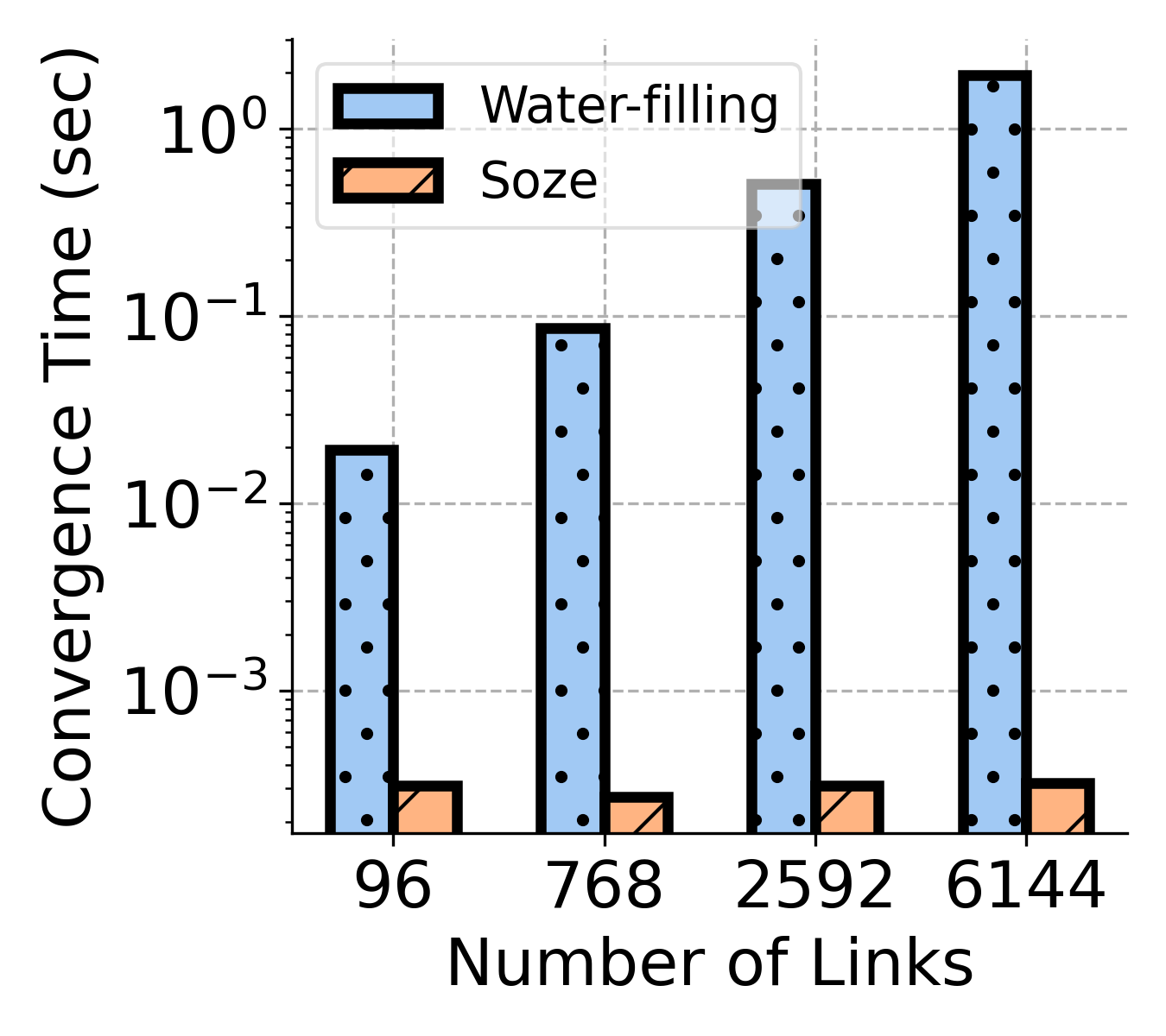}
         \vspace{-7mm}
         \caption{Increasing topology size.}
         \label{fig:scale-k}
     \end{subfigure}
     \hfill
     \begin{subfigure}[b]{0.235\textwidth}
         \centering
         \includegraphics[width=\textwidth]{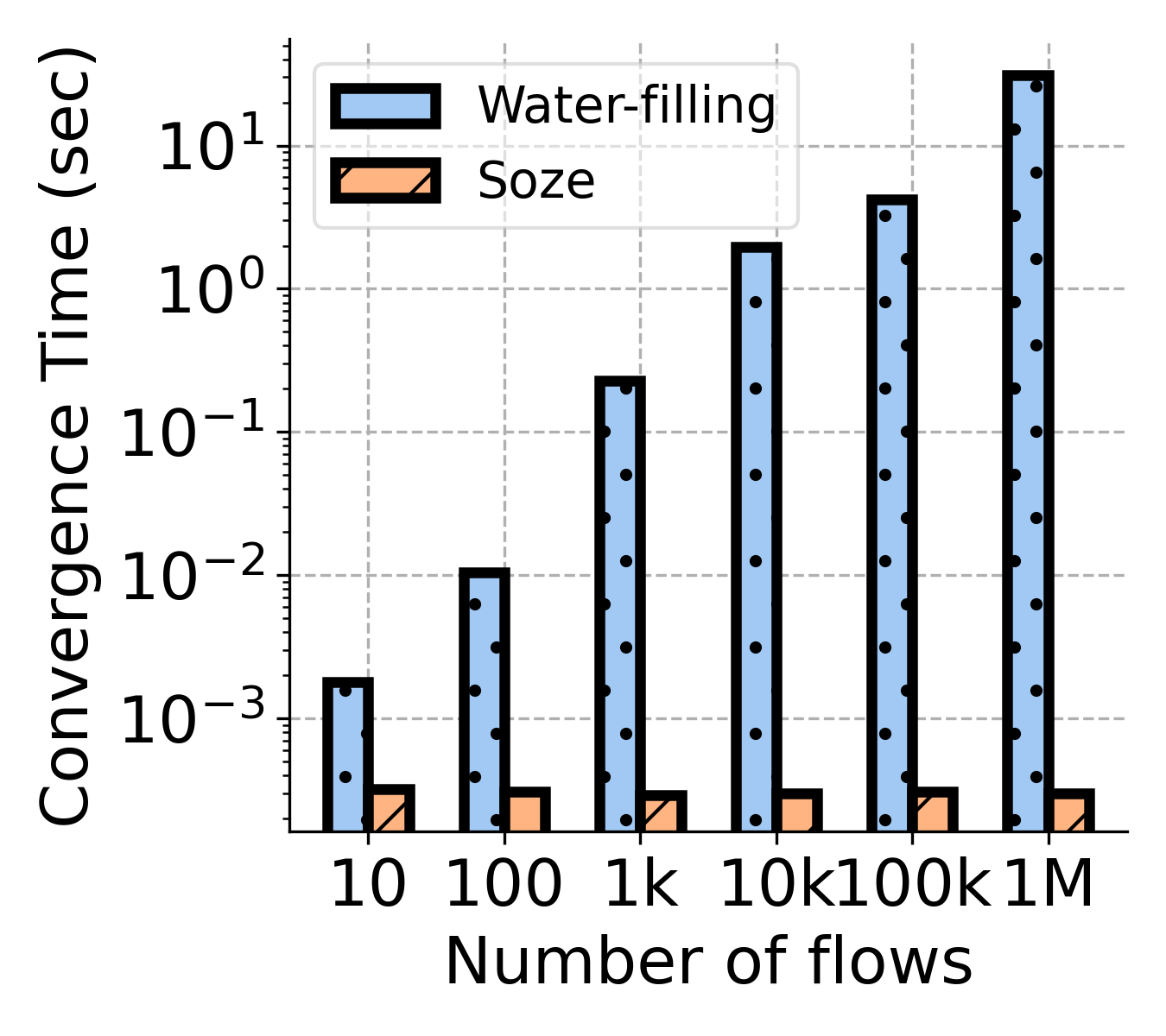}
         \vspace{-7mm}
         \caption{Increasing flow count.}
         \label{fig:scale-f}
     \end{subfigure}
     \hfill
\vspace{-6mm}
\caption{The solving time for the water-filling algorithm increases drastically with either increasing topology size or increasing number of flows; while \oursystemsoze provides consistent convergence speed toward the global weighted allocation state despite of the increasing topology size or flow count.}
\label{fig:scale}
\vspace{-6mm}
\end{figure}

Firstly, we maintain the same number of flows and change the fat-tree network topology sizes by parameter $K$. We chose the number of flows to be 10k so there are roughly 10 flows per server when $K=16$. In \fref{fig:scale-k}, when the topology size increases from $K=4$ (16 hosts) to $K=16$ (1024 hosts), the solving time of the water filling algorithm increases drastically.
In contrast, for \oursystemsoze, once applications have determined the weights, the convergence time towards the global weighted bandwidth allocation is always around the same magnitude. The average convergence time for \oursystemsoze is 0.3 milliseconds, which is around $10$ RTTs.
    
Moreover, we also tested the convergence time for \oursystemsoze and the water filling algorithm with increasing number of flows under the same topology. As shown in \fref{fig:scale-f}, the network topology is always a fat-tree with 1024 servers ($K=16$), and the number of flows in the network increases from 10 flows to 1 million flows. As more and more flows are added to the network, the solving time for the water filling algorithms increases from 1.81 milliseconds with 10 flows to 31.1 seconds with 1 million flows. While \oursystemsoze still provides consistent convergence towards the global weighted allocation within 10 RTTs.
Thus, the water-filling algorithm has inherent drawbacks to support large-scale data center; while \oursystemsoze benefits from the decentralized design and can be scale up with nearly no performance degradation.

\vspace{-1mm}
\section{Related Work} \label{sec:relatedworks}
\vspace{-2mm}

One line of work focuses on using weighted fair queueing (WFQ) on the switches to achieve weighted allocation.
NumFabric~\cite{nagaraj2016numfabric} assumes switches support WFQ in hardware, then provides algorithms for setting the weight parameters in switches to optimize for a utility objective. NumFabric is complementary to \oursystemsoze, because \oursystemsoze replaces switch WFQ; weight parameters provided by NumFabric can be used to implement the desired utility objective on \oursystemsoze even more easily, because there is no talking to switches’ control planes. 

Besides using weighted fair-queueing algorithms on switches~\cite{demers1989analysis,parekh1993generalized,shreedhar1995efficient,golestani1994self,bennett1996wf} to allocate bandwidth, a recent trend is to use programmable switches to approximate weighted fair-queueing~\cite{sharma2018approximating,gao2022gearbox,yu2022cebinae,sivaraman2016programmable,yu2021programmable}. 
    Such switch-based solutions are hard to scale because of the resource limitations on the switches. To enforce the per-flow weighted allocation accurately among many flows, the switch data plane needs to keep a large amount of information, which may exceed the memory capacity.
    Moreover, the delay and the complexity of the control plane for such a scheme are both high, which leads to a lack of agility. To add a new flow or to change the weight, the flow sender needs to inform all the switches along the flow's path individually.
    Besides, those works heavily rely on programmable switches, which are not widely used in production, and those switches are usually less cost-effective than commodity switches with fixed telemetry functions.
PERC~\cite{jose2015high} proposes an alternative approach that departs from the existing DCN service model; it requires switches to run a distributed algorithm to continuously compute the max-min fair shares for all flows and communicate them to end hosts; the algorithm requires control packets that must be processed by switches, requiring multiplication and division operations. In contrast, \oursystemsoze only requires INT from the switches, and end hosts can do the rest.

As for solutions that do not rely on switches, a bandwidth allocator \cite{kumar2015bwe,guo2010secondnet,shieh2011sharing,lam2010netshare} is also capable of achieving weighted max-min fairness for all flows.
    By aggregating the sources, destinations, and demands of all flows, the allocator can calculate a rate allocation plan for each flow. 
    However, this solution requires aggregating information and calculating the allocation plan at the controller, which may lead to a long solving time when the scale of the network increases. Moreover, such solutions are not agile. Each time a flow needs to change weight or a new flow arrives, the whole rate allocation algorithm may need to be executed, leading to high overhead and long control plane delay. 

There are also proposals that aim to manipulate the congestion window to allow a flow to take up more bandwidth when needed. For example,
D2TCP~\cite{vamanan2012deadline} proposes to adjust the congestion window size more or less aggressively based on how close the time is to the deadline. 
MulTCP~\cite{crowcroft1998differentiated} heuristically lets a flow act like N flows in adjusting its congestion window more aggressively. Both solutions build on top of the TCP algorithm, but suffer from the inherited drawbacks, like slow convergence and throughput jitter. Weighted allocation is not the goal of D2TCP, while in MulTCP the effect of increasing the N parameter is not linear and may change for different situations. In contrast, \oursystemsoze achieves fast convergence, low throughput jitter, and accurate weighted fair allocation.

Besides weighted allocation, priority-based allocation solutions \cite{montazeri2018homa,alizadeh2013pfabric} are also used to provide service differentiation between flows, but such solutions have some intrinsic disadvantages over weighted bandwidth allocation, such as head-of-line blocking if preemption is not allowed, and starvation risk. 
Moreover, priority-based differentiation schemes are a larger departure from the existing DCN service model that’s based on fair allocation. \oursystemsoze represents a potentially more compatible evolutionary direction with flexible and controllable allocation for general-purpose DCN.

\vspace{-2mm}
\section{Conclusion} \label{sec:conclusion}
\vspace{-2mm}

In this paper, we propose \oursystemsoze, a simple and efficient weighted bandwidth allocation system for data center networks. 
    \oursystemsoze designs the maxQD INT signal to not only indicate network status but also provide a control knob to coordinate multiple flow senders. 
    Furthermore, \oursystemsoze co-designs the decentralized rate update algorithm with the maxQD signal, which allows each flow to independently move toward the weighted max-min fair allocation in arbitrary networks. 
    We prototype \oursystemsoze across various platforms and demonstrate that \oursystemsoze can achieve weighted bandwidth allocation with fine granularity, high agility, and high scalability.

\section*{Acknowledgment}
We thank the reviewers and the shepherd Ratul Mahajan for their insightful
feedback. This work is partially supported by the NSF under CNS-2214272.

\newpage

\bibliographystyle{plain}
\bibliography{weighted_poseidon}

\begin{thebibliography}{10}

\bibitem{TPCH}
Tpc benchmark h, 2001.
\newblock \url{http://www.tpc.org/tpch/}.

\bibitem{ns3}
Ns-3, 2023.
\newblock \url{https://www.nsnam.org/}.

\bibitem{tpugoogle}
Accelerate ai development with google cloud tpus, 2024.
\newblock \url{https://cloud.google.com/tpu/}.

\bibitem{tpugoogle2}
Google tpu: Architecture and performance best practices, 2024.
\newblock \url{https://www.run.ai/guides/cloud-deep-learning/google-tpu}.

\bibitem{INTbare}
{In-band Network Telemetry in Barefoot Tofino}, 2024.
\newblock \url{https://www.opencompute.org/files/INT-In-Band-Network-Telemetry-A-\ PowerfulAnalytics-Framework-for-\ your-Data-Center-OCP-Final3.pdf}.

\bibitem{INTtomo}
{In-band Network Telemetry in Broadcom Tomahawk 3}.
\newblock \url{https://www.broadcom.com/company/news/product-releases/2372840}, 2024.

\bibitem{INTtri}
{In-band Network Telemetry in Broadcom Trident3}.
\newblock \url{https://www.broadcom.com/blog/new-trident-3-switch-delivers-\ smarterprogrammability-for-enterprise-\ and-service-provider-datacenters}, 2024.

\bibitem{gpugaming}
Nvidia rtx server: Powering the future of cloud gaming and ar/vr, 2024.
\newblock \url{https://www.nvidia.com/en-us/data-center/rtx-server-gaming/}.

\bibitem{tofino}
Tofino switches by intel, 2024.
\newblock \url{https://www.intel.com/content/www/us/en/products/details/network-io/intelligent-fabric-processors.html}.

\bibitem{abts2022high}
Dennis Abts and John Kim.
\newblock {\em High performance datacenter networks: Architectures, algorithms, and opportunities}.
\newblock Springer Nature, 2022.

\bibitem{achiam2023gpt}
Josh Achiam, Steven Adler, Sandhini Agarwal, Lama Ahmad, Ilge Akkaya, Florencia~Leoni Aleman, Diogo Almeida, Janko Altenschmidt, Sam Altman, Shyamal Anadkat, et~al.
\newblock Gpt-4 technical report.
\newblock {\em arXiv preprint arXiv:2303.08774}, 2023.

\bibitem{alizadeh2010data}
Mohammad Alizadeh, Albert Greenberg, David~A Maltz, Jitendra Padhye, Parveen Patel, Balaji Prabhakar, Sudipta Sengupta, and Murari Sridharan.
\newblock Data center tcp (dctcp).
\newblock In {\em Proceedings of the ACM SIGCOMM 2010 Conference}, pages 63--74, 2010.

\bibitem{alizadeh2013pfabric}
Mohammad Alizadeh, Shuang Yang, Milad Sharif, Sachin Katti, Nick McKeown, Balaji Prabhakar, and Scott Shenker.
\newblock pfabric: Minimal near-optimal datacenter transport.
\newblock {\em ACM SIGCOMM Computer Communication Review}, 43(4):435--446, 2013.

\bibitem{allalouf2008centralized}
Miriam Allalouf and Yuval Shavitt.
\newblock Centralized and distributed algorithms for routing and weighted max-min fair bandwidth allocation.
\newblock {\em IEEE/ACM Transactions on networking}, 16(5):1015--1024, 2008.

\bibitem{barroso2022datacenter}
Luis~Andre Barroso and Jimmy Clidaras.
\newblock {\em The datacenter as a computer: An introduction to the design of warehouse-scale machines}.
\newblock Springer Nature, 2022.

\bibitem{bennett1996wf}
Jon~CR Bennett and Hui Zhang.
\newblock Wf/sup 2/q: worst-case fair weighted fair queueing.
\newblock In {\em Proceedings of IEEE INFOCOM'96. Conference on Computer Communications}, volume~1, pages 120--128. IEEE, 1996.

\bibitem{casini2020predictable}
Daniel Casini, Paolo Pazzaglia, Alessandro Biondi, Marco Di~Natale, and Giorgio Buttazzo.
\newblock Predictable memory-cpu co-scheduling with support for latency-sensitive tasks.
\newblock In {\em 2020 57th ACM/IEEE Design Automation Conference (DAC)}, pages 1--6. IEEE, 2020.

\bibitem{chowdhury2015efficient}
Mosharaf Chowdhury and Ion Stoica.
\newblock Efficient coflow scheduling without prior knowledge.
\newblock {\em ACM SIGCOMM Computer Communication Review}, 45(4):393--406, 2015.

\bibitem{chowdhury2014efficient}
Mosharaf Chowdhury, Yuan Zhong, and Ion Stoica.
\newblock Efficient coflow scheduling with varys.
\newblock In {\em ACM SIGCOMM}, 2014.

\bibitem{crowcroft1998differentiated}
Jon Crowcroft and Philippe Oechslin.
\newblock Differentiated end-to-end internet services using a weighted proportional fair sharing tcp.
\newblock {\em ACM SIGCOMM Computer Communication Review}, 28(3):53--69, 1998.

\bibitem{demers1989analysis}
Alan Demers, Srinivasan Keshav, and Scott Shenker.
\newblock Analysis and simulation of a fair queueing algorithm.
\newblock {\em ACM SIGCOMM Computer Communication Review}, 19(4):1--12, 1989.

\bibitem{gao2022gearbox}
Peixuan Gao, Anthony Dalleggio, Yang Xu, and H~Jonathan Chao.
\newblock Gearbox: A hierarchical packet scheduler for approximate weighted fair queuing.
\newblock In {\em 19th USENIX Symposium on Networked Systems Design and Implementation (NSDI 22)}, pages 551--565, 2022.

\bibitem{golestani1994self}
S~Jamaloddin Golestani.
\newblock A self-clocked fair queueing scheme for broadband applications.
\newblock In {\em Proceedings of INFOCOM'94 Conference on Computer Communications}, pages 636--646. IEEE, 1994.

\bibitem{gos2020comparison}
Konrad Gos and Wojciech Zabierowski.
\newblock The comparison of microservice and monolithic architecture.
\newblock In {\em 2020 IEEE XVIth International Conference on the Perspective Technologies and Methods in MEMS Design (MEMSTECH)}, pages 150--153. IEEE, 2020.

\bibitem{grosvenor2015queues}
Matthew~P Grosvenor, Malte Schwarzkopf, Ionel Gog, Robert~NM Watson, Andrew~W Moore, Steven Hand, and Jon Crowcroft.
\newblock Queues don’t matter when you can jump them!
\newblock In {\em 12th USENIX Symposium on Networked Systems Design and Implementation (NSDI 15)}, pages 1--14, 2015.

\bibitem{gujarati2020serving}
Arpan Gujarati, Reza Karimi, Safya Alzayat, Wei Hao, Antoine Kaufmann, Ymir Vigfusson, and Jonathan Mace.
\newblock Serving $\{$DNNs$\}$ like clockwork: Performance predictability from the bottom up.
\newblock In {\em 14th USENIX Symposium on Operating Systems Design and Implementation (OSDI 20)}, pages 443--462, 2020.

\bibitem{guo2010secondnet}
Chuanxiong Guo, Guohan Lu, Helen~J Wang, Shuang Yang, Chao Kong, Peng Sun, Wenfei Wu, and Yongguang Zhang.
\newblock Secondnet: a data center network virtualization architecture with bandwidth guarantees.
\newblock In {\em Proceedings of the 6th International COnference}, pages 1--12, 2010.

\bibitem{hashemi2019tictac}
Sayed~Hadi Hashemi, Sangeetha Abdu~Jyothi, and Roy Campbell.
\newblock Tictac: Accelerating distributed deep learning with communication scheduling.
\newblock {\em Proceedings of Machine Learning and Systems}, 1:418--430, 2019.

\bibitem{jayarajan2019priority}
Anand Jayarajan, Jinliang Wei, Garth Gibson, Alexandra Fedorova, and Gennady Pekhimenko.
\newblock Priority-based parameter propagation for distributed dnn training.
\newblock {\em Proceedings of Machine Learning and Systems}, 1:132--145, 2019.

\bibitem{jose2015high}
Lavanya Jose, Lisa Yan, Mohammad Alizadeh, George Varghese, Nick McKeown, and Sachin Katti.
\newblock High speed networks need proactive congestion control.
\newblock In {\em Proceedings of the 14th acm workshop on hot topics in networks}, pages 1--7, 2015.

\bibitem{kalia2019datacenter}
Anuj Kalia, Michael Kaminsky, and David Andersen.
\newblock Datacenter $\{$RPCs$\}$ can be general and fast.
\newblock In {\em 16th USENIX Symposium on Networked Systems Design and Implementation (NSDI 19)}, pages 1--16, 2019.

\bibitem{kumar2015bwe}
Alok Kumar, Sushant Jain, Uday Naik, Anand Raghuraman, Nikhil Kasinadhuni, Enrique~Cauich Zermeno, C~Stephen Gunn, Jing Ai, Bj{\"o}rn Carlin, Mihai Amarandei-Stavila, et~al.
\newblock Bwe: Flexible, hierarchical bandwidth allocation for wan distributed computing.
\newblock In {\em Proceedings of the 2015 ACM Conference on Special Interest Group on Data Communication}, pages 1--14, 2015.

\bibitem{kumar2020swift}
Gautam Kumar, Nandita Dukkipati, Keon Jang, Hassan~MG Wassel, Xian Wu, Behnam Montazeri, Yaogong Wang, Kevin Springborn, Christopher Alfeld, Michael Ryan, et~al.
\newblock Swift: Delay is simple and effective for congestion control in the datacenter.
\newblock In {\em Proceedings of the Annual conference of the ACM Special Interest Group on Data Communication on the applications, technologies, architectures, and protocols for computer communication}, pages 514--528, 2020.

\bibitem{lam2010netshare}
Terry Lam, Sivasankar Radhakrishnan, Amin Vahdat, and George Varghese.
\newblock Netshare: Virtualizing data center networks across services.
\newblock 2010.

\bibitem{li2019hpcc}
Yuliang Li, Rui Miao, Hongqiang~Harry Liu, Yan Zhuang, Fei Feng, Lingbo Tang, Zheng Cao, Ming Zhang, Frank Kelly, Mohammad Alizadeh, et~al.
\newblock Hpcc: High precision congestion control.
\newblock In {\em Proceedings of the ACM Special Interest Group on Data Communication}, pages 44--58. 2019.

\bibitem{liang2021joint}
Jie Liang, Kenli Li, Chubo Liu, and Keqin Li.
\newblock Joint offloading and scheduling decisions for dag applications in mobile edge computing.
\newblock {\em Neurocomputing}, 424:160--171, 2021.

\bibitem{ma2014resource}
Tinghuai Ma, Ya~Chu, Licheng Zhao, and Otgonbayar Ankhbayar.
\newblock Resource allocation and scheduling in cloud computing: Policy and algorithm.
\newblock {\em IETE Technical review}, 31(1):4--16, 2014.

\bibitem{marbach2002priority}
Peter Marbach.
\newblock Priority service and max-min fairness.
\newblock In {\em Proceedings. Twenty-First Annual Joint Conference of the IEEE Computer and Communications Societies}, volume~1, pages 266--275. IEEE, 2002.

\bibitem{mittal2015timely}
Radhika Mittal, Vinh~The Lam, Nandita Dukkipati, Emily Blem, Hassan Wassel, Monia Ghobadi, Amin Vahdat, Yaogong Wang, David Wetherall, and David Zats.
\newblock Timely: Rtt-based congestion control for the datacenter.
\newblock {\em ACM SIGCOMM Computer Communication Review}, 45(4):537--550, 2015.

\bibitem{montazeri2018homa}
Behnam Montazeri, Yilong Li, Mohammad Alizadeh, and John Ousterhout.
\newblock Homa: A receiver-driven low-latency transport protocol using network priorities.
\newblock In {\em Proceedings of the 2018 Conference of the ACM Special Interest Group on Data Communication}, pages 221--235, 2018.

\bibitem{moon2000scalable}
Sung-Whan Moon, Jennifer Rexford, and Kang~G Shin.
\newblock Scalable hardware priority queue architectures for high-speed packet switches.
\newblock {\em IEEE Transactions on computers}, 49(11):1215--1227, 2000.

\bibitem{nagaraj2016numfabric}
Kanthi Nagaraj, Dinesh Bharadia, Hongzi Mao, Sandeep Chinchali, Mohammad Alizadeh, and Sachin Katti.
\newblock Numfabric: Fast and flexible bandwidth allocation in datacenters.
\newblock In {\em Proceedings of the 2016 ACM SIGCOMM Conference}, pages 188--201, 2016.

\bibitem{nishtala2013scaling}
Rajesh Nishtala, Hans Fugal, Steven Grimm, Marc Kwiatkowski, Herman Lee, Harry~C Li, Ryan McElroy, Mike Paleczny, Daniel Peek, Paul Saab, et~al.
\newblock Scaling memcache at facebook.
\newblock In {\em 10th USENIX Symposium on Networked Systems Design and Implementation (NSDI 13)}, pages 385--398, 2013.

\bibitem{oudghiri1992global}
Houria Oudghiri and Bozena Kaminska.
\newblock Global weighted scheduling and allocation algorithms.
\newblock In {\em Proceedings The European Conference on Design Automation}, pages 491--492. IEEE Computer Society, 1992.

\bibitem{pan2022efficient}
Rui Pan, Yiming Lei, Jialong Li, Zhiqiang Xie, Binhang Yuan, and Yiting Xia.
\newblock Efficient flow scheduling in distributed deep learning training with echelon formation.
\newblock In {\em Proceedings of the 21st ACM Workshop on Hot Topics in Networks}, pages 93--100, 2022.

\bibitem{parekh1993generalized}
Abhay~K Parekh and Robert~G Gallager.
\newblock A generalized processor sharing approach to flow control in integrated services networks: the single-node case.
\newblock {\em IEEE/ACM transactions on networking}, 1(3):344--357, 1993.

\bibitem{bytescheduler}
Yanghua Peng, Yibo Zhu, Yangrui Chen, Yixin Bao, Bairen Yi, Chang Lan, Chuan Wu, and Chuanxiong Guo.
\newblock A generic communication scheduler for distributed dnn training acceleration.
\newblock In {\em ACM SOSP}, 2019.

\bibitem{ramezani2021dynamic}
Reza Ramezani.
\newblock Dynamic scheduling of task graphs in multi-fpga systems using critical path.
\newblock {\em The Journal of Supercomputing}, 77(1):597--618, 2021.

\bibitem{roy2015inside}
Arjun Roy, Hongyi Zeng, Jasmeet Bagga, George Porter, and Alex~C Snoeren.
\newblock Inside the social network's (datacenter) network.
\newblock In {\em Proceedings of the 2015 ACM Conference on Special Interest Group on Data Communication}, pages 123--137, 2015.

\bibitem{sharma2018approximating}
Naveen~Kr Sharma, Ming Liu, Kishore Atreya, and Arvind Krishnamurthy.
\newblock Approximating fair queueing on reconfigurable switches.
\newblock In {\em 15th USENIX Symposium on Networked Systems Design and Implementation (NSDI 18)}, pages 1--16, 2018.

\bibitem{shieh2011sharing}
Alan Shieh, Srikanth Kandula, Albert Greenberg, Changhoon Kim, and Bikas Saha.
\newblock Sharing the data center network.
\newblock In {\em 8th USENIX Symposium on Networked Systems Design and Implementation (NSDI 11)}, 2011.

\bibitem{shreedhar1995efficient}
Madhavapeddi Shreedhar and George Varghese.
\newblock Efficient fair queueing using deficit round robin.
\newblock In {\em Proceedings of the conference on Applications, technologies, architectures, and protocols for computer communication}, pages 231--242, 1995.

\bibitem{sigelman2010dapper}
Benjamin~H Sigelman, Luiz~Andr{\'e} Barroso, Mike Burrows, Pat Stephenson, Manoj Plakal, Donald Beaver, Saul Jaspan, and Chandan Shanbhag.
\newblock Dapper, a large-scale distributed systems tracing infrastructure.
\newblock 2010.

\bibitem{sivaraman2016programmable}
Anirudh Sivaraman, Suvinay Subramanian, Mohammad Alizadeh, Sharad Chole, Shang-Tse Chuang, Anurag Agrawal, Hari Balakrishnan, Tom Edsall, Sachin Katti, and Nick McKeown.
\newblock Programmable packet scheduling at line rate.
\newblock In {\em Proceedings of the 2016 ACM SIGCOMM Conference}, pages 44--57, 2016.

\bibitem{touvron2023llama}
Hugo Touvron, Thibaut Lavril, Gautier Izacard, Xavier Martinet, Marie-Anne Lachaux, Timoth{\'e}e Lacroix, Baptiste Rozi{\`e}re, Naman Goyal, Eric Hambro, Faisal Azhar, et~al.
\newblock Llama: Open and efficient foundation language models.
\newblock {\em arXiv preprint arXiv:2302.13971}, 2023.

\bibitem{vamanan2012deadline}
Balajee Vamanan, Jahangir Hasan, and TN~Vijaykumar.
\newblock Deadline-aware datacenter tcp (d2tcp).
\newblock {\em ACM SIGCOMM Computer Communication Review}, 42(4):115--126, 2012.

\bibitem{villamizar2016infrastructure}
Mario Villamizar, Oscar Garces, Lina Ochoa, Harold Castro, Lorena Salamanca, Mauricio Verano, Rubby Casallas, Santiago Gil, Carlos Valencia, Angee Zambrano, et~al.
\newblock Infrastructure cost comparison of running web applications in the cloud using aws lambda and monolithic and microservice architectures.
\newblock In {\em 2016 16th IEEE/ACM International Symposium on Cluster, Cloud and Grid Computing (CCGrid)}, pages 179--182. IEEE, 2016.

\bibitem{wang2021mxdag}
Weitao Wang, Sushovan Das, Xinyu~Crystal Wu, Zhuang Wang, Ang Chen, and TS~Eugene Ng.
\newblock Mxdag: A hybrid abstraction for emerging applications.
\newblock In {\em Proceedings of the Twentieth ACM Workshop on Hot Topics in Networks}, pages 221--228, 2021.

\bibitem{yu2022cebinae}
Liangcheng Yu, John Sonchack, and Vincent Liu.
\newblock Cebinae: scalable in-network fairness augmentation.
\newblock In {\em Proceedings of the ACM SIGCOMM 2022 Conference}, pages 219--232, 2022.

\bibitem{yu2021programmable}
Zhuolong Yu, Chuheng Hu, Jingfeng Wu, Xiao Sun, Vladimir Braverman, Mosharaf Chowdhury, Zhenhua Liu, and Xin Jin.
\newblock Programmable packet scheduling with a single queue.
\newblock In {\em Proceedings of the 2021 ACM SIGCOMM 2021 Conference}, pages 179--193, 2021.

\bibitem{zhu2015congestion}
Yibo Zhu, Haggai Eran, Daniel Firestone, Chuanxiong Guo, Marina Lipshteyn, Yehonatan Liron, Jitendra Padhye, Shachar Raindel, Mohamad~Haj Yahia, and Ming Zhang.
\newblock Congestion control for large-scale rdma deployments.
\newblock {\em ACM SIGCOMM Computer Communication Review}, 45(4):523--536, 2015.

\bibitem{zhu2016ecn}
Yibo Zhu, Monia Ghobadi, Vishal Misra, and Jitendra Padhye.
\newblock Ecn or delay: Lessons learnt from analysis of dcqcn and timely.
\newblock In {\em Proceedings of the 12th International on Conference on emerging Networking EXperiments and Technologies}, pages 313--327, 2016.

\end{thebibliography}

\clearpage
\appendix

\newpage

\section{Notations}
\label{appendix:notations}

We list all the symbols used in the main body and the appendix as follow:

\begin{table}[h]
\centering
\resizebox{\columnwidth}{!}{%
\begin{tabular}{|c|l|}
\hline
\rowcolor[HTML]{C0C0C0} 
Symbol & \multicolumn{1}{c|}{\cellcolor[HTML]{C0C0C0}Explanation} \\ \hline
$f_i$ & The flow with index $i$ \\ \hline
$w_i$ & The weight of flow $f_i$ \\ \hline
$r_i$ & The rate of flow $f_i$ \\ \hline
$r_i(t)$ & The rate of flow $f_i$ at time $t$ \\ \hline
$r_i(\infty)$ & The final rate of flow $f_i$ after convergence \\ \hline
$s_i$ & The flow $f_i$'s rate divided by weight, equals to $\frac{r_i}{w_i}$ \\ \hline
$s_i(t)$ & The flow's rate divided by weight at time $t$ \\ \hline
$D(t)$ & The queueing delay signal at time $t$ \\ \hline
$D(\infty)$ & Final queueing delay signal after convergence \\ \hline
$B$ & Bandwidth \\ \hline
$W$ & Total weight on the link: $w_0+w_1+...+w_n$ \\ \hline
$R$ & Aggregated link arrival rate \\ \hline
$R(t)$ & Aggregated link arrival rate at time $t$ \\ \hline
$T(s)$ & \begin{tabular}[c]{@{}l@{}}Target delay function, \\ take $s_i$ as input and give target queueing as output\end{tabular} \\ \hline
$U(s, D)$ & \begin{tabular}[c]{@{}l@{}}Rate update function, \\ rate and queueing as input and output update ratio\end{tabular} \\ \hline
$\alpha$ & Upperbound rate (the highest link bandwidth) \\ \hline
$\beta$ & Expected lowest rate \\ \hline
$k$ & Base queueing delay in target function \\ \hline
$q$ & Increment queueing delay in target function \\ \hline
$m$ & Convergence control parameter in PID control \\ \hline
$wfs$ & weighted fair-share rate \\ \hline
$\Delta t$ & the rate update period \\ \hline
\end{tabular}%
}
\caption{Symbol table.}
\label{tab:symbols}
\end{table}
\section{Proof for Goal Transformation}
\label{appendix:goal-transform}

The goal of weighted bandwidth allocation is to achieve for any flow $f_i$:

\begin{equation}
    r_i= \frac{w_i}{w_0 + w_1 + ... + w_n} \cdot B
\end{equation}

In the paper, we gave another goal:

\begin{equation}
\begin{dcases*}
    r_0 + r_1 + ... + r_n = B \\
    \frac{r_0}{w_0} = \frac{r_1}{w_1} = ... = \frac{r_n}{w_n}
\end{dcases*}
\end{equation}

Here we can prove that those two goals are equivalent by proving one goal can derive the other one.

\textbf{Original goal to proposed goal:}

If we assume for any flow $f_i$:

\begin{equation}
    r_i= \frac{w_i}{w_0 + w_1 + ... + w_n} \cdot B
\end{equation}

Then we have:

\begin{equation}
\begin{aligned}
    r_0 + r_1 + ... + r_n &= \frac{w_0 + w_1 + ... + w_n}{w_0 + w_1 + ... + w_n} \cdot B \\
    &= B
\end{aligned}
\end{equation}

For any flow $f_i$, we also have:

\begin{equation}
\begin{aligned}
    \frac{r_i}{w_i} &= \frac{w_i}{w_0 + w_1 + ... + w_n} \cdot B \cdot \frac{1}{w_i} \\
    & = \frac{B}{w_0 + w_1 + ... + w_n}
\end{aligned}
\end{equation}

For flows with specific weights, $\frac{B}{w_0 + w_1 + ... + w_n}$ is a constant. In conclusion, original goal can derive the proposed goal.

\textbf{Proposed goal to original goal:}

If we assume:

\begin{equation}
\begin{dcases*}
    r_0 + r_1 + ... + r_n = B \\
    \frac{r_0}{w_0} = \frac{r_1}{w_1} = ... = \frac{r_n}{w_n}
\end{dcases*}
\end{equation}

For any flow $f_i$, we can have:

\begin{equation}
    r_i = \frac{r_0}{w_0} \cdot w_i = \frac{w_i}{w_0} \cdot r_0
\end{equation}

So that we have:

\begin{equation}
\begin{aligned}
    r_0 + r_1 + ... + r_n &= B \\
    \frac{w_0}{w_0} \cdot r_0 + \frac{w_1}{w_0} \cdot r_0 + ... + \frac{w_n}{w_0} \cdot r_0 &= B \\
    \frac{w_0 + w_1 + ... + w_n}{w_0} \cdot r_0 &= B \\
    r_0 &= B \cdot \frac{w_0}{w_0 + w_1 + ... + w_n} \\
\end{aligned}
\end{equation}

For any flow $f_i$, we can have:

\begin{equation}
    r_i= \frac{w_i}{w_0 + w_1 + ... + w_n} \cdot B
\end{equation}

In conclusion, the proposed goal can also derive the original goal. Thus, those two goals are equivalent.

\section{Proof of Convergence to Weighted Fairness on a Single Switch}
\label{appendix:fairness-single}

In this section, we will prove that the weighted fairness will be achieved with \oursystemsoze.

\begin{lemapp}[Convergence to Weighted Fairness]{}
    \oursystemsoze converges to the weighted fairness on a link if and only if $0 < m < 2$ in the update function.
\end{lemapp}

To formally prove the above lemma, consider the scenario as follows. 
Assume a link with two flows A and B with current rates $r_a$ and $r_b$ and weights $w_a$ and $w_b$ respectively where $\frac{r_a}{w_a}<\frac{r_b}{w_b}$. The weighted fairness can be verified with the angle between the actual bandwidth share and the weighted fair-share.
    We define the update function $U(s, D)$ as the update ratio between the current rate and the new rate, i.e., $r'= r \cdot U(s, D)$. We also denote that $s_a = \frac{r_a}{w_a}$ and $s_b = \frac{r_b}{w_b}$

To ensure that fairness improves, the updated rates of flows A and B (i.e., $s'_a$ and $s'_b$ respectively) should reduce the angle with the weighted fairness line.
Therefore, the lower bound of the new allocation's angle with weighted fair is defined by the current ratio of rates (i.e., having the slope $\frac{s_a}{s_b}$), and the upper bound is symmetric across the weighted fair-share line (i.e., having the slope $\frac{s_b}{s_a}$).
Assuming $s_a<s_b$, the requirement to converge to fair share can be written as follow:

\begin{equation}
\begin{aligned}
    \frac{s_a}{s_b} < \frac{s_b \cdot U(s_b, D)}{s_a \cdot U(s_a,D)} < \frac{s_b}{s_a}, \forall s_a<s_b, \forall D > 0
\end{aligned}
\label{equ:more-fair}
\end{equation}

Without loss of generality, we assume $s_a < s_b$. 

For the RHS:

\begin{equation}
\begin{aligned}
    \frac{U(s_b,D)}{U(s_a,D)}
    &=\frac{\left(\frac{T\left(D\right)}{s_b}\right)^{m}}{\left(\frac{T\left(D\right)}{s_a}\right)^{m}} \\
    &=\left(\frac{s_a}{s_b}\right)^{m} \\
    &<1
\end{aligned}
\end{equation}

For the LHS:

\begin{equation}
\begin{aligned}
    \frac{U(s_b,D)}{U(s_a,D)}
    &=\frac{\left(\frac{T\left(D\right)}{s_b}\right)^{m}}{\left(\frac{T\left(D\right)}{s_a}\right)^{m}} \\
    &=\left(\frac{s_a}{b}\right)^{m}
\label{equ:mimd_intuition}
\end{aligned}
\end{equation}

So as long as $m < 2$, we can have

\begin{equation}
\begin{aligned}
    \frac{U(s_b,D)}{U(s_a,D)}
    = \left(\frac{s_a}{s_b}\right) ^ {m}
    > \left(\frac{s_a}{s_b}\right) ^ 2
\end{aligned}
\end{equation}

Thus, as long as $0 < m < 2$, \eref{equ:update} satisfies \eref{equ:more-fair}.
\section{Proof of Convergence to Target Queueing Delay on a Single Switch}
\label{appendix:util-single}

Since convergence to weighted fairness has been proven to be achieved no matter whether the target queueing delay is achieved, we could assume that fairness has been achieved and that all flows have the same rate. 

\begin{lemapp}[Convergence to Target Queueing]{}
    \oursystemsoze converges to the target queueing delay level, if and only if $p>\frac{\Delta t}{2}\cdot\left[\ln(\alpha) - \ln(\beta)\right]$.
\end{lemapp}

Denote the link bandwidth as $B$ and the total weight on the link as $W$. For any flow $f_i$ on the link, its weighted fair share rate of flows as $s_i(\infty) = \frac{r_i(\infty)}{w_i}$. Since we assume all the flows converge to weighted fair-share, so the total arrival rate for the link can be calculated as:

\begin{equation}
    R(i) = s(i) \cdot W
\end{equation}

From the target function~\ref{equ:target}, we have the following:

\begin{equation}
    s(i+1) = s(i) * \left[\frac{T^{-1}\left[D(i)\right]}{s(i)}\right]^m
\end{equation}

\begin{equation}
\begin{aligned}
    D(i+1) &= D(i) + \frac{\left[R(i) - B\right] \cdot \Delta t}{B} \\
    &= D(i) + \frac{\left[s(i)\cdot W - B\right] \cdot \Delta t}{B}
\end{aligned}
\end{equation}

In which, the $\Delta t$ is the rate update period. For per-packet update, $\Delta t = \frac{RTT}{CWND}$.

To simplify the target function and the inverse target function, we have the following:

\begin{equation}
\begin{aligned}
    T\left(s\right) &= p \cdot \frac{ln(\alpha)-ln(s)}{ln(\alpha)-ln(\beta)} \\
    &= p \cdot \frac{ln(\frac{\alpha}{s})}{ln(\frac{\alpha}{\beta})} \\
    &= \frac{\ln(\frac{\alpha}{s})}{-\ln\left((\frac{\beta}{\alpha})^\frac{1}{p}\right)} \\
\end{aligned}
\end{equation}

\begin{equation}
\begin{aligned}
    T^{-1}(D) &= \exp\left[ln(\alpha) - \frac{D}{p}\cdot (ln(\alpha)-ln(\beta))\right] \\
    &= \exp\left[ln(\alpha) - \frac{ln(\alpha)-ln(\beta)}{p}\cdot D\right] \\
    &= \alpha \cdot \frac{\exp[\frac{ln(\beta)}{p}\cdot D]}{\exp[\frac{ln(\alpha)}{p}\cdot D]} \\
    &= \alpha \cdot \frac{\beta ^ {\frac{D}{p}}}{\alpha ^ {\frac{D}{p}}} \\
    &= \alpha \cdot \left(\frac{\beta}{\alpha}\right) ^ {\frac{D}{p}}
\end{aligned}
\end{equation}

Denote $(\frac{\beta}{\alpha})^\frac{1}{p} = A$, we have:

\begin{equation}
\begin{aligned}
    T\left(s\right) = \log_A^{\frac{s}{\alpha}} \\
\end{aligned}
\end{equation}

\begin{equation}
\begin{aligned}
    T^{-1}(D) = \alpha \cdot A ^ {D}\\
\end{aligned}
\end{equation}

With the simplified function, we can calculate the updated $D(i+1)$ according to $D(i)$. Here we assume $m=1$ for a simpler representation. 

\begin{equation} 
    s(i) = \alpha \cdot A ^ {D(i)}
\end{equation} 

\begin{equation} 
\begin{aligned} 
    D(i+1) &= D(i) + \frac{\Delta t}{B} \cdot \left[W\cdot s(i) - B\right]
\end{aligned} 
\end{equation} 


\subsection{Convergence with Oscillation}

\begin{figure}[h]
    \centering
    \includegraphics[width=0.4\textwidth]{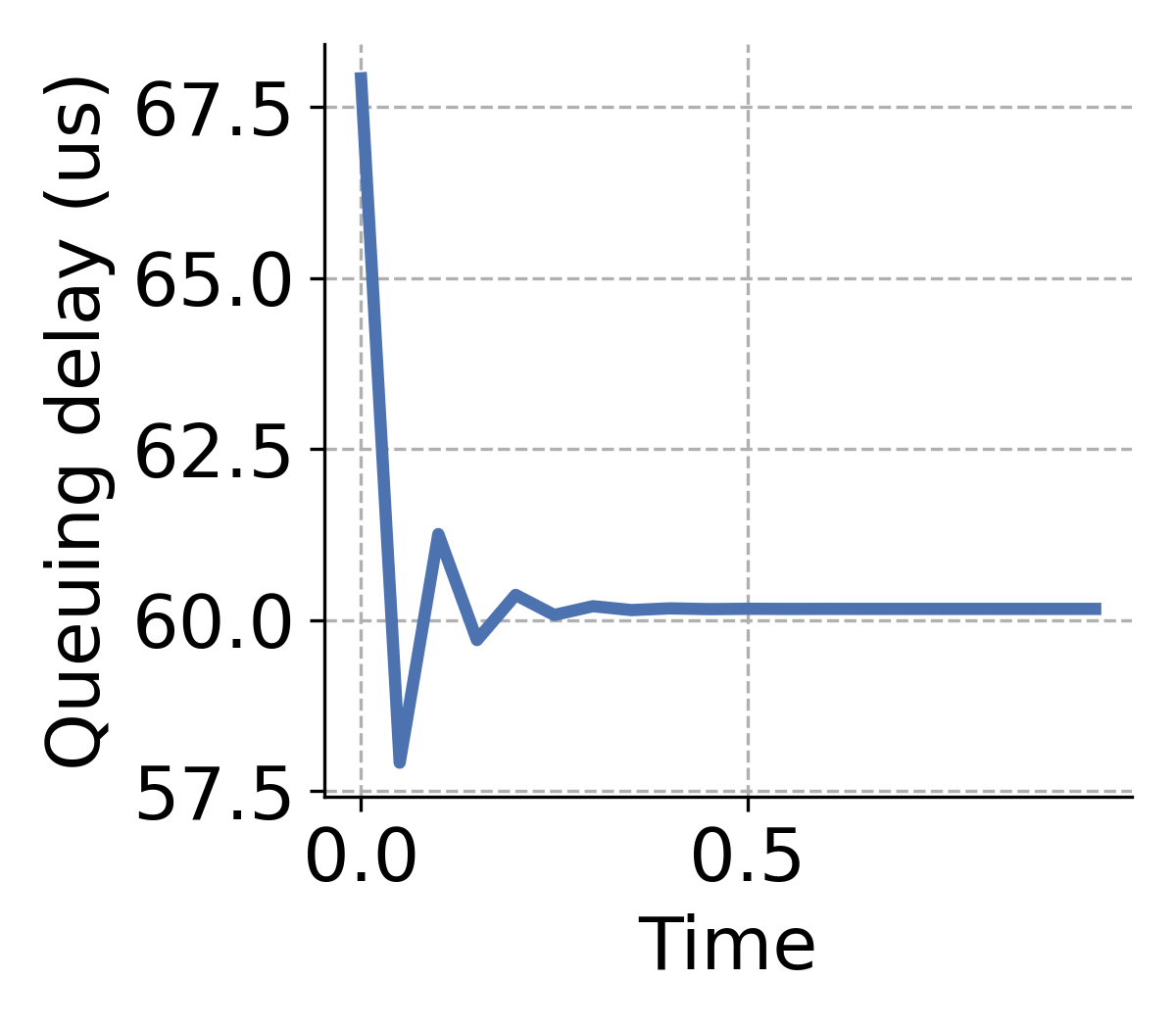}
    \caption{Convergence with oscillation}
    \label{fig:conv_with_osci}
\end{figure}

If the queuing delay is allowed to converge to the target delay with oscillation as \fref{fig:conv_with_osci}, then we require:

\begin{equation}
    - |D(i) - D(\infty)| < D(i+1) - D(\infty) < |D(i) - D(\infty)|
\end{equation}

Without loss of generality, we assume $D(i) > D(\infty)$, namely, $s(i) < s(\infty)$:

\begin{equation}
    2\cdot D(\infty) - D(i) < D(i+1) < D(i)
\label{equ:with_osci}
\end{equation}

For RHS:

\begin{equation}
\begin{aligned}
    D(i+1) &< D(i) \\
    D(i) + \frac{\Delta t}{B} \cdot \left[W\cdot s(i) - B\right] &< D(i) \\
    \frac{\Delta t}{B} \cdot \left[W\cdot s(i) - B\right] &< 0 \\
    W\cdot s(i) - B &< 0 \\
    s(i) &< \frac{B}{W} \\
    s(i) &< s(\infty) \\
\end{aligned}
\end{equation}

Thus, we can prove that the RHS is satisfied.

For LHS:

\begin{equation}
\begin{aligned}
    D(i+1) &> 2\cdot D(\infty) - D(i) \\
    D(i) + \frac{\Delta t}{B} \cdot \left[W\cdot s(i) - B\right] &> 2\cdot D(\infty) - D(i) \\
    \frac{\Delta t}{B} \cdot \left[W\cdot s(i) - B\right] &> 2\cdot \left[D(\infty) - D(i)\right] \\
    W\cdot s(i) - B &> \frac{2\cdot B}{\Delta t} \cdot \left[D(\infty) - D(i)\right]
\end{aligned}
\end{equation}

We know in the converged state, we have $B = W\cdot s(\infty)$. Thus, we can have:

\begin{equation}
\begin{aligned}
    W\cdot s(i) - B &> \frac{2\cdot B}{\Delta t} \cdot \left[D(\infty) - D(i)\right] \\
    W\cdot s(i) - W\cdot s(\infty) &> \frac{2\cdot B}{\Delta t} \cdot \left[D(\infty) - D(i)\right] \\
    W\cdot [s(i) - s(\infty)] &> \frac{2\cdot B}{\Delta t} \cdot \left[D(\infty) - D(i)\right] \\
    W\cdot \alpha\cdot [A^{D(i)}- A^{D(\infty)}] &> \frac{2\cdot B}{\Delta t} \cdot \left[D(\infty) - D(i)\right] \\
\end{aligned}
\end{equation}

Thus, we have: 

\begin{equation}
\begin{aligned}
    \frac{A^{D(i)} - A ^ {D(\infty)}}{D(i)- D(\infty)} &> -\frac{2\cdot B}{\alpha \cdot W\cdot \Delta t} \\
    \frac{A^{D(i)} - A ^ {D(\infty)}}{D(i)- D(\infty)} &> -\frac{2\cdot  A^{D(\infty)}}{\Delta t} \\
    \frac{\frac{A^{D(i)}}{A^{D(\infty)}} - 1}{D(i)- D(\infty)} &> -\frac{2}{\Delta t} \\
    \frac{A^{D(i)-D(\infty)} - 1}{D(i)- D(\infty)} &> -\frac{2}{\Delta t} \\
\end{aligned}
\end{equation}


We can observe that $\frac{A^{D(i)-D(\infty)} - 1}{D(i)- D(\infty)}$ is a function with variable $D(i)-D(\infty)>0$. Denote function $K(x) = \frac{A^x-1}{x}$. Easily, we can know that the value of $K(x)$ is minimum when $x$ is close to 0.

Because $A^x-1=0$ when $x=0$, so the value of $K(\epsilon)$ should be just the derivative of $A^x-1$:

\begin{equation}
\begin{aligned}
    K(\epsilon) &= \frac{A^\epsilon-1}{\epsilon} \\
    &= (A^\epsilon-1)' \\
    &= \ln(A)\cdot A^\epsilon \\
    &= \ln(A)
\end{aligned}
\end{equation}

Thus, we only need to satisfy that:

\begin{equation}
\begin{aligned}
    \ln(A) &> -\frac{2}{\Delta t} \\
    \frac{1}{p}\cdot \ln\left(\frac{\beta}{\alpha}\right) &> -\frac{2}{\Delta t} \\
    \frac{p}{\Delta t} &> \frac{1}{2} \cdot \ln\left(\frac{\alpha}{\beta}\right) \\
\end{aligned}
\end{equation}

If we assume $\alpha = 100 Gbps$ and $\beta = 0.1 Gbps$, which is a wide range of flow rates:

\begin{equation}
\begin{aligned}
    \frac{p}{\Delta t} > 0.5 \cdot \ln(1000) = 3.45
\end{aligned}
\end{equation}

Thus, as long as $p$, which is the range of the target delay values (3 to 40 us in our previous equation), is large enough or the update interval $\Delta t$ is small enough, the LHS can be satisfied.

\subsection{Convergence without Oscillation}

\begin{figure}[h]
    \centering
    \includegraphics[width=0.4\textwidth]{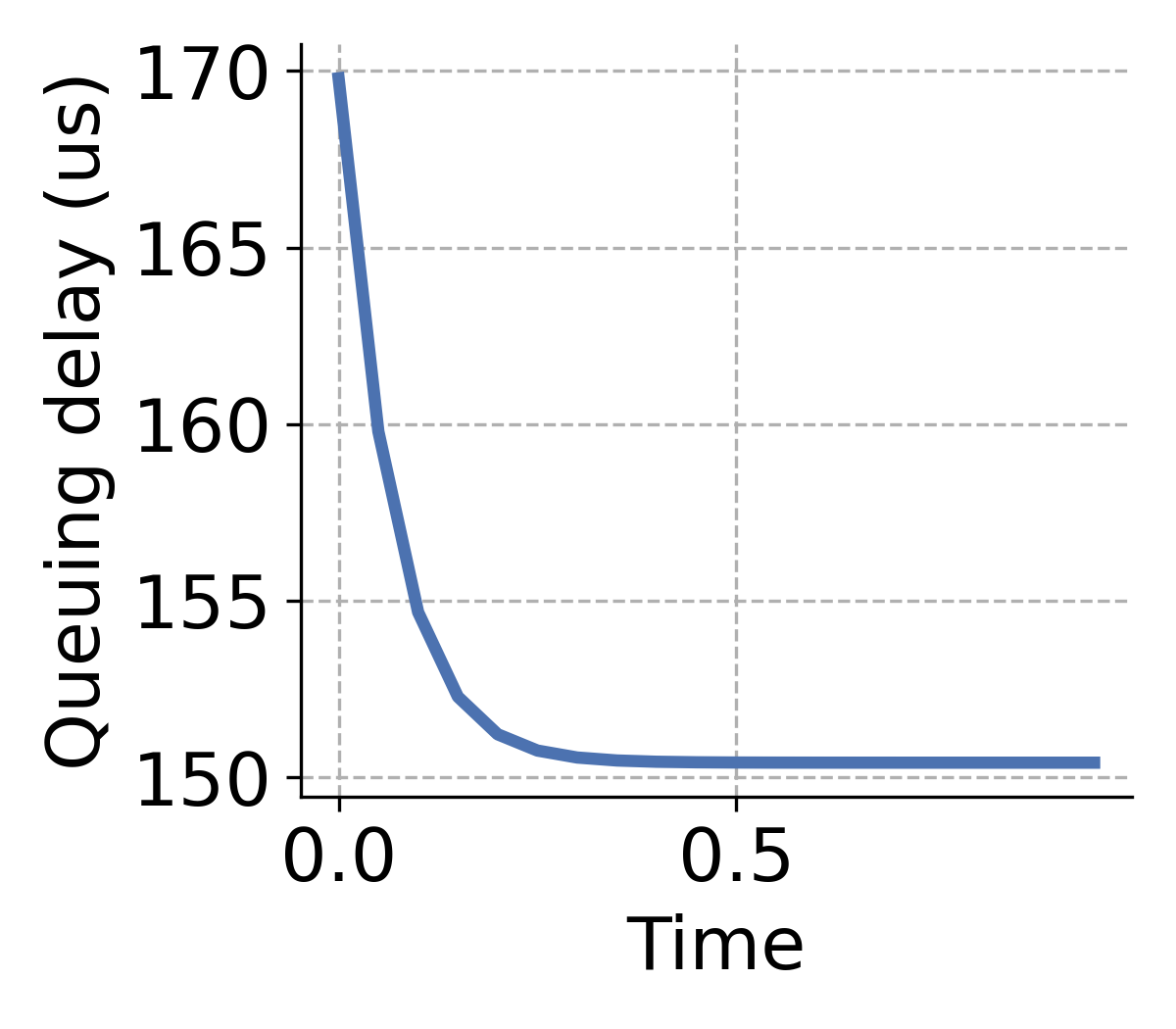}
    \caption{Convergence with oscillation}
    \label{fig:conv_without_osci}
\end{figure}

As \fref{fig:conv_without_osci} shows, convergence without oscillation is a stronger requirement than convergence with oscillation. 

Instead of satisfying the inequality \eref{equ:with_osci}, we need to satisfy the following inequality (without loss of generality, we assume $D(i) > D(\infty)$):

\begin{equation}
    D(\infty) < D(i+1) < D(i)
\end{equation}

For RHS, the proof is similar:

\begin{equation}
\begin{aligned}
    D(i+1) &< D(i) \\
    D(i) + \frac{\Delta t}{B} \cdot \left[W\cdot s(i) - B\right] &< D(i) \\
    \frac{\Delta t}{B} \cdot \left[W\cdot s(i) - B\right] &< 0 \\
    W\cdot s(i) - B &< 0 \\
    s(i) &< \frac{B}{W} \\
    s(i) &< s(\infty) \\
\end{aligned}
\end{equation}

We proved that the RHS is satisfied.

For LHS:

\begin{equation}
\begin{aligned}
    D(i+1) &> D(\infty) \\
    D(i) + \frac{\Delta t}{B} \cdot \left[W\cdot s(i) - B\right] &> D(\infty) \\
    \frac{\Delta t}{B} \cdot \left[W\cdot s(i) - B\right] &> D(\infty) - D(i) \\
    W\cdot s(i) - B &> \frac{B}{\Delta t} \cdot \left[D(\infty) - D(i)\right]
\end{aligned}
\end{equation}

We know in the converged state, we have $B = W\cdot s(\infty)$. Thus, we can have:

\begin{equation}
\begin{aligned}
    W\cdot s(i) - B &> \frac{B}{\Delta t} \cdot \left[D(\infty) - D(i)\right] \\
    W\cdot s(i) - W\cdot s(\infty) &> \frac{B}{\Delta t} \cdot \left[D(\infty) - D(i)\right] \\
    W\cdot [s(i) - s(\infty)] &> \frac{B}{\Delta t} \cdot \left[D(\infty) - D(i)\right] \\
    W\cdot \alpha\cdot [A^{D(i)}- A^{D(\infty)}] &> \frac{B}{\Delta t} \cdot \left[D(\infty) - D(i)\right] \\
\end{aligned}
\end{equation}

Thus, we have: 

\begin{equation}
\begin{aligned}
    \frac{A^{D(i)} - A ^ {D(\infty)}}{D(i)- D(\infty)} &> -\frac{B}{\alpha \cdot W\cdot \Delta t} \\
    \frac{A^{D(i)} - A ^ {D(\infty)}}{D(i)- D(\infty)} &> -\frac{A^{D(\infty)}}{\Delta t} \\
    \frac{\frac{A^{D(i)}}{A^{D(\infty)}} - 1}{D(i)- D(\infty)} &> -\frac{1}{\Delta t} \\
    \frac{A^{D(i)-D(\infty)} - 1}{D(i)- D(\infty)} &> -\frac{1}{\Delta t} \\
\end{aligned}
\end{equation}


We can observe that $\frac{A^{D(i)-D(\infty)} - 1}{D(i)- D(\infty)}$ is a function with variable $D(i)-D(\infty)>0$. Denote function $K(x) = \frac{A^x-1}{x}$. Easily, we can know that the value of $K(x)$ is minimum when $x$ is close to 0.

Because $A^x-1=0$ when $x=0$, so the value of $K(\epsilon)$ should be just the derivative of $A^x-1$:

\begin{equation}
\begin{aligned}
    K(\epsilon) &= \frac{A^\epsilon-1}{\epsilon} \\
    &= (A^\epsilon-1)' \\
    &= \ln(A)\cdot A^\epsilon \\
    &= \ln(A)
\end{aligned}
\end{equation}

Thus, we only need to satisfy that:

\begin{equation}
\begin{aligned}
    \ln(A) &> -\frac{1}{\Delta t} \\
    \frac{1}{p}\cdot \ln\left(\frac{\beta}{\alpha}\right) &> -\frac{1}{\Delta t} \\
    \frac{p}{\Delta t} &> \ln\left(\frac{\alpha}{\beta}\right) \\
\end{aligned}
\end{equation}

Thus, as long as $p$, which is the range of the target delay values (3 to 40 us in our previous equation), is large enough or the update interval $\Delta t$ is small enough, the LHS can be satisfied.

Compared to convergence with oscillation, convergence without oscillation has a more strict requirement on $\frac{p}{\Delta t}$, which is twice higher than convergence with oscillation.
\section{Proof of Convergence to Weighted Max-min Fairness in Arbitrary Network}
\label{appendix:bottleneck-signal}

Firstly, we give the definition of max-min fairness:

\begin{defnapp}[Weighted Max-min Fair~\cite{marbach2002priority, allalouf2008centralized}]{} 
For all flows $\{f1, ..., fn\}$ in the network, denote their weight to be $\{w_{f1}, ..., w_{fn}\}$. A rate allocation $\{r_{f1}, ..., r_{fn}\}$ is weighted max-min fair when for each flow $f$, any increase in $r_f$ would cause a decrease in the transmission rate for some flow $f'$ satisfying $\frac{r_{f'}}{w_{f'}} \le \frac{r_f}{w_f}$.
\end{defnapp}

In this section, we will prove that the maxQD signal must come from the bottleneck hop, and thus we achieve weighted max-min fair allocation.

To prove the above lemma, we need to prove that the allocation for any arbitrary network is unique and the bottleneck hop can be identified by the maxQD signal. Thus, the proof is divided into three parts: 
    1) Prove that the bandwidth allocation is unique for any arbitrary network; 
    2) Prove that in weighted max-min fairness, a flow must have the largest rate on its bottleneck hop;
    3) Prove that only the signal from the bottleneck hop will be collected in \oursystemsoze, which is required to achieve weighted max-min fairness.

Once we have proved that the maxQD signal must only come from the bottleneck hop, the proofs for the single-switch scenario in Appendix \ref{appendix:fairness-single} and Appendix \ref{appendix:util-single} can be directly used to show the properties can be achieved on the bottleneck hop in an arbitrary network, namely, the weighted max-min fairness can be achieved.

\subsection{Weighted Max-min Fair is Unique}

Assume a feasible weighted max-min fair allocation of rates $\vec{x}$ exists, so that every flow has a specific bottleneck link, which is the link with the highest queueing signal.

\begin{lemapp}[Unique Bandwidth Allocation]{lem:unique}
    For the weighted max-min fair bandwidth allocation, each link in the network has a specific weighted fair-share rate.
\end{lemapp}

Assume there exists another allocation $\vec{y}$, which changes the bottleneck hop for a flow $i$. Its rate was previously $x_i$, but now is $y_i$ in the new allocation. Denote the bottleneck hop for this flow in $\vec{x}$ as $\gamma$ and the new bottleneck hop in $\vec{y}$ as $\gamma'$. Denote the weighted fair-share rate for hop $\gamma$ in allocation $\vec{x}$ as $R_x(\gamma)$. 

Then in allocation $\vec{x}$, we can easily conclude that 

\begin{equation}
    \frac{x_i}{w_i} = R_x(\gamma) < R_x(\gamma')
\end{equation}

In the new allocation, WLOG, we assume that the new flow rate $y_i$ is smaller than $x_i$, namely $y_i < x_i$. Because the flow rate $i$' decreases, the rate of the other flows must increase. Thus, we can easily have the following.

\begin{equation}
    R_x(\gamma') < \frac{y_i}{w_i} = R_y(\gamma') < \frac{x_i}{w_i}
\end{equation}

With the contradiction that $\frac{x_i}{w_i} < R_x(\gamma')$ and $\frac{x_i}{w_i} > R_x(\gamma')$ cannot be satisfied at the same time, we proved that the bottleneck hop for each flow is unique and the weighted max-min fair allocation is unique. During the whole proof, the link bandwidth is not relevant, so the conclusion can be generalized to links with heterogeneous bandwidth.

\subsection{Bottleneck Hop Properties}

From the definition of the weighted max-min fair, we can derive a lemma that reveals that each flow's rate-per-weight is the largest among flows sharing its bottleneck hop.

\begin{lemapp}[Bottleneck Hop Properties]{}
    When achieving weighted max-min fair, each flow will have the largest rate-per-weight among all flows on its bottleneck hop and not on any other saturated hop. 
\end{lemapp}

Formally, for the ``weighted max-min fair'' allocation $\vec{x}$, for any flow $f$, denote the flows shared the same bottleneck with $f$ as ${b_1,b_2,...,b_k}$. For any flow $b_i$, $\frac{x_f}{w_f} \ge \frac{x_{b_i}}{w_{b_i}}$. Denote the flow's share on the saturated non-bottleneck hop of $f$ as ${c_1,c_2,...,c_k}$, then there must exist some $c_j$ such that $\frac{x_{c_j}}{w_{c_j}} > \frac{x_f}{w_f}$.

Assume that there exists a flow $f$ that has reached its weighted fair-share $\frac{r}{w}$, and there is another flow $f'$ on its bottleneck hop with an even larger rate-per-weight $\frac{r'}{w}$. But this state is not max-min fair because flow $f$ could get some bandwidth from flow $f'$ and let them have the same rate-per-weight. By contradiction, the flow $f$ has the largest rate on its bottleneck hop. 

On the other hand, assume there exists a flow $f$, which is the fastest flow, with weighted fair share $\frac{r}{w}$, on one of the non-bottleneck hops. However, given that this link is congested, its fair-share flows with a weight per weight rate $\frac{r'}{w}$ could obtain the bandwidth from the flow $s$ and increase their fair-share rate to at least $\frac{r+W\cdot r'}{w\cdot (W+1)}$, where $W$ is the sum of the flow weights. By contradiction, the flow $f$ cannot be the largest flow on its non-bottleneck hop.

\subsection{maxQD must Come from Bottleneck}

With the above lemma, we can prove that the maximum queueing signal must come from the bottleneck hop, and weighted fairness and target queueing are achieved on the bottleneck hop. 

\begin{lemapp}[Bottleneck Signal]{}
    In \oursystemsoze, the maxQD signal must come from the bottleneck hop of each flow.
\end{lemapp}

In the converged state, only on the bottleneck hop, a flow has the largest sending rate. So, on all other hops, this flow is not the fastest flow. Given that the queueing signal decreases when the fair-share rate increases, all other hops have a smaller queueing signal than the bottleneck hop. Thus, the maximum queueing signal must come from the bottleneck hop.

With the queueing signal from the bottleneck hop, only the flows that are bottlenecked by that hop will share the available bandwidth fairly. Thus, for each flow, \oursystemsoze identifies the bottleneck hop from all other hops and achieves fair allocations on the bottleneck.

\subsection{Achieve Weighted Max-min Fairness}

\begin{theoapp}[Weighted Max-min Fairness]{}
    For every flow in an arbitrary network, \oursystemsoze converges to a weighted max-min fair allocation, if and only if $0 < m < 2$ in the update function and $p>\frac{\Delta t}{2}\cdot\left[\ln(\alpha) - \ln(\beta)\right]$ in the target function.
\end{theoapp}

Since the queueing signal always comes from the bottleneck hop and \oursystemsoze reacts only to the bottleneck signal, the weighted fairness will be achieved among all the flows sharing the same bottleneck. And if all the flows converge to the weighted fairness on their bottleneck, according to the definition, the weighted max-min fair has been achieved.

\end{document}